\documentclass[12pt]{iopart}
\usepackage{iopams} 
\usepackage{graphics}
\usepackage{pennames}
\usepackage{amssymb}
\usepackage{setstack}
\usepackage{color}

\def\mt{m_{\tt T}}
\def\Mt{m_{\tt T}}
\def\pt{p_{\tt T}}
\def\Kt{K_{\tt T}}
\def\Ro{R_{\tt out}}
\def\Rs{R_{\tt side}}
\def\Rl{R_{\tt long}}
\def\Rol{R_{\tt ol}}
\def\Qo{q_{\tt o}}
\def\Qs{q_{\tt s}}
\def\Ql{q_{\tt l}}
\def\hm{\tt h^-}
\def\diffD{{\rm d}}
\def\Bt{\langle\beta_\perp\rangle}

\begin{document}
\title[Expansion dynamics of Pb--Pb collisions at 40 $A$\ GeV/$c$]
{Expansion dynamics  of   
 Pb--Pb collisions at 40 $A$\ GeV/$c$\ 
 viewed by 
 negatively charged hadrons} 
\author{  
F~Antinori$^{1}$,
P~Bacon$^{2}$,
A~Badal{\`a}$^{3}$,
R~Barbera$^{3}$,
A~Belogianni$^{4}$,
I~J~Bloodworth$^{2}$,
M~Bombara$^{5}$,
G~E~Bruno$^{6}$,
S~A~Bull$^{2}$,
R~Caliandro$^{6}$,
M~Campbell$^{7}$,
W~Carena$^{7}$,
N~Carrer$^{7}$,
R~F~Clarke$^{2}$,
A~Dainese$^{1}$\footnote[1]{Present address: Laboratori Nazionali di Legnaro, Legnaro, Italy},
D~Di~Bari$^{6}$,
S~Di~Liberto$^{8}$,
R~Divi\`a$^{7}$,
D~Elia$^{6}$,
D~Evans$^{2}$,
G~A~Feofilov$^{9}$,
R~A~Fini$^{6}$,
P~Ganoti$^{4}$,
B~Ghidini$^{6}$,
G~Grella$^{10}$,
H~Helstrup$^{11}$,
K~F~Hetland$^{11}$,
A~K~Holme$^{12}$,
A~Jacholkowski$^{3}$,
G~T~Jones$^{2}$,
P~Jovanovic$^{2}$,
A~Jusko$^{2}$,
R~Kamermans$^{13}$,
J~B~Kinson$^{2}$,
K~Knudson$^{7}$,
V~Kondratiev$^{9}$,
I~Kr\'alik$^{5}$,
A~Krav\v c\'akov\'a$^{14}$,
P~Kuijer$^{13}$,
V~Lenti$^{6}$,
R~Lietava$^{2}$,
G~L\o vh\o iden$^{12}$,
V~Manzari$^{6}$,
M~A~Mazzoni$^{8}$,
F~Meddi$^{8}$,
A~Michalon$^{15}$,
M~Morando$^{1}$,
P~I~Norman$^{2}$,
A~Palmeri$^{3}$,
G~S~Pappalardo$^{3}$,
B~Pastir\v c\'ak$^{5}$,
R~J~Platt$^{2}$,
E~Quercigh$^{1}$,
F~Riggi$^{3}$,
D~R\"ohrich$^{16}$,
G~Romano$^{10}$,
R~Romita$^{6}$,
K~\v{S}afa\v{r}\'{\i}k$^{7}$,
L~\v S\'andor$^{5}$,
E~Schillings$^{13}$,
G~Segato$^{1}$,
M~Sen\'e$^{17}$,
R~Sen\'e$^{17}$,
W~Snoeys$^{7}$,
F~Soramel$^{1}$\footnote[2]{Permanent
address: University of Udine, Udine, Italy},
M~Spyropoulou-Stassinaki$^{4}$,
P~Staroba$^{18}$,
R~Turrisi$^{1}$,
T~S~Tveter$^{12}$,
J~Urb\'{a}n$^{14}$,
P~van~de~Ven$^{13}$,
P~Vande~Vyvre$^{7}$,
A~Vascotto$^{7}$,
T~Vik$^{12}$,
O~Villalobos~Baillie$^{2}$,
L~Vinogradov$^{9}$,
T~Virgili$^{10}$,
M~F~Votruba$^{2}$,
J~Vrl\'{a}kov\'{a}$^{14}$\ and
P~Z\'{a}vada$^{18}$
}
\address{
$^{1}$ University of Padua and INFN, Padua, Italy\\
$^{2}$ University of Birmingham, Birmingham, UK\\
$^{3}$ University of Catania and INFN, Catania, Italy\\
$^{4}$ Physics Department, University of Athens, Athens, Greece\\
$^{5}$ Institute of Experimental Physics, Slovak Academy of Science,
              Ko\v{s}ice, Slovakia\\
$^{6}$ Dipartimento IA di Fisica dell'Universit{\`a}
       e del Politecnico di Bari and INFN, Bari, Italy \\
$^{7}$ CERN, European Laboratory for Particle Physics, Geneva, Switzerland\\
$^{8}$ University ``La Sapienza'' and INFN, Rome, Italy\\
$^{9}$ State University of St. Petersburg, St. Petersburg, Russia\\
$^{10}$ Dipartimento di Scienze Fisiche ``E.R. Caianiello''
       dell'Universit{\`a} and INFN, Salerno, Italy\\
$^{11}$ H{\o}gskolen i Bergen, Bergen, Norway\\
$^{12}$ Fysisk Institutt, Universitetet i Oslo, Oslo, Norway\\
$^{13}$ Utrecht University and NIKHEF, Utrecht, The Netherlands\\
$^{14}$ P.J. \v{S}af\'{a}rik University, Ko\v{s}ice, Slovakia\\
$^{15}$ IReS/ULP, Strasbourg, France\\
$^{16}$ Fysisk Institutt, Universitetet i Bergen, Bergen, Norway\\
$^{17}$ Coll\`ege de France, Paris, France\\
$^{18}$ Institute of Physics, Prague, Czech Republic 
}
\ead{Giuseppe.Bruno@ba.infn.it, Rossella.Romita@ba.infn.it}
\begin{abstract}
In this paper we present results on transverse mass spectra and Hanbury-Brown and Twiss correlation 
functions of negatively charged hadrons, which are expected to be mostly \Pgpm, 
measured in Pb--Pb collisions at 40 $A$\ GeV/$c$\ beam momentum. 
Based on these data, the collision dynamics and the space-time extent of the system at the 
thermal freeze-out are studied over a centrality range corresponding to the most central 53\%
of the Pb--Pb inelastic cross section. Comparisons with freeze-out conditions of strange particles 
and HBT results from other experiments are discussed.  
\end{abstract}
%
%
%
\section{Introduction} 
Ultra-relativistic collisions between 
heavy ions are used to study the properties of nuclear matter 
at high energy density. In particular, 
lattice QCD calculations predict a transition from confined hadronic 
matter to a state of deconfined quarks and gluons at a critical energy density 
around 1 GeV/fm$^3$~\cite{lattice}.   
For recent reviews of experimental results and theoretical developements 
see references~\cite{QM04-QM05}.  


The momentum distributions of the particles emerging from the Pb--Pb interactions  
are expected to be sensitive to the collision dynamics. In particular,
collective dynamics in the transverse direction is of major interest 
since it can only arise by the buildup of a pressure gradient in 
that direction; this in turn would be strongly suggestive of thermal  
equilibration of the nuclear matter. Indeed such an effect has already been 
observed at the highest SPS and RHIC energies.   
The shapes of the $m_{\tt T}=\sqrt{\pt^2+m^2}$\ distributions are expected to be 
determined by 
an interplay between two effects: 
the thermal motion of the
particles in the fireball and a pressure-driven radial flow, induced by the
fireball expansion.  
In reference~\cite{Blast40} we have analyzed the $\mt$\ distributions  
of strange particles (\PgL, \PgXm, \PgOm, their anti-hyperons and \PKzS)   
based on the blast-wave model~\cite{BlastRef}, a parameterized model inspired to hydro-dynamics.  
Due to the large number of particle species considered in that analysis, a simultaneous fit to 
the strange particle spectra allowed to disentangle the effect of the thermal motion
from that of the collective expansion. With one particle species only, e.g. the negative pion,
an oblongated confidence region   
can be obtained for the pair of freeze-out parameters 
temperature ($T$) and average transverse flow velocity ($\Bt$). 

A systematic study of the space-time extent and the dynamical behavior of the fireball  
at thermal freeze-out can be obtained via the  
identical particle interferometry technique, first used by 
Hanbury-Brown and Twiss (HBT)~\cite{HBT}.   
The width of the correlation peak at vanishing relative momenta reflects the so-called 
length of homogeneity (also called the ``HBT radius'') of the particle emitting source. 
Only in static sources can the homogeneity length be interpreted as the true geometrical 
size of the system. In a dynamic system, the occurrence of space-momentum correlations 
of the emitted particles due to collective expansion generally leads to a 
reduction of the observed HBT radii. The degree of reduction depends on the gradients of the 
collective expansion velocity and on the thermal velocity of the particles at thermal freeze-out.  
A differential analysis of the HBT  correlations in bins of the pair momentum 
(in the longitudinal and transverse directions) thus provides valuable  information both 
on the spatial extent and on the properties of the collective expansion of the system.  
Combining single particle spectra and two particle HBT correlations 
allows us to disentangle the collective dynamics from the thermal motion 
relying on one particle species only.  

In this paper we study the transverse mass spectra and the HBT correlation functions 
of unidentified negatively charged hadrons ($\hm$), which consist mainly of    
negative pions.   

\section{\label{setup} The NA57 set-up and the data sample}
Descriptions of the NA57 apparatus   
can be found in references~\cite{enh160,BlastPaper,MANZ}, and in
reference~\cite{Blast40} in particular for the 40 $A$\ GeV/$c$\ set-up.

The tracking device of the NA57 experiment consisted of  
a telescope made of  an array of silicon detector planes of 5x5 cm$^2$\ cross-section
placed in an approximately uniform magnetic field of 1.4 Tesla perpendicular to the 
beam line;  
the bulk of the detectors was closely packed in an approximately 30 cm long
compact part used for pattern recognition.
To improve the momentum resolution of high momentum tracks a
lever arm detector (an array of four double-sided silicon micro-strip detectors)
was placed downstream of the tracking telescope.

The centrality of the Pb-Pb collisions is determined (off-line) by analyzing the
charged particle multiplicity measured by two stations of micro-strip silicon
detectors (MSD) which sample the pseudo-rapidity intervals $1.9<\eta<3$\ and $2.4<\eta<3.6$.  

The results presented in this paper are based on the analysis of the  
data sample collected in Pb--Pb collisions at 40 $A$\ GeV/$c$.  
The selected sample of events corresponds to the 
most central 53\% of the inelastic Pb--Pb cross-section and   
has been divided into five centrality classes (labelled with integers 0, 1, 2, 3, and 4,   
class 4 being the most central) according to the value of the charged particle 
multiplicity  measured by the MSD. 
The procedure for the measurement of the multiplicity distribution and
the determination of the collision centrality for each class
is described in reference~\cite{Multiplicity}.
The fractions of the inelastic cross-section for the five classes, calculated 
assuming 
an  inelastic Pb--Pb 
cross-section of 7.26 barn, are the same as those 
used for the study of strange particles~\cite{Blast40}  
and 
are given in table~\ref{tab:centrality}.  
\begin{table}[h]
\caption{Centrality ranges for the five classes.
\label{tab:centrality}}
\begin{center}
\begin{tabular}{llllll}
\hline
 Class &   $0$   &   $1$   &   $2$   &  $3$   &   $4$ \\ \hline
 $\sigma/\sigma_{inel}$\ \; (\%)   & 40 to 53 & 23 to 40 & 11 to 23& 4.5 to 11 & 0 to 4.5 \\
 \hline
\end{tabular}
\end{center}
\end{table}
\noindent

Negatively charged tracks have been selected by requiring them to have
clusters in more than 80\% of the telescope planes and 
less than 30\% of the clusters shared with other
tracks, and 
using an impact parameter cut\footnote{
The impact parameter is approximated as the distance from the primary vertex 
of the intersection of the measured particle trajectory
with a plane transverse to the beam line passing through the target position.
} 
to ensure they come from the main  interaction vertex.

\section{\label{single} Single particles $\mt$\ spectra}  

\begin{figure}[hbt]
\centering
\resizebox{0.34\textwidth}{!}{%
\includegraphics{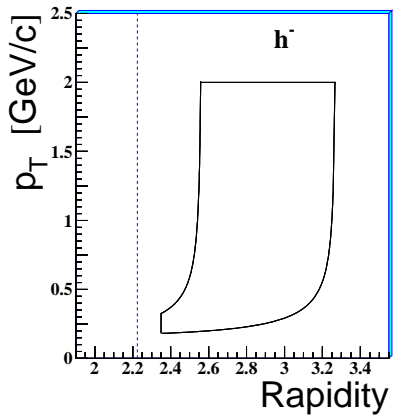}}\\
\caption{The $y$--$\pt$\ acceptance window of negative hadrons.
         Dashed lines show the position of mid-rapidity ($y_{\tt cm}=2.225$).}
\label{fig:acceptance}
\end{figure}

The acceptance region in the transverse momentum ($\pt$) versus rapidity ($y$) plane
is shown in figure~\ref{fig:acceptance}\footnote{The rapidity has been evaluated 
assuming the pion mass.}. 
The limits of this window have been
defined in order to exclude from the final sample the particles whose
lines of flight are very close to the border limits of the telescope, where the 
systematic errors are more difficult to evaluate.
                                                                                                                                      
A weight is assigned to each reconstructed $\hm$\ to correct for
acceptance and reconstruction inefficiencies. The computational algorithm
of the event weight is the same as that used for strange  
particles~\cite{enh160,BlastPaper,Blast40}: a number of Monte Carlo events are generated, 
each event consisting of one simulated particle, with the same momentum 
of the real particle, merged with  
a real event of similar telescope hit multiplicity as the original event, and they
are reconstructed with the same analysis tools as for real events.  
A total of about $10,000$\ 
$\hm$,  
sampled uniformly over the full data taking periods, have  
been individually weighted with this method.  
                                                                                             
In order to check the stability of the results,   
the selection criteria (i.e. the impact parameter cut and the number  
of clusters associated to the track or shared with another track) have been  
varied, by changing their values one at a time.  
As a result of these studies we can estimate the contribution of the selection and
correction procedure to the systematic errors on the slope of the   
$1/\mt \, \diffD N/\diffD\mt$\
distributions of $\hm$\ to be about 7\%.  

The experimental procedure for the determination of the $m_{\tt T}$\
distribution is described in detail in references~\cite{BlastPaper,Blast40},  
where the results for strange particles are shown.  

The distribution of the double differential invariant cross-section
$\frac{1}{\mt}\frac{d^2N}{dm_{\tt T} dy} $\ has been assumed to factorize into
a $y$\ and an $\mt$\ dependent part:
\begin{equation}
\label{eq:factorize}
\frac{1}{\mt}\frac{\diffD^2N}{\diffD\mt\,\diffD y }(y,\mt) = f(y) \cdot
    \frac{1}{\mt}\frac{ \diffD N }{\diffD \mt} (\mt)   .
\end{equation}
This assumption has been verified by considering
the $\mt$\ and $y$\  distributions for different slices, respectively, in rapidity  
and transverse mass.  
The shape of the rapidity distribution ($f(y)$\ in equation~\ref{eq:factorize})
has been found to be well described by a Gaussian within our limited range.  
The hypotheses on the factorization of the double differential invariant
cross-section  (equation~\ref{eq:factorize}) and on the shape of the
rapidity distributions ($f(y)$) can introduce a contribution to the 
systematic error on the slope of the
$\frac{1}{\mt}\frac{ \diffD N }{\diffD \mt} (\mt) $\
distributions which has been estimated to be about 5\%.

\begin{figure}[hbt]
\centering
\resizebox{0.40\textwidth}{!}{%
\includegraphics{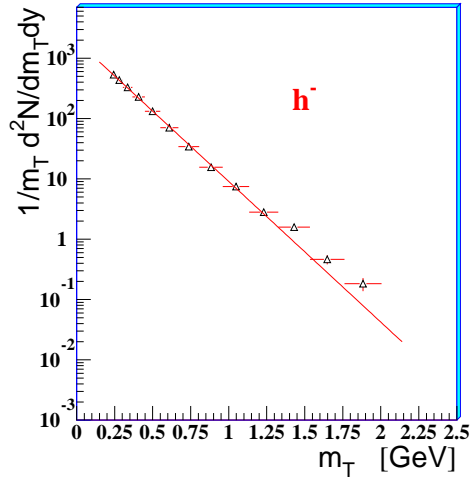}}\\
\caption{Transverse mass spectrum of negatively charged particles 
 for the most central 53\% of the Pb--Pb inelastic cross-section.
 The superimposed exponential function has inverse slope equal to the
 $T_{\tt app}$\ value obtained from the maximum likelihood fit.}
\label{fig:all_spectra}
\end{figure}

The $\mt$\ distribution has been parameterized as 
$\frac{1}{\mt}\frac{ \diffD N }{\diffD \mt} \propto \exp\left(-\frac{m_{\tt T}}{T_{\tt app}}\right)$.  
The inverse slope parameter $T_{\tt app}$\ (``apparent temperature'')  
has been extracted by means of a maximum likelihood fit 
of the measured double differential invariant cross-section 
to the formula  
\begin{equation}
\frac{1}{\mt}\frac{d^2N}{dm_{\tt T} dy} = 
   f(y) \hspace{1mm} \exp\left(-\frac{m_{\tt T}}{T_{\tt app}}\right) 
\label{eq:InvSlope}
\end{equation}
obtaining $T_{\tt app}=186 \pm 2 (\rm stat) \pm 16 (\rm syst)$\ MeV    
for the most central 53\% of the inelastic Pb--Pb cross-section.     
The differential invariant cross-section distribution is shown in 
figure~\ref{fig:all_spectra} as a function of $\mt$\ 
with superimposed the likelihood fit result.   

In the hydro-dynamical view, the apparent temperature 
is interpreted as due to the thermal motion coupled with a collective transverse flow 
of the fireball components~\cite{BlastRef}, and it 
depends on $\mt$\footnote{In this view, 
the graphical interpretation of $T_{\tt app}$\ would be the inverse of the 
local tangent to the invariant $1/\mt \, \diffD N / \diffD \mt$\ distribution.  
See reference~\cite{BlastPaper} for a detailed discussion.}.  
At a given ${\mt}_0$\ value, it can be calculated according to the 
formula~\cite{BlastRef}:  
\begin{equation}
T_{\tt app}({\mt}_0)=\left[\lim_{\mt \rightarrow {\mt}_0}
\frac{\diffD}{\diffD\mt}(\log \frac{\diffD N}{\diffD \mt^2})  \right]^{-1}
\label{eg:Tapp}
\end{equation} 
This  
expression simplifies  
for $\mt \gg m$: the apparent temperature   
is simply blue-shifted by the collective dynamics 
\begin{equation}
T_{\tt app}= T \sqrt{\frac{1+\Bt}{1-\Bt}}  
\label{eq:BlueShift}
\end{equation}  
where $T$\ is the  freeze-out temperature and $\Bt$\ is the average  
transverse flow velocity.  

\subsection{Centrality dependence}
\begin{figure}[hbt]
\centering
\resizebox{0.40\textwidth}{!}{%
\includegraphics{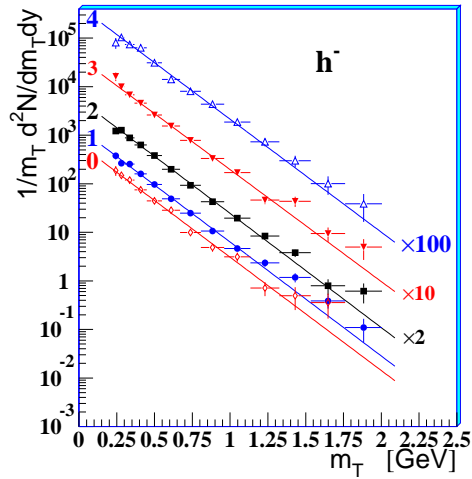}}\\
\caption{  Transverse mass spectra of negatively charged particles
           in Pb--Pb collisions at 40 $A$\ GeV/$c$\ for
           the five centrality classes of table~\ref{tab:centrality}.
           The spectra of class $2$, $3$\ and $4$ have been scaled by factors $2$, $10$\
           and $100$, respectively, for display purposes.
}
\label{fig:mt_centr}
\end{figure}

The transverse mass spectra of negative hadrons are shown in figure~\ref{fig:mt_centr} 
for the five centrality classes defined in table~\ref{tab:centrality}:  
the inverse slope parameter $T_{\tt app}$\ 
does not depend on centrality and its values are given in table~\ref{tab:InvMSD}.
\begin{table}[h]
\caption{
         Inverse slopes (MeV) of the  
         $m_{\tt T}$\ distributions of  $\hm$\ in Pb--Pb collisions at 40 $A$\ GeV/$c$\ 
         for the five centrality classes defined in table~\ref{tab:centrality}.  
	 Systematic errors are estimated to be 8.5\% for all centralities.   
\label{tab:InvMSD}}
\begin{center}
\footnotesize{
\begin{tabular}{|c|c|c|c|c|} 
\hline
     0      &    1      &    2      &    3    &    4   \\ \hline
 $186\pm6$ & $185\pm3$ & $185\pm3$ & $188\pm3$ & $187\pm4$ \\
\hline
\end{tabular}
}
\end{center}
\end{table}

\section{\label{hbt} Two-particle correlation}
In this section we present the study of the Hanbury-Brown and Twiss correlation for 
negatively charged hadrons assumed to be pions (see section~\ref{conta}).  
An introduction to this topic and recent 
reviews of experimental results and theoretical  
developements 
can be found in references~\cite{RevHBT1,RevHBT2}.  

The data will be analyzed in the framework of the hydro-dynamical inspired  
blast-wave model~\cite{BlastRef}, which assumes cylindrical symmetry for an 
expanding fireball in local thermal equilibrium.   
Hydro-dynamics or parameterized models inspired by hydro-dynamics
have shown to give a successful description 
of a number of observables, i.e. transverse momentum ($p_{\tt T}$) and 
rapidity ($y$) distributions, direct and elliptical flow, 
two-particle correlation functions (for recent reviews see,  e.g.,  
references~\cite{ReviewHydro}).
The single set of free parameters of the model to be obtained from the HBT study is:   
the kinetic freeze-out temperature ($T$), the average transverse 
flow velocity ($\Bt$), the Gaussian radius of the cylindrical
system ($R_{\rm G}$), the chaoticity of the pion emission ($\lambda$),  
the proper time  of the freeze-out  ($\tau_f=\sqrt{t^2_f-x^2_f}$)  
and the emission duration ($\Delta\tau$).

Experimentally, the correlation function can be defined as  
\begin{equation}
C_2(q)= N \frac{S(q)}{B(q)}
\nonumber
\end{equation}
where the signal $S(q)$ is the measured distribution  
of the relative four-momentum $q=p_1-p_2$\ of two identical particles in one event  
and the background $B(q)$\ 
is the reference distribution built by pairing two particles taken from
different events. For each pair of negative hadrons in the signal,  
there are 
about 15 pairs in the background, formed from events  
of similar multiplicity, so that the error on $C_2$ is statistically 
dominated by the signal. The normalization factor $N$\ is obtained by 
imposing that the integral of $C_2(q)$\ be equal to unit in the region of 
large $q$, where there is neither 
quantum (i.e. Bose-Einstein)  
nor other (e.g. Coulomb) correlations.  

The relative momentum $q$\ is measured on a pair-by-pair basis 
relative to the {\em out--side--long} reference,  
which is a Cartesian system defined  by choosing the {\em long} axis   
along the beam direction; the {\em out} and {\em side} axes lay in the transverse plane with the 
former aligned with the average transverse momentum of the pair and the latter 
perpendicular to the other two axes.   
With respect to the laboratory system, the {\em out--side--long} reference is boosted pair-by-pair 
along the {\em long} axis in such a way to bring to rest the pair in that direction:  
in the literature this boosted system is usually referred to as the Longitudinally Co-Moving 
System (LCMS).
In this reference system, the correlation function is parameterized according to a three-dimensional 
Gaussian function modified by the addition of an {\em out--long} cross-term:   
\begin{equation}
C_2=1+\lambda\exp\left[ 
 -\Ro^2\Qo^2 - \Rs^2 \Qs^2 - \Rl^2 \Ql^2  -2|\Rol|\Rol \Qo \Ql  
\right]
\label{eq:Cartesian}
\end{equation} 
Such a parameterization is suited for central collisions which are azimuthally 
symmetric. In case of a peripheral collision, the azimuthal symmetry is evidently 
broken; however it is recovered when building the correlation function from 
particles accumulated over 
many events with random impact parameters, which is our approach.  
The $\lambda$\ parameter in equation~\ref{eq:Cartesian} is referred to as the 
chaoticity parameter and it is expected to range in the interval $[0,1]$. 
\subsection{Study of the event sample properties}
\subsubsection{\label{sec:Accept} Acceptance region}
\begin{figure}[t]
\centering
\resizebox{0.35\textwidth}{!}{%
\includegraphics{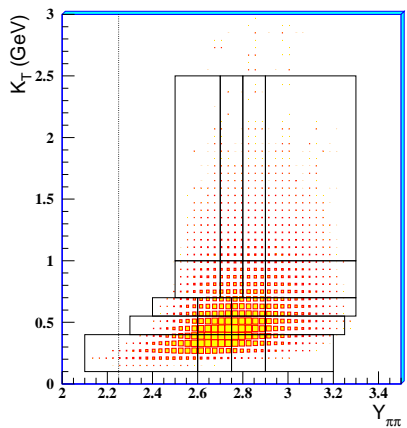}}
\caption{Two-particle acceptance window 
used in the Hanbury-Brown and Twiss 
analysis. The correlation function has been measured in each of the 20 rectangles patching the window 
together. The $\Kt$\ and $Y_{\pi\pi}$ average values, measured over the rectangles, 
are reported in table~\ref{tab:Acc}. The dotted line is drawn at $Y_{\pi\pi} = y_{\tt cm}$.}
\label{fig:2pAcc}
\end{figure}

The determination of the size and dynamical 
evolution  of the system  
at freeze-out requires information about the dependence of the HBT radii  
on the mean momentum $\vec{K}=\frac{1}{2}(\vec{p}_1+\vec{p}_2)$ of the pair.
This can be parameterized by its transverse component and the pair rapidity 
(computed assuming the pion mass): 
\begin{eqnarray}
\Kt=\frac{1}{2}\sqrt{(p_{y1}+p_{y2})^2+(p_{z1}+p_{z2})^2} \label{eq:Kt}\\
Y_{\pi\pi}=\frac{1}{2}\log\frac{E_1+E_2+p_{x1}+p_{x2}} 
{E_1+E_2-p_{x1}-p_{x2}} \label{eq:Y}
\end{eqnarray}
With this in mind, our (two-particle) acceptance window, which
is shown in figure~\ref{fig:2pAcc},
has been binned in 20 rectangles in each of which 
the correlation function $C_2(q)$\ has been measured independently. 
Table~\ref{tab:Acc} displays the average values of
the $\Kt$\ and $Y_{\pi\pi}$\ distributions in each of the 20 bins used
for the analysis. 
\begin{table}[hbt]
\caption{Averages of the
 $\Kt$\ (uppermost number in the cells, in ${\rm GeV}/c$\ units) and
 $Y_{\pi\pi}$\ (bottom numbers, in blue on-line) distributions  in the 20 
 rectangles of figure~\ref{fig:2pAcc}
\label{tab:Acc}}
\begin{center}
\footnotesize{
\begin{tabular}{|l|c|c|c|c|} \hline 
& $\begin{array}{cc} 1.157 \\ \textcolor{blue}{2.65} \end{array} $ &
  $\begin{array}{cc} 1.162 \\ \textcolor{blue}{2.75} \end{array} $ &
  $\begin{array}{cc} 1.163 \\ \textcolor{blue}{2.85} \end{array} $ &
  $\begin{array}{cc} 1.175 \\ \textcolor{blue}{2.99} \end{array} $  \\ \cline{2-5}
& $\begin{array}{cc} 0.804 \\ \textcolor{blue}{2.64} \end{array} $ &
  $\begin{array}{cc} 0.808 \\ \textcolor{blue}{2.75} \end{array} $ &
  $\begin{array}{cc} 0.809 \\ \textcolor{blue}{2.85} \end{array} $ &
  $\begin{array}{cc} 0.813 \\ \textcolor{blue}{2.98} \end{array} $  \\ \cline{2-5}
{\bf $\uparrow $} &
  $\begin{array}{cc} 0.610 \\ \textcolor{blue}{2.57} \end{array} $ &
  $\begin{array}{cc} 0.615 \\ \textcolor{blue}{2.68} \end{array} $ &
  $\begin{array}{cc} 0.617 \\ \textcolor{blue}{2.82} \end{array} $ &
  $\begin{array}{cc} 0.620 \\ \textcolor{blue}{2.97} \end{array} $  \\ \cline{2-5}
%
{\bf $\Kt$}  &
  $\begin{array}{cc} 0.460 \\ \textcolor{blue}{2.55} \end{array} $ &
  $\begin{array}{cc} 0.468 \\ \textcolor{blue}{2.68} \end{array} $ &
  $\begin{array}{cc} 0.472 \\ \textcolor{blue}{2.82} \end{array} $ &
  $\begin{array}{cc} 0.478 \\ \textcolor{blue}{2.97} \end{array} $  \\ \cline{2-5}
& $\begin{array}{cc} 0.305 \\ \textcolor{blue}{2.51} \end{array} $ &
  $\begin{array}{cc} 0.332 \\ \textcolor{blue}{2.67} \end{array} $ &
  $\begin{array}{cc} 0.345 \\ \textcolor{blue}{2.81} \end{array} $ &
  $\begin{array}{cc} 0.358 \\ \textcolor{blue}{2.95} \end{array} $  \\ \hline
\multicolumn{1}{c|}{ } & \multicolumn{4}{c|}
{\textcolor{blue}{\bf $Y_{\pi\pi} \, \longrightarrow$}} \\ \cline{2-5}
\end{tabular}
}
\end{center}
\end{table}

Additionally, in order to investigate the dynamics of the collision as a function of
centrality, the whole analysis has been repeated in three centrality classes obtained from 
those defined in table~\ref{tab:centrality}: 
central ($3+4$),  semi-central ($2$) and semi-peripheral ($0+1$), which 
correspond to the ranges of centrality 0--11\%, 11--23\% and 23--53\%, 
respectively\footnote{For reasons of statistics  
the centrality classes 0 and 1, as well as 3 and 4,  
have been combined.}.  

\subsubsection{\label{conta} Contamination from non $\pi^-$--$\pi^-$\ pairs }  
The correlation analysis presented here is based on pairs of negative hadrons  
($\hm$--$\hm$), which are expected to be dominated by pairs of identical pions. 
The purity  decreases with increasing $\Kt$: based on the results of NA49~\cite{NA49yield}   
we estimate that at $\Kt \approx 0.3$\ GeV/$c$\ it is about 90\%, 
at $\Kt \approx 0.8$\ (1.2) GeV/$c$\ it goes down to  about  80\% (65\%).  
The main contamination comes from \Pgpm--\PKm pairs 
(from about 5\% at $\Kt \approx 0.3$\ GeV/$c$\ to about 25\% at $\Kt \approx1.2 $ GeV/$c$) 
and also from \PKm--\PKm pairs (up to $\approx$ 10\% at $\Kt \approx1.2 $ GeV/$c$). Other 
combinations, e.g. \Pgpm--\Pap\ or \PKm--\Pap, are negligible,  
the largest being about a few per cent 
for \Pgpm--\Pap\ at $\Kt \approx 1.2 $ GeV/$c$.  

\begin{figure}[hbt]
\centering
\resizebox{0.33\textwidth}{!}{%
\includegraphics{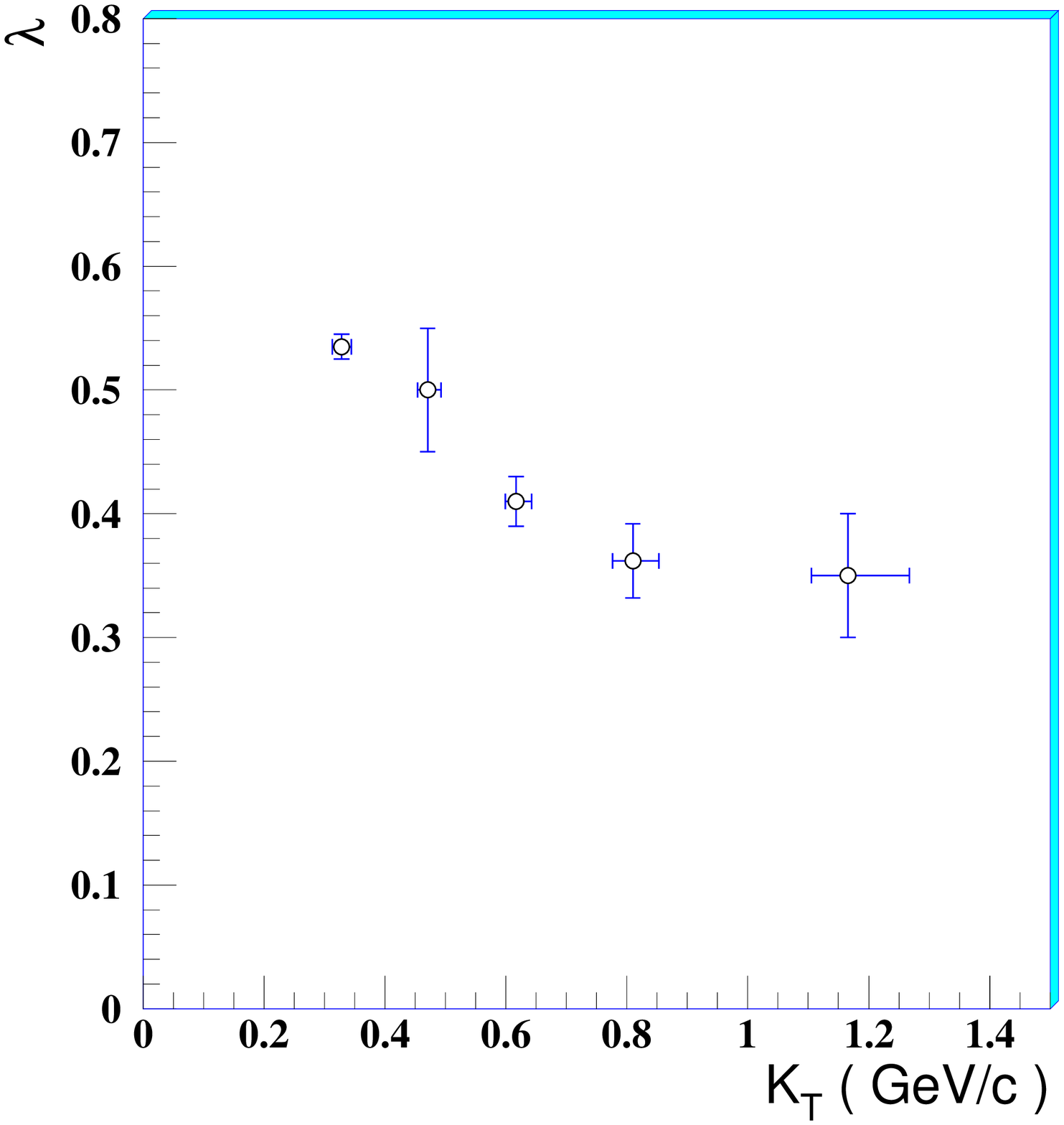}}
\caption{The chaoticity parameter $\lambda$\ as a function of $\Kt$\ for the
most central 53\% of the Pb--Pb cross-section. Statistical errors are shown.}
\label{fig:Lambda}
\end{figure}

Excepting  
\PKm--\PKm pairs at high $\Kt$,  
misidentified particles thus lead to counting of unlike pairs 
which do not give rise to Bose-Einstein correlations. Moreover, 
in Pb--Pb collisions at 
$E_{\rm beam} = 158$\ GeV per nucleon the measured HBT radius 
parameters   
of the \PKm--\PKm\ (and \PKp--\PKp) correlation 
functions~\cite{NA49HBTkaon} (NA49 Collaboration)
are fully consistent with the 
published pion results and the hydro-dynamic expansion model.  
It has been shown in~\cite{NA35} (NA35 experiment), in~\cite{NA49HBT} (NA49 experiment) and 
in~\cite{WA97HBT} (WA97 experiment)  
that the main effect of particle misidentification is to reduce the value of 
the chaoticity parameter $\lambda$.   
Apart from particle misidentification which produces the strongest bias, 
the $\lambda$\ parameter, shown in figure~\ref{fig:Lambda} as a function of $\Kt$, 
is also affected  
in a non-trivial way by resonance decays and other effects~\cite{ref:17},  
which are expected to depend on the pair momentum.  
The $\lambda$\ parameter, however, is not used for the reconstruction of  
the size and dynamical state of the source; 
the relevant source parameters are affected to the level of a few per cent only.  

\subsubsection{Two-track resolution}
The detector's ability to distinguish a pair of close tracks from
a single track decreases with decreasing track separation, depending
on the Si-pixel size.  
The correction for this inefficiency, applied pair-by-pair to the signal 
distribution $S(q)$,  
was calculated via a Monte Carlo simulation 
by reconstructing
a sample of (generated) two-tracks events embedded in real ones. 
The maximum inefficiency was found to be $\approx$10\% 
(relative to high $q$\ pairs) for  pairs populating the 10-MeV-wide  
bin of the correlation functions centred at $q=0$\ and it goes   
to zero outside that bin.  
\subsubsection{Track splitting}
The possibility of ghost tracks in the data sample has been investigated.  
Fake tracks might be produced by the reconstruction algorithm due to, e.g., 
allowing track candidates to share several clusters. 
To avoid such a bias, tracks have been selected by requiring them to have  
clusters in more than 80\% of the telescope planes and 
less than 30\% of the clusters shared with other 
tracks\footnote{The same requirement was applied to the tracks used for the 
transverse mass spectra analysis discussed in section~\ref{single}, as 
already mentioned.}.
As a final check, Monte Carlo events have been simulated 
by generating 
a sample of  \Pgpm\ distributed in phase-space according to the measured 
double differential cross-section of equation~\ref{eq:InvSlope}.   
Then, the simulated events have been reconstructed using the same algorithm  
as for real events. Three tracks only have been reconstructed   
with one associated ghost out of 250K generated tracks which 
pass through the pixel telescope. This effect is therefore negligible.  
                                                                                                                             
\subsubsection{Momentum resolution}  
The quality of a correlation measurement is determined by the
resolution of the two-particle momentum difference (``relative momentum'') $q$.  
The relative momentum resolution was studied using a 
Monte Carlo chain based on the simulation code GEANT~\cite{GEANT}.  
As listed in table~\ref{tab:Resolution},  
we have estimated that the relative momentum projections $\Qs$ and $\Ql$\ 
(the latter evaluated in the LCMS system)  
are measured in this analysis with an error not larger than about $20 \, {\rm MeV}/c$\
over our two-particle acceptance window. The $\Qo$\ projection, on the other 
hand, shows a stronger $\Kt$\ dependence, which can introduce an important 
contribution to the systematics errors on the extracted HBT radii, as discussed 
below.  
\begin{table}
\caption{The resolution of relative momentum components $\Qo$, $\Qs$\ and $\Ql$\ (MeV/$c$)  
in the LCMS system  as a function of $\Kt$ (GeV/$c$).  
\label{tab:Resolution}}
\begin{center}
\begin{tabular}{|c|ccc|}  \hline
  &  $\Kt < 0.5$ & $0.5<\Kt<1.0$ & $\Kt>1.0$ \\
\hline\hline
$\Qo$       &  22   & 52    & 110  \\ \hline
$\Qs$       &  15  & 18    & 21  \\ \hline
$\Ql$       &  10   & 11    & 13   \\ \hline
\end{tabular}
\end{center}
\end{table}

\subsubsection{\label{secCoul} Coulomb correction}
The observed two-particle correlation is expected to result from two  
different contributions, the Bose-Einstein effect and the Coulomb  
interaction. The Coulomb interaction between particles of same charge sign  
is repulsive, 
thus depleting the  
two-particle correlation function at small relative momenta.   
Two methods have been considered to correct for the Coulomb interaction. 

A standard procedure consists in applying $q_{\rm inv}$-dependent weights to each pair 
in the background distribution $B(q)$\ to get the {\em Coulomb corrected correlation 
function}:
\begin{eqnarray}
\fl
C_2^{corr}(q)=N \frac{S(q)}{K_{\rm Coul}(q_{\rm inv}) \cdot B(q)}= \label{eq:C2corr} \\
 1+\lambda\exp\left[
 -\Ro^2\Qo^2 - \Rs^2 \Qs^2 - \Rl^2 \Ql^2  -2|\Rol|\Rol \Qo \Ql
\right] 
\nonumber
\end{eqnarray}
where $K_{\rm Coul}$\ is the squared Coulomb wave-function integrated over the whole source 
and $q_{\rm inv}=\sqrt{\vec{q}^2-q_0^2}$.   

Bowler and Sinyukov pointed out~\cite{Bowler-Sinyukov} that the correction procedure 
should be different in  
the case where not 
all particle pairs in the signal distribution are subject to the Coulomb correlation.  
The argument here is that if physics (or detector) related effects lead to a reduction of 
the observed quantum correlation strength, then the same effects also lead 
to a similar reduction of the Coulomb repulsion. Therefore, the strength of the 
Coulomb correction applied to the data should be linked to the experimentally 
observed $\lambda$\ parameter. 
This method has been implemented recently by the CERES~\cite{CEREShbt}, PHOBOS~\cite{PHOBOShbt} 
and STAR~\cite{STARhbt} Collaborations. 
The correlation function in this procedure is fitted to
\begin{eqnarray}
\fl
C_2(q)=\frac{S(q)}{B(q)}=(1-\lambda) +
 \label{C2Bowler} \\
 \lambda \cdot K_{\rm Coul}(q_{\rm inv}) \cdot \left[
  1+\exp(
  -\Ro^2\Qo^2 - \Rs^2 \Qs^2 - \Rl^2 \Ql^2  -2|\Rol|\Rol \Qo \Ql)
  \right]  \nonumber
\end{eqnarray}
where $K_{\rm Coul}(q_{\rm inv})$\ is the same as in the {\em standard} procedure.
We argue that this alternative method is 
appropriate  
when the correlation function is measured with 
a negligible contamination of non-identical particle pairs; in fact the particles of these pairs 
would Coulomb interact in the signal and would not in the background distribution. Therefore we consider 
it only as an alternative method to control possible systematic errors on the HBT radii due to the Coulomb 
correction, as discussed later.  

For both procedures, three kinds of pion emitting sources have been considered with a method  
similar to that described in reference~\cite{WA97HBT}: 
\begin{itemize}
\item a point-like source (in space--time); 
\item a static Gaussian source; 
\item an expanding source, parameterized according to the same 
      blast-wave model~\cite{BlastRef} used to describe the data.  
\end{itemize}

The point-like source was used for comparison with the analytical Gamow factor~\cite{Gamov,Gyulassy} 
for the sake of  validation of the method.  
A static Gaussian source with a radius of 5 fm yields a Coulomb correction smaller than 
the Gamow factor and 
slightly larger than that obtained for the expanding source.  
The computation of the Coulomb correction for the expanding source requires knowledge of 
the source parameters, which are not known {\em a priori} but should be obtained from the 
correlation study. Therefore an iterative procedure has been used, as discussed in reference~\cite{WA97HBT}, 
starting with the following values for the blast-wave model freeze-out parameters (to be discussed later):  
$T=120$\ MeV, $\Bt=0.4$, $R_G=5$\ fm, $\tau_{\rm f}=5$ fm/$c$\ and $\Delta\tau=1$\ fm/$c$. 
The results for the model parameters become stable after the first 
iteration\footnote{Varying the blast-wave model 
parameters by 20\% produces a maximum variation of the Coulomb correction of about 5\%.}. 

\begin{figure}
\centering
\resizebox{0.39\textwidth}{!}{%
\includegraphics{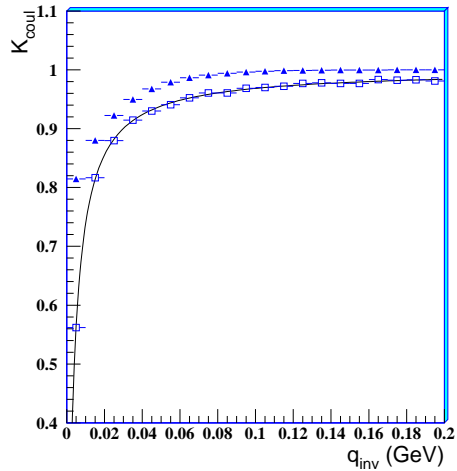}}
\caption{Coulomb correction for an expanding source   
         (full triangles, see text for details) and for a point-like 
         source (open squares) as compared to the Gamow factor (full line).
\label{fig:Coul}
}
\end{figure}

In figure~\ref{fig:Coul} the correction factor $K_{\rm Coul}$\ for an expanding 
source, which has been used in this analysis, is shown superimposed to 
the correction for a point-like source and to the Gamow factor.    

\subsection{Multidimensional fit}
Equation~\ref{eq:Cartesian} has been fitted to the corrected correlation 
functions by  
maximizing the negative logarithmic likelihood function:  
\begin{equation}
-2 \log \ L(\vec{R})=2 \sum_{i} [C_i B_i - S_i \log(C_i B_i)+\log(S_i!)]
\label{likelihood}
\end{equation}
where $C_i$\ is the theoretical value of the correlation function for a given  
set of parameters $\vec{R}= (\lambda,\Ro,\Rl,\Rs,\Rol)$\ in the $i^{\rm th}$ bin of $q$\  
and $S_i$\ and $B_i$ are, respectively, 
the distribution of signal and background in that bin.

\begin{figure}[hbt]
\centering
\resizebox{0.95\textwidth}{!}{%
\includegraphics{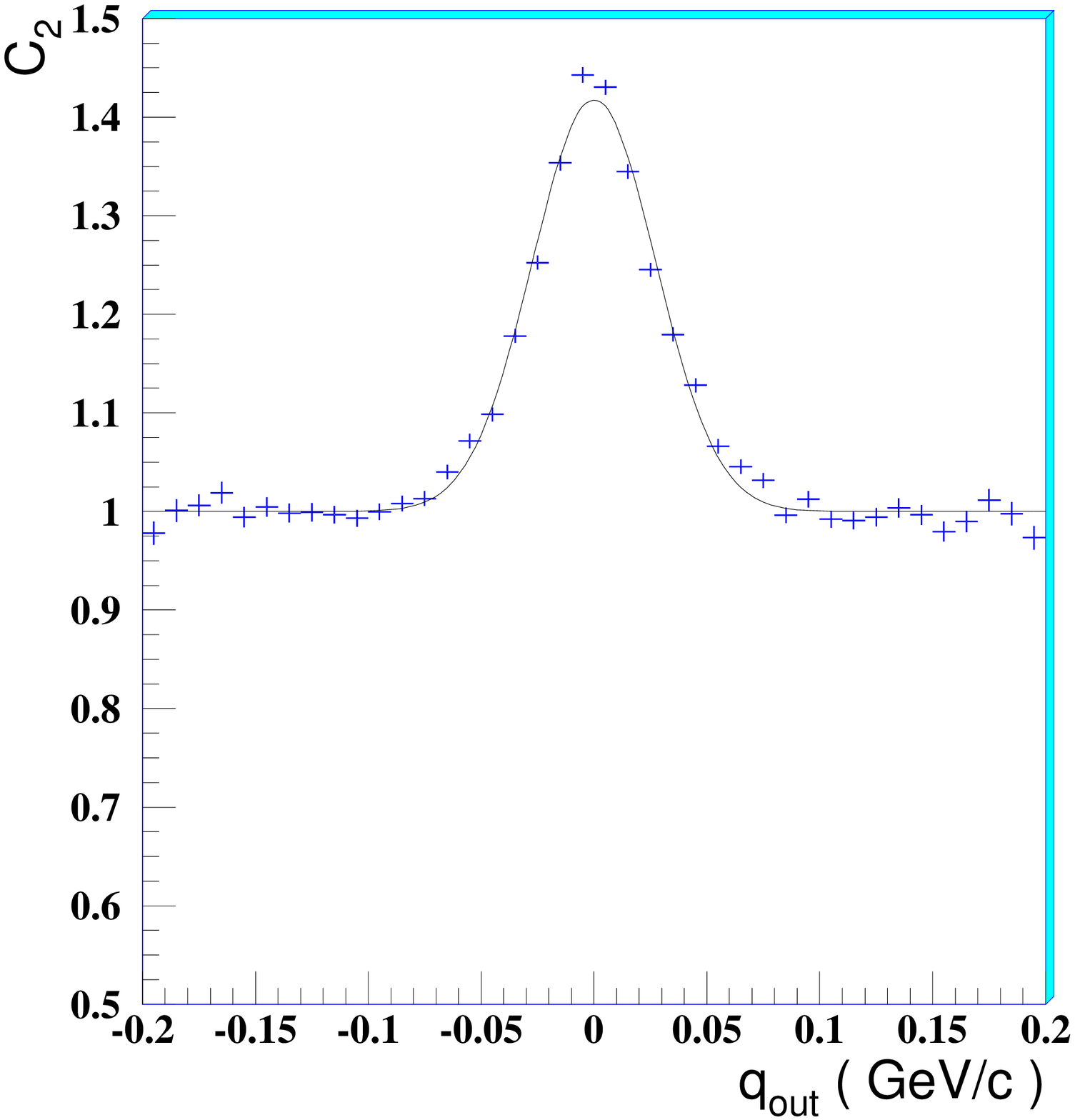}   
\includegraphics{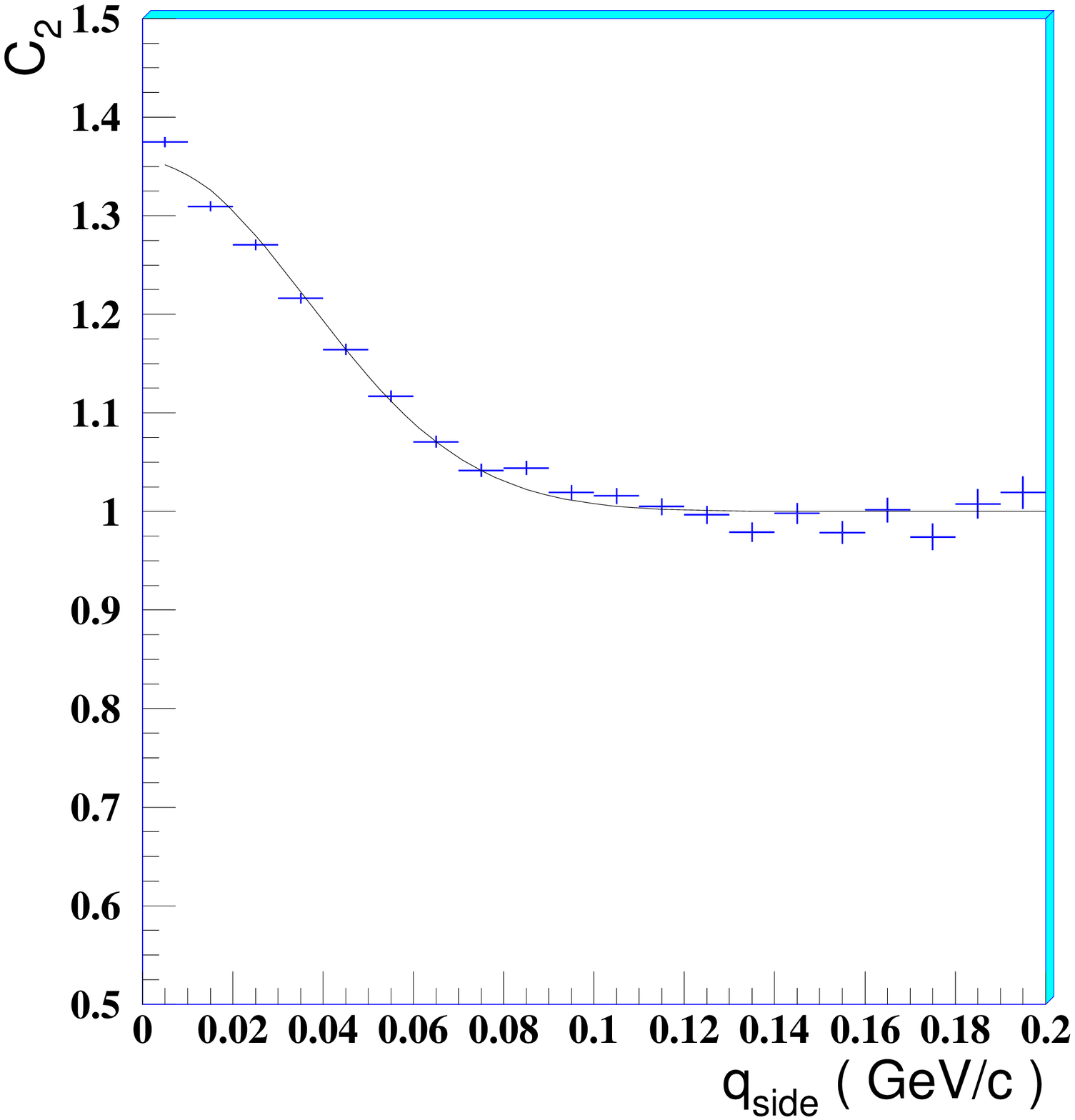}   
\includegraphics{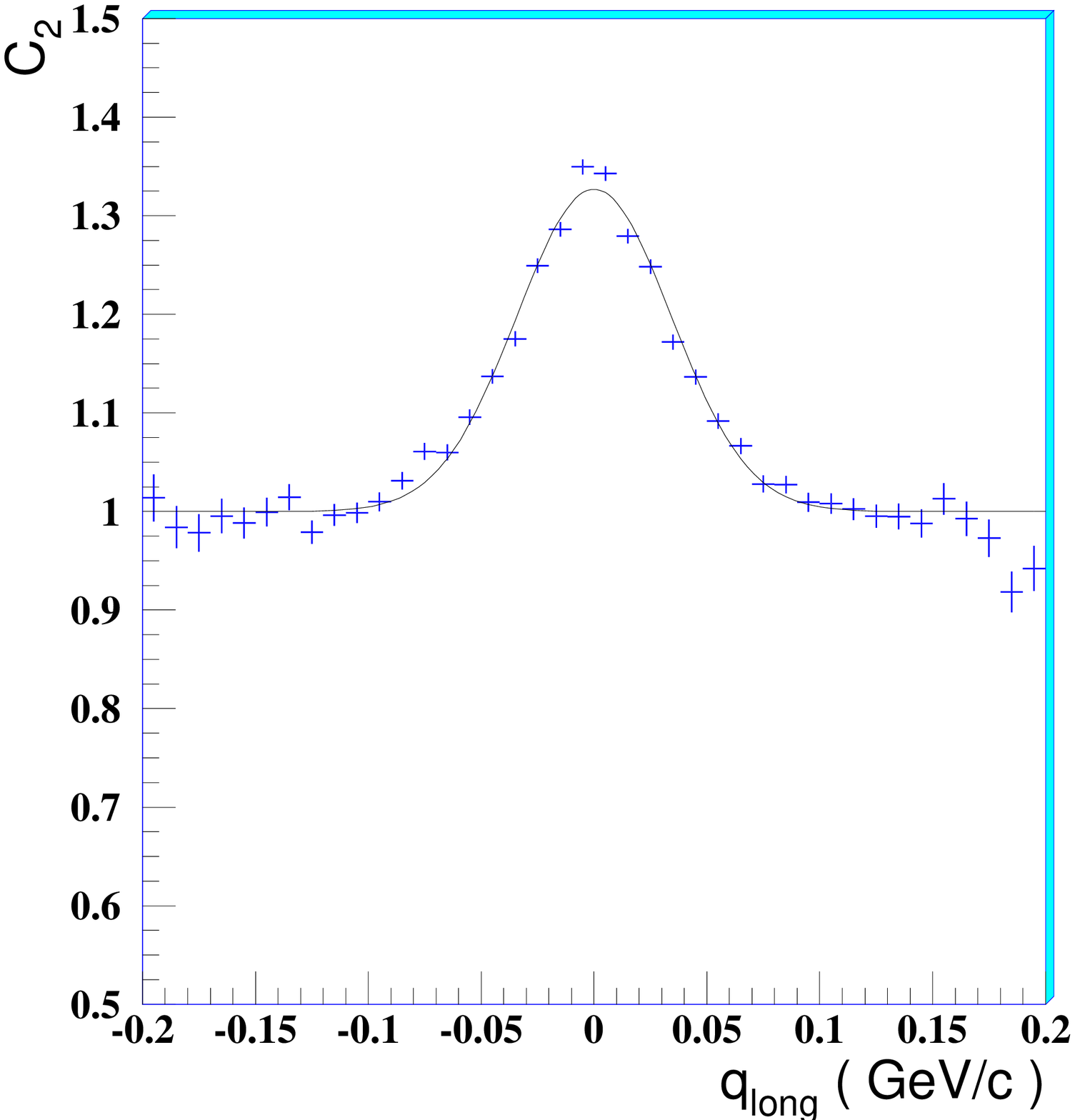}}  
\caption{Projections of the $\hm$--$\hm$\ correlation function measured using Cartesian 
parameterization for pair rapidities $2.30 < Y_{\pi\pi} < 3.25$\ and pair transverse momenta
$0.40 < \Kt < 0.55$\ GeV/c. Full curves are the result of a least-squares fit to a Gaussian 
function (see the text).}
\label{fig:projec}
\end{figure}

Two checks on the fit quality have been performed.
First, the $\chi^2/{\rm ndf}$--values are calculated 
on the three-dimensional correlation function using 
the parameters corresponding to the 
maximum  
likelihood fit and are
found to be distributed around unity. Second, the projections 
of the 3-dimensional correlation function onto each momentum 
difference component, with narrow cuts on the other components,
have been fitted by the least-squares method to a Gaussian function,
yielding results consistent with the results of the
3-dimensional fits. A typical sample of these projections is
shown in figure~\ref{fig:projec}
where the correlation function has been
integrated  along the other $q_j$\ components
in the intervals $| q_j |  <  30\, {\rm MeV}$.  
\subsection{Systematic errors}
The main sources of systematic uncertainties on the HBT radii are attributed to  
three previously mentioned effects, namely, with decreasing importance,  
{\em (i)}~the momentum resolution,  
{\em (ii)}~the Coulomb correction procedure and {\em (iii)}~the double-track resolution.  

The systematic error due to the finite momentum resolution, which depends on the transverse 
momentum of the pair,  has been evaluated  by a Monte Carlo simulation with the following 
approach. A sample of pion pairs was generated, whose particles are 
extracted from the phase-space 
$\frac{{\rm d}N}{{\rm d}p^\mu \, {\rm d}x^\nu}$
of the expanding model with the parameters specified in section~\ref{secCoul}.  
The momentum-dependent part of this 
particle phase-space distribution  
is then  
propagated  
through   
the apparatus to take into account  
the experimental acceptance and reconstruction inefficency,  
in order to end up with the raw measured momentum distribution  
$\frac{{\rm d} N}{dp^\mu}^{\rm raw}$.   
Quantum and other correlations are introduced in a second stage using the  
method outlined in reference~\cite{Crab}.   
The same procedure is repeated for a second sample of pairs, 
obtained from the first one by smearing the momentum distribution according to 
the experimental momentum resolution. The variations of the extracted HBT radii 
provide an estimate of the systematic errors.  

As discussed in section~\ref{secCoul}, two Coulomb correction procedures  
have been implemented, each with different assumptions about the source properties  
(its space-time extent, static or expanding source, etc.).  The differences in the HBT radii 
obtained with the different methods allow us to estimate a contribution to  the systematic 
error of 10\% for $\Ro$, $\Rs$\ and $\Rl$, and of 5\% for the cross-term $\Rol$.  

The last source of systematical errors can be associated to the double-track 
resolution and arises from the finite statistics of Monte Carlo events generated 
to correct for this effect.  It is of the order of a few per cent for all the radii.   

In table~\ref{tab:SysTot} the total systematic errors are given in percentages as a function of $\Kt$.  
%
\begin{table}
\caption{Percentages of systematic errors on the HBT radii for three intervals 
of $\Kt$.  
\label{tab:SysTot}}
\begin{center}
\begin{tabular}{|c|ccc|}  \hline
  &  $\Kt < 0.5$ & $0.5<\Kt<1.0$ & $\Kt>1.0$ \\ 
\hline\hline
$\Ro$       &  12   & 14   & 16   \\ \hline  
$\Rs$       &  10   & 10.5 & 11   \\ \hline   
$\Rl$       &  11   & 11.5 & 13   \\ \hline    
$\Rol$      &  11.5 & 15   & 28   \\ \hline   %
\end{tabular}
\end{center}
\end{table}
%
%
\section{Results and discussion}
\subsection{Longitudinal expansion}
\begin{figure}[hbt]
\centering
\resizebox{0.70\textwidth}{!}{%
\includegraphics{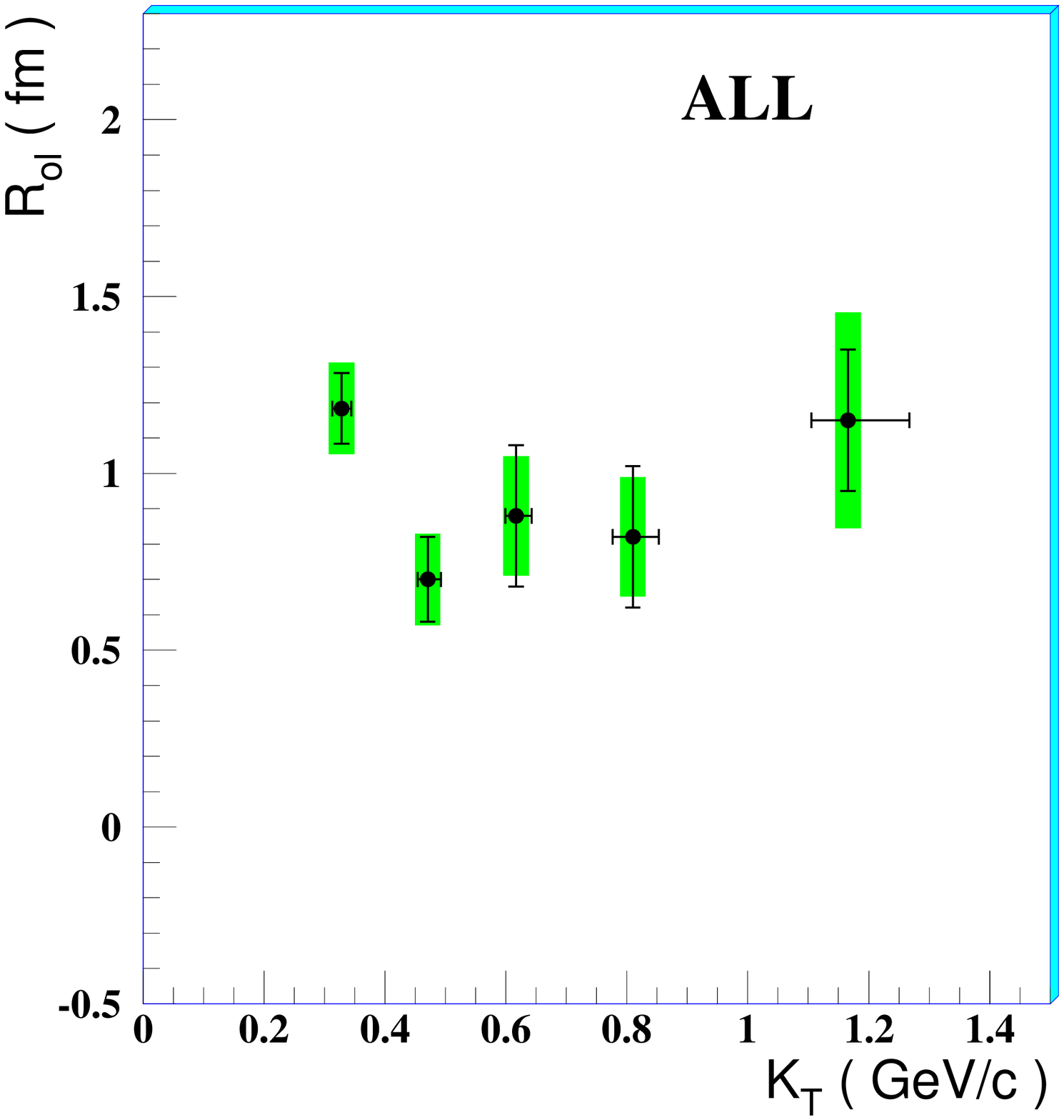}
\includegraphics{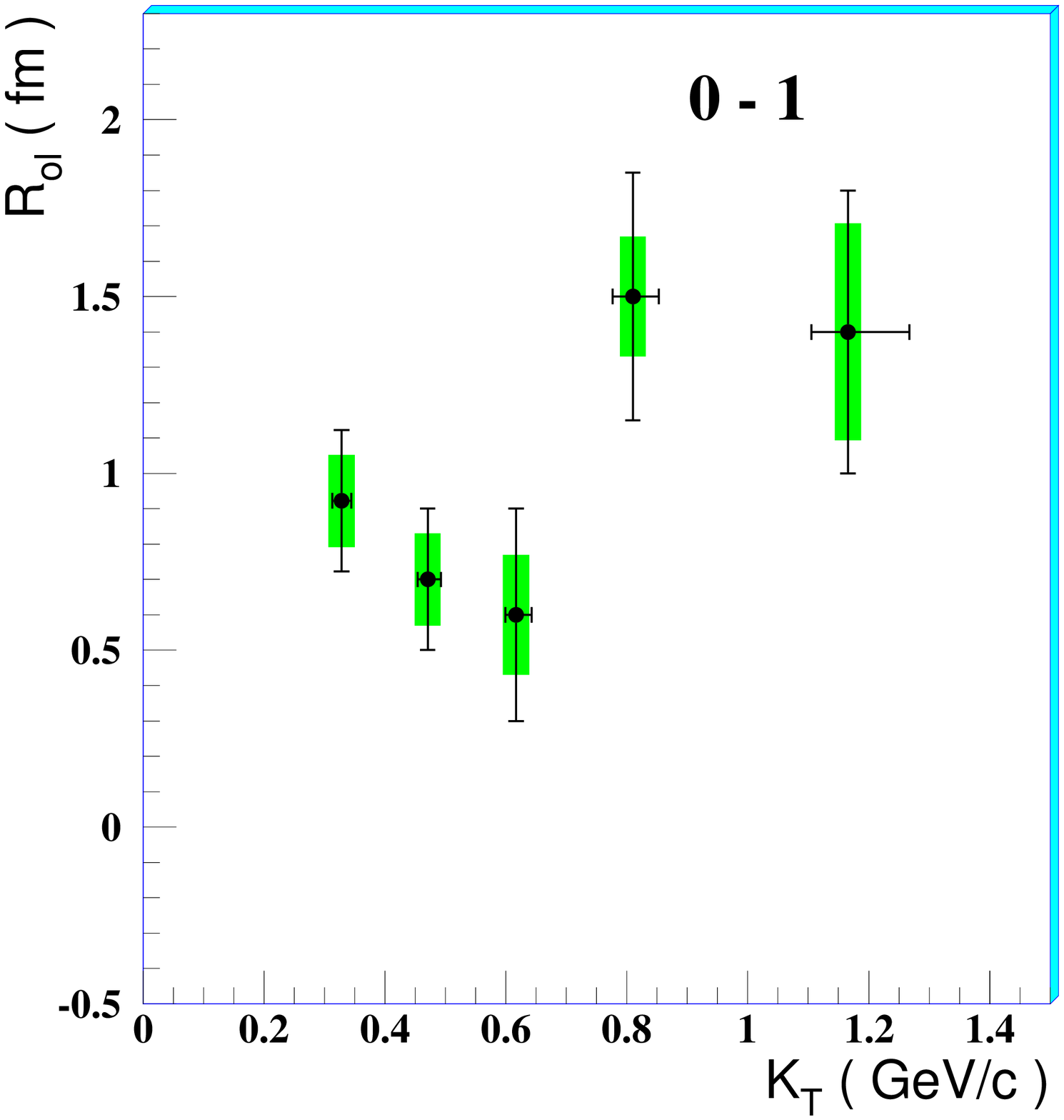}}\\
\resizebox{0.70\textwidth}{!}{%
\includegraphics{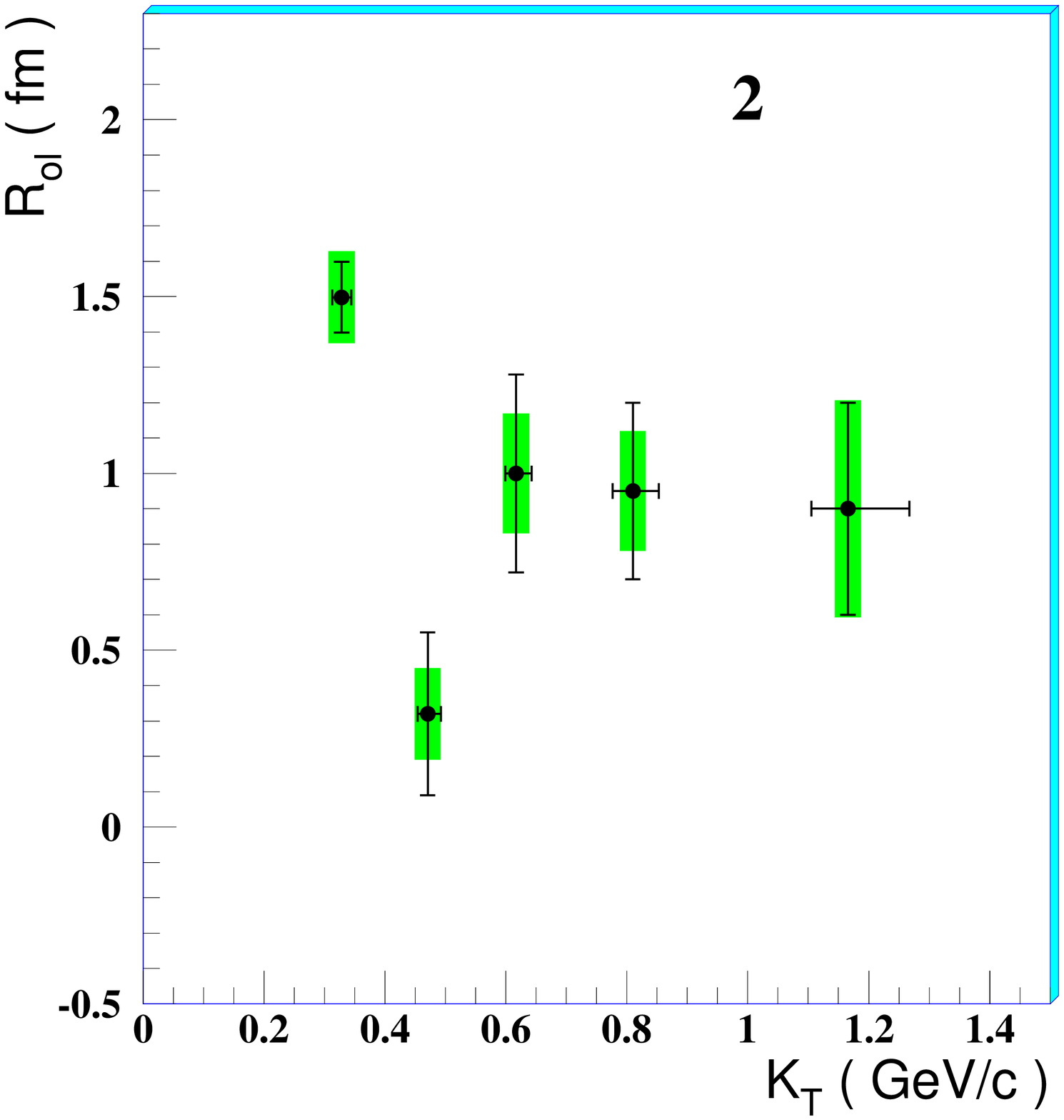}
\includegraphics{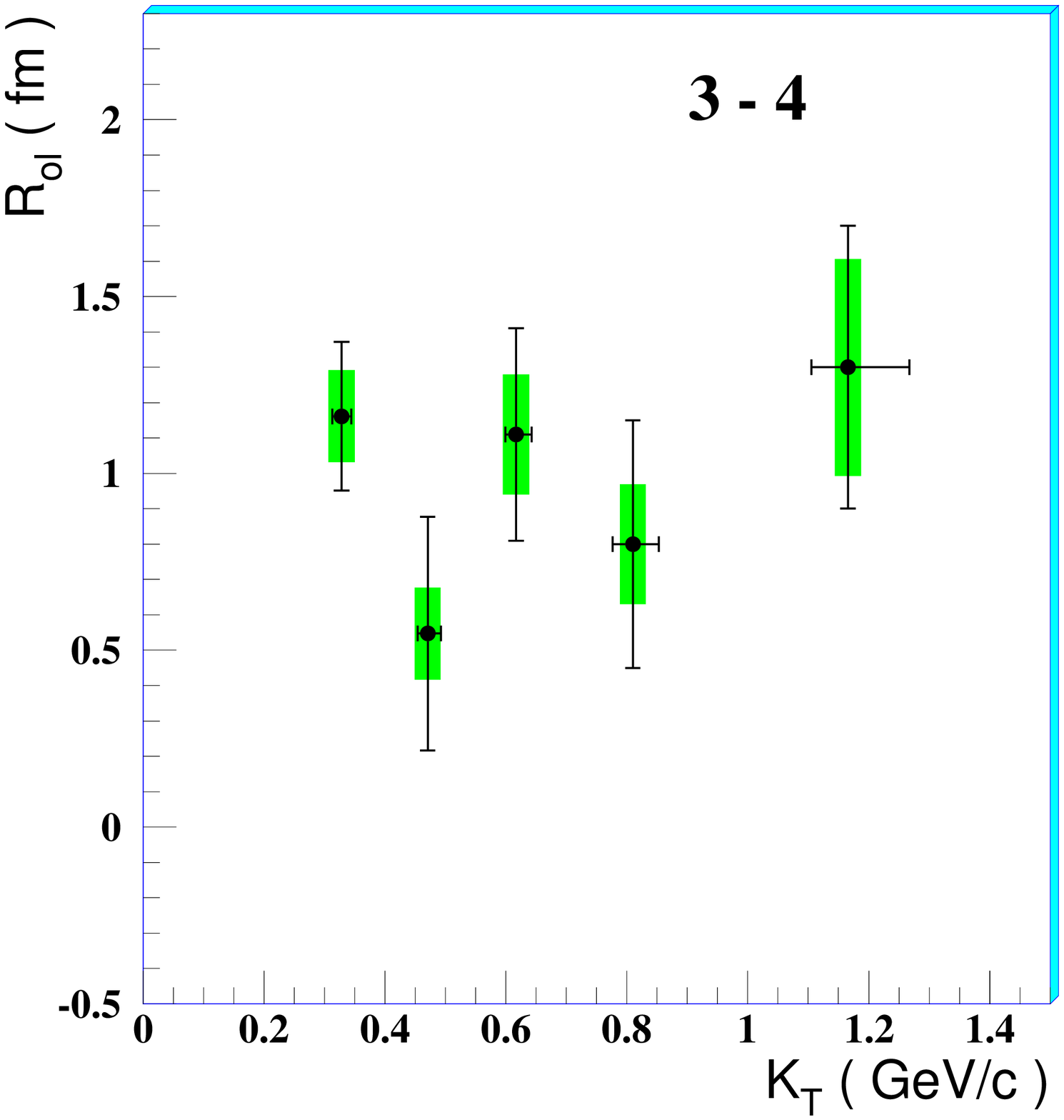}}
\caption{The $\Rol$\ cross-term as a function of $\Kt$\ for
         the most central 53\% of the Pb--Pb cross-section (top-left panel) and for individual
         centrality classes 0--1 (top-right), 2 (bottom-left) and 3-4 (bottom right) of table 1.
         The bars and the boxes show the statistical and systematic errors, respectively.
         }
\label{fig:Rol}
\end{figure}

The cross-term $\Rol$ provides information about the  
longitudinal expansion of the system; e.g. for a longitudinal  
Bjorken-like expansion~\cite{Bjorken} $\Rol$\ should vanish in 
the LCMS system at mid-rapidity~\cite{HeinzLisa}.  
Figure~\ref{fig:Rol} shows the $\Kt$\ dependence of the $\Rol$\ parameters for the most
central 53\% of the Pb--Pb inelastic cross-section (``all") and for the three centrality classes specified
at the end of section~\ref{sec:Accept}. The statistical and systematic errors are indicated in these
and following plots with bars and shadow boxes (green on-line), respectively.
Unfortunately, in most of the recent experimental works about HBT correlations in 
heavy ion collisions 
scarce consideration is given to the study of this parameter.  
Indeed, the cross term is often not 
used in the Cartesian parameterization (equation~\ref{eq:Cartesian})  
or, when used, it is frequently assumed to be of a  given sign (e.g. positive in the LCMS system).  
Such an approach could be justified at RHIC energies near mid-rapidity,  
where the Bjorken model should provide a good description of the longitudinal expansion;   
it is not at lower energy.  
Instead, 
the longitudinal dynamics is derived from the Yano-Koonin-Podgoretskii parameterization~\cite{YKparam} of 
the HBT correlation function      
\begin{equation}
\label{YKP}
\fl
C_2(q) = 1+\lambda \exp \left[
 -R^2_{\perp}q^2_{\perp}-
   \gamma^2_{\rm yk}(q_{\parallel}-
   v_{\rm yk}q_{\rm 0})^2 R^2_{\parallel}
   -\gamma^2_{\rm yk}(q_{\rm 0}-v_{\rm yk}q_{\parallel})^2R^2_{\rm 0} \right]
\end{equation}
where $R_{\perp}$, $R_{\parallel}$, 
$R_{\rm 0}$\ are called the ``YKP radii''\footnote{The subscripts $\perp$\ and $\parallel$\ stand 
for ``perpendicular'' and  ``parallel'' 
to the beam direction, respectively.},  
although $R_{\rm 0}$\  has temporal dimension (${\rm fm}/c$);
$v_{\rm yk}$, called  ``Yano-Koonin velocity'',
is measured in units of {\em c} and
$\gamma_{\rm yk}=(1-v_{\rm yk}^2)^{-1/2}$.
In fact, in this parameterization the $v_{\rm yk}$\ parameter can be easily  
related, under some approximations, to the longitudinal flow velocity of the 
expanding source~\cite{SuperHeinz,RecProgr2}.   

The set of parameters of the Yano-Koonin parameterization can be computed from 
that of the Cartesian parameterization (and vice versa) via analytical 
relations~\cite{SuperHeinz,RecProgr2,Anal}.  
In particular, for the Yano-Koonin velocity the relation is:
\begin{equation}
v_{\rm yk}=  
\frac{A+B}{2C}\left[  1-\sqrt{1-\left(\frac{2C}{A+B}\right)^2} 
 \right]  
\label{Relation}
\end{equation}
where  
      in the LCMS system, due to the vanishing of the longitudinal pair velocity, 
      $A=\frac{\Ro^2 - \Rs^2 }{\beta^2_\perp}$,  
      $B=\Rl^2$\ and $C=-|\Rol|\Rol$;   
      finally $\beta_{\rm T}$\ 
      is the transverse component of  
      the pair velocity $\vec{\beta} \simeq \vec{K}/K^0 $.

\begin{figure}[hbt]
\centering
\resizebox{0.74\textwidth}{!}{%
\includegraphics{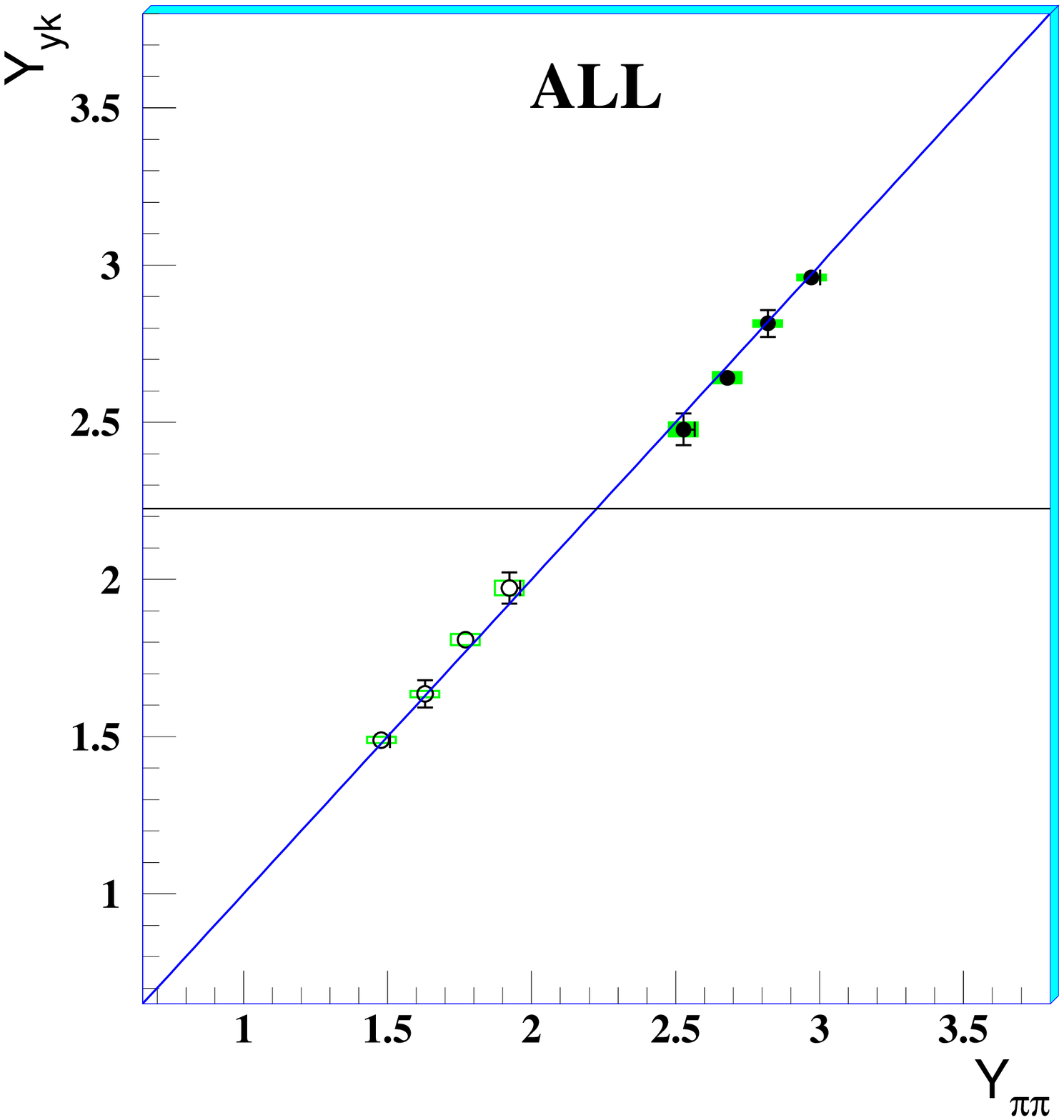}
\includegraphics{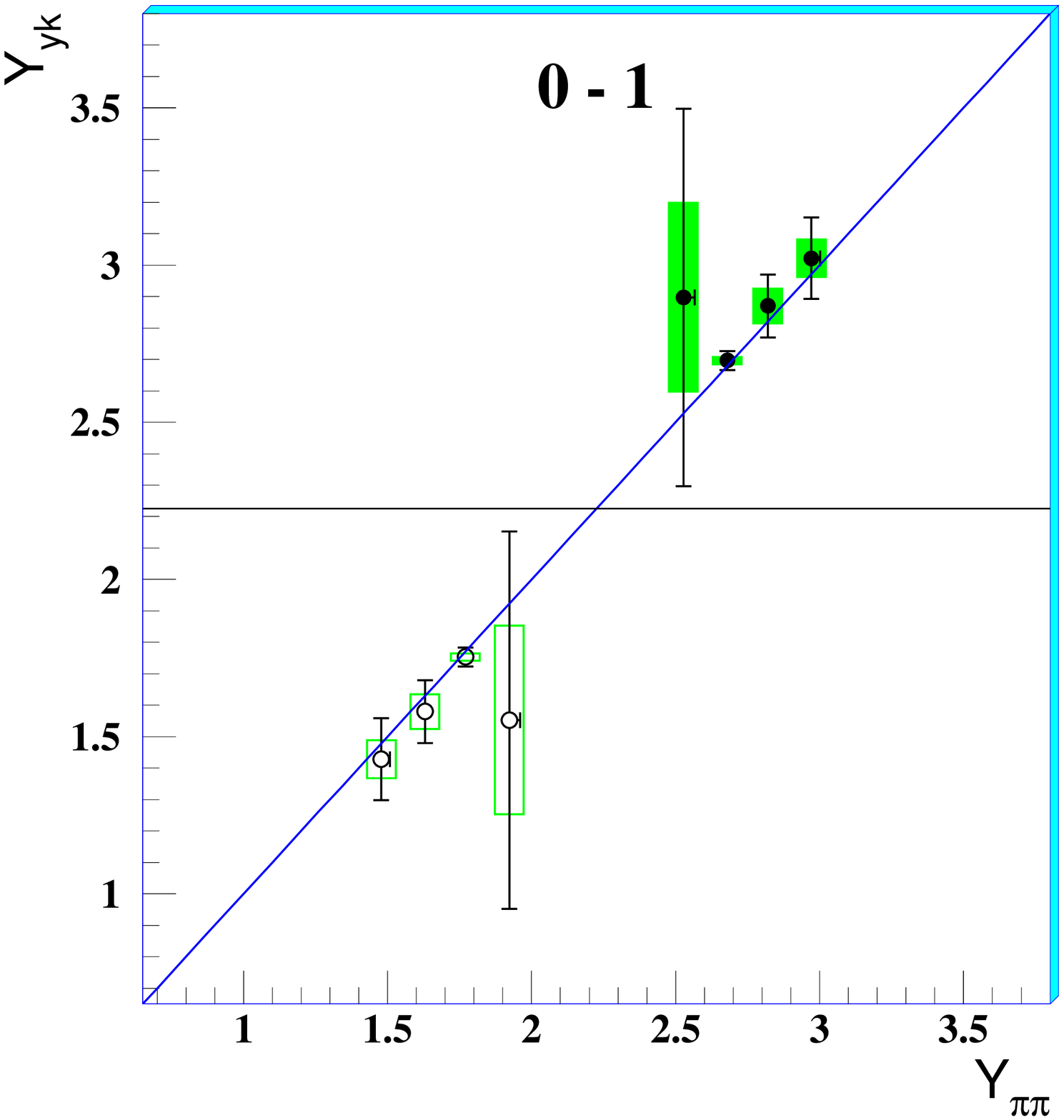}}\\
\resizebox{0.74\textwidth}{!}{%
\includegraphics{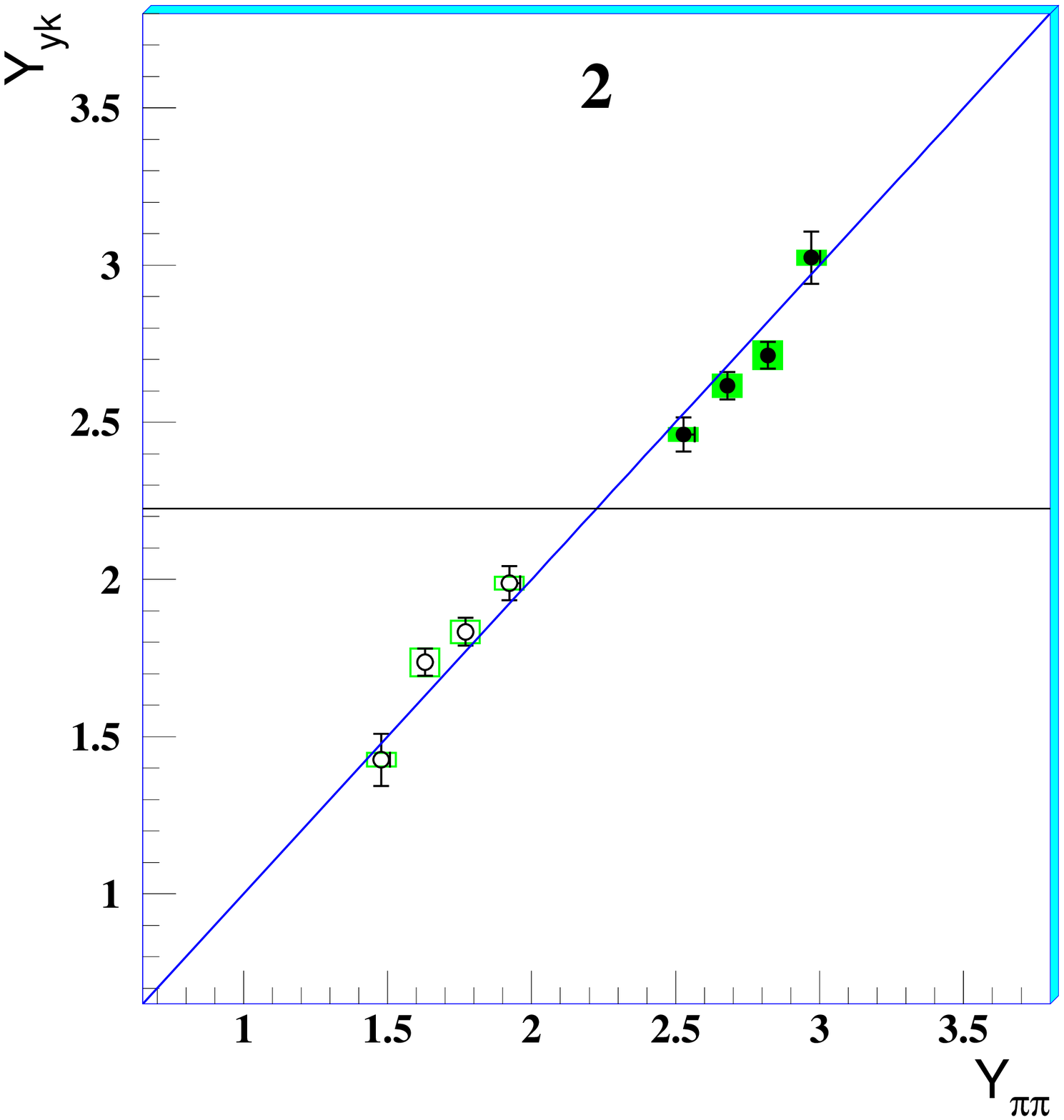}
\includegraphics{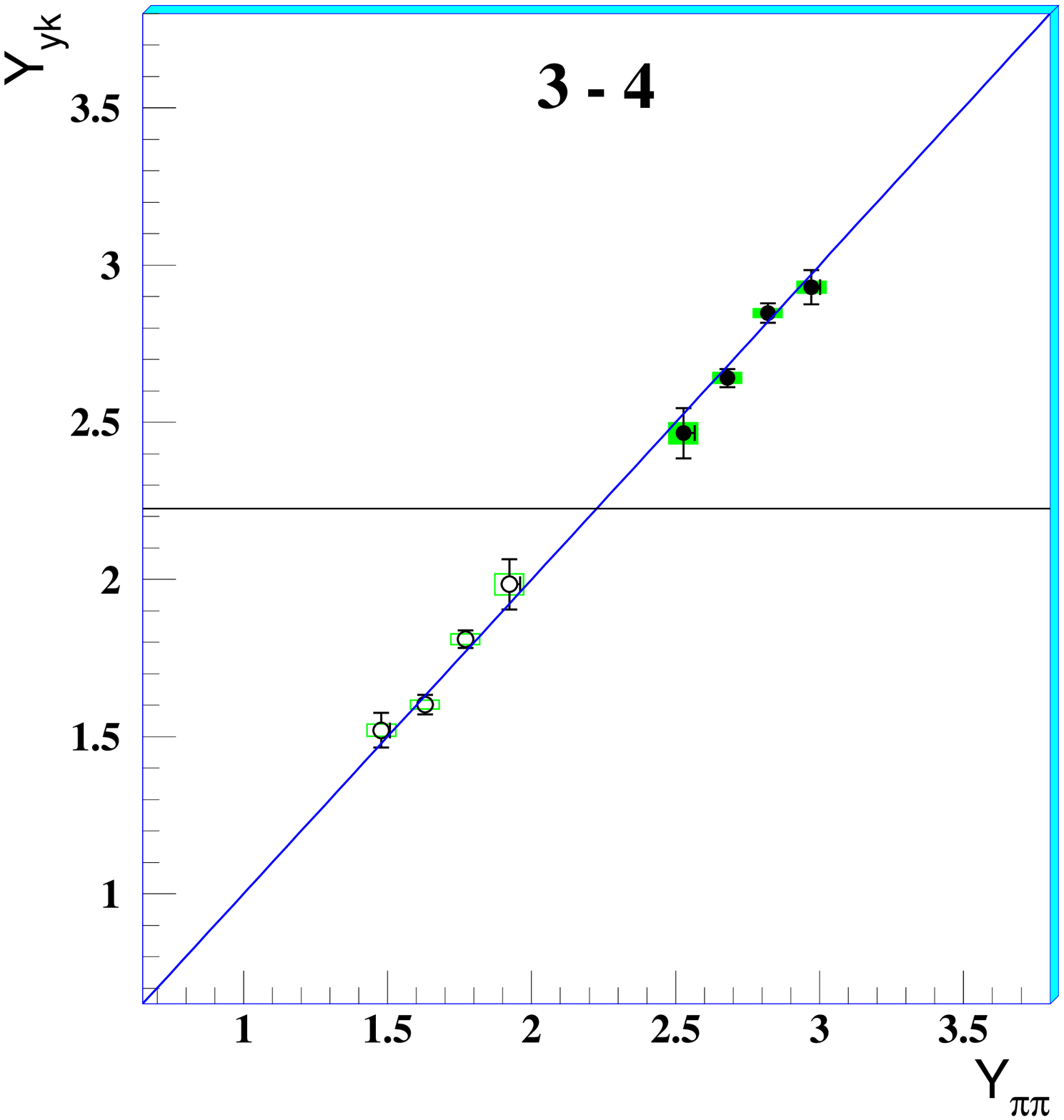}}
\caption{The Yano-Koonin rapidity $Y{\tt yk}$\ as a function of the pair rapidity $Y_{\pi\pi}$\
         measured in the interval $0.1< \Kt < 0.7$\ GeV/$c$ (see text for the averages of $\Kt$\ in
         each bin).
         The most central 53\% of the Pb--Pb cross-section (top-left panel) and the individual
         centrality classes 0--1 (top-right), 2 (bottom-left) and 3-4 (bottom right) of table 1
         are shown.
         Full circles are data, open circles are data reflected about mid-rapidity.
         Statistical and systematic errors are indicated with line bars and shaded 
         boxes (green on-line), respectively.}
\label{fig:YKRap}
\end{figure}

From the Yano-Koonin velocity, one can compute its associated rapidity 
$Y_{\rm yk}=\frac{1}{2}\ln(\frac{1+v_{\rm yk}}{1-v_{\rm yk}})$, which is often 
plotted as a function of the pair rapidity to 
study the longitudinal expansion (see, e.g., references~\cite{NA49HBT,PHOBOShbt}).   
In figure~\ref{fig:YKRap} we plot this quantity, evaluated in the laboratory system,  
as a function of the pair rapidity $Y_{\pi\pi}$\ also given in the same frame:  
$Y_{\rm yk}^{\rm lab}=Y_{\rm yk}^{\rm LCMS}+Y_{\pi\pi}$.   
Statistical and systematic errors, shown at the 1$\sigma$\ confidence level,
have been propagated from those of the out-side-long HBT radii.   
In these plots the correlation functions have been integrated in the 
range $0.1<\Kt<0.7$ GeV/$c$\ where the $\langle\Kt\rangle $\ has been evaluated to be  
0.36, 0.41, 0.51 and 0.58  GeV/$c$, for the data points at $Y_{\pi\pi}=$ 
2.53, 2.68, 2.82 and 2.97, respectively.   


The source rapidity scales with the rapidity of the pair, 
indicating the presence of strong position-momentum
correlations. A static source would exhibit no correlation
and would correspond to a horizontal line 
at the rapidity of the center of mass.    
($Y_{\tt yk} = y_{\tt cm}$). 
A source with strong dynamical correlations would correspond
to a straight line along $Y_{\tt yk} = Y_{\pi\pi}$.
The data are consistent with the latter scenario: 
particles emitted at a given rapidity are produced by a source moving 
collectively at the same rapidity.   

\subsection{\label{transv} Transverse expansion}

The $\vec{K}$-dependence of $\Rs$\ provides information about the transverse expansion
of the source and its geometrical transverse size at freeze-out~\cite{Trovami}. 

\begin{figure}[hbt]
\centering
\resizebox{0.70\textwidth}{!}{%
\includegraphics{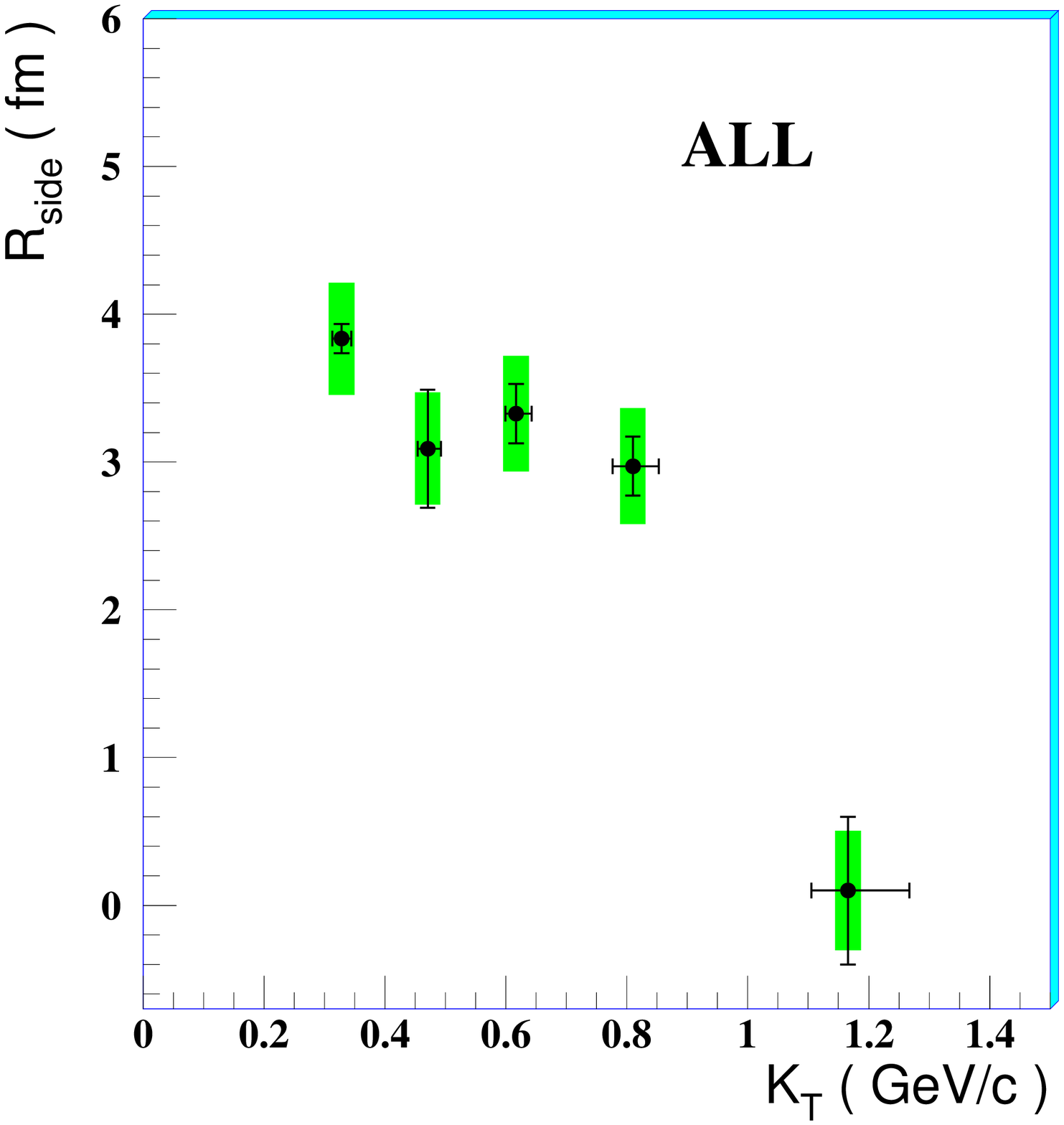}
\includegraphics{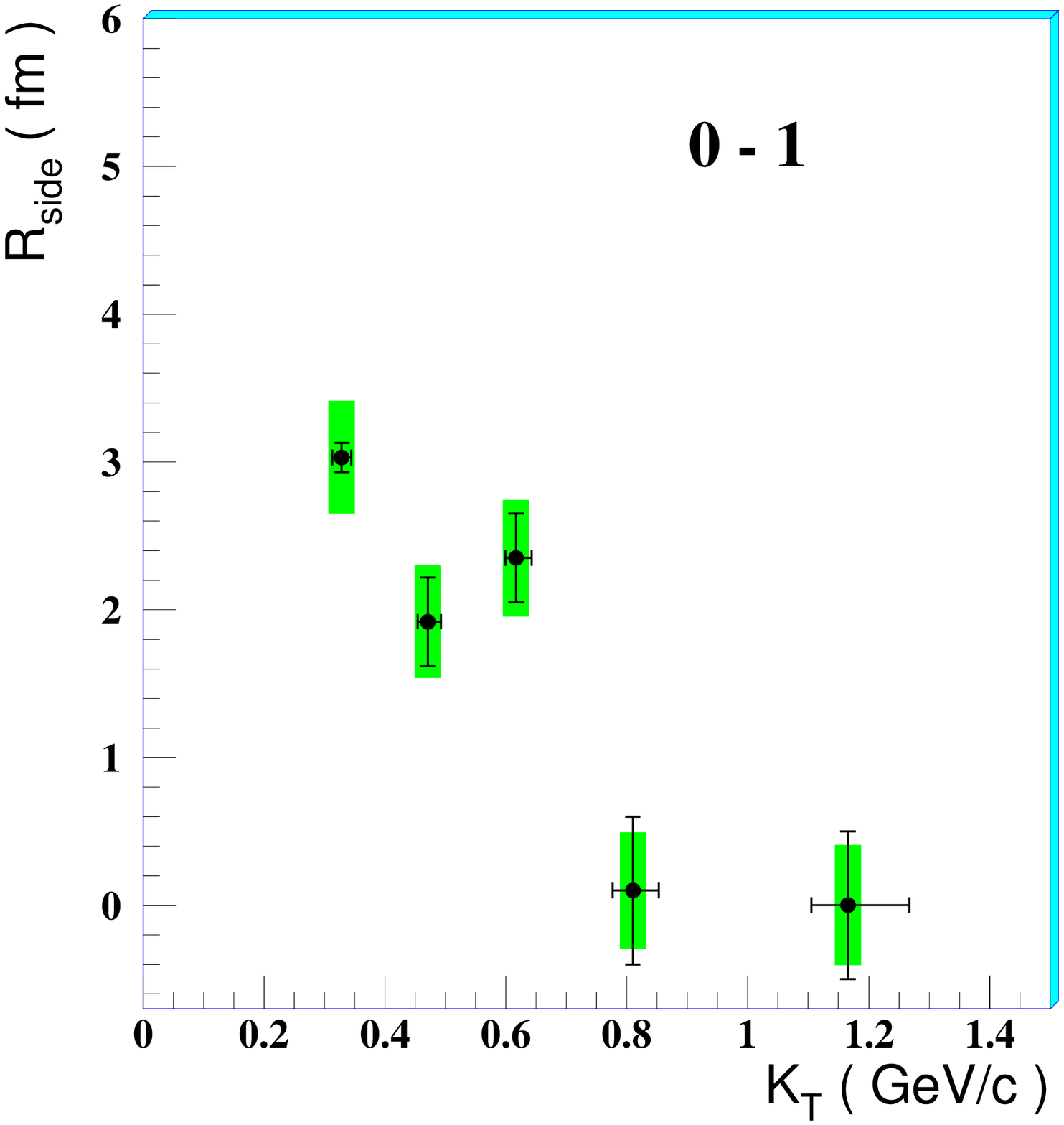}}\\
\resizebox{0.70\textwidth}{!}{%
\includegraphics{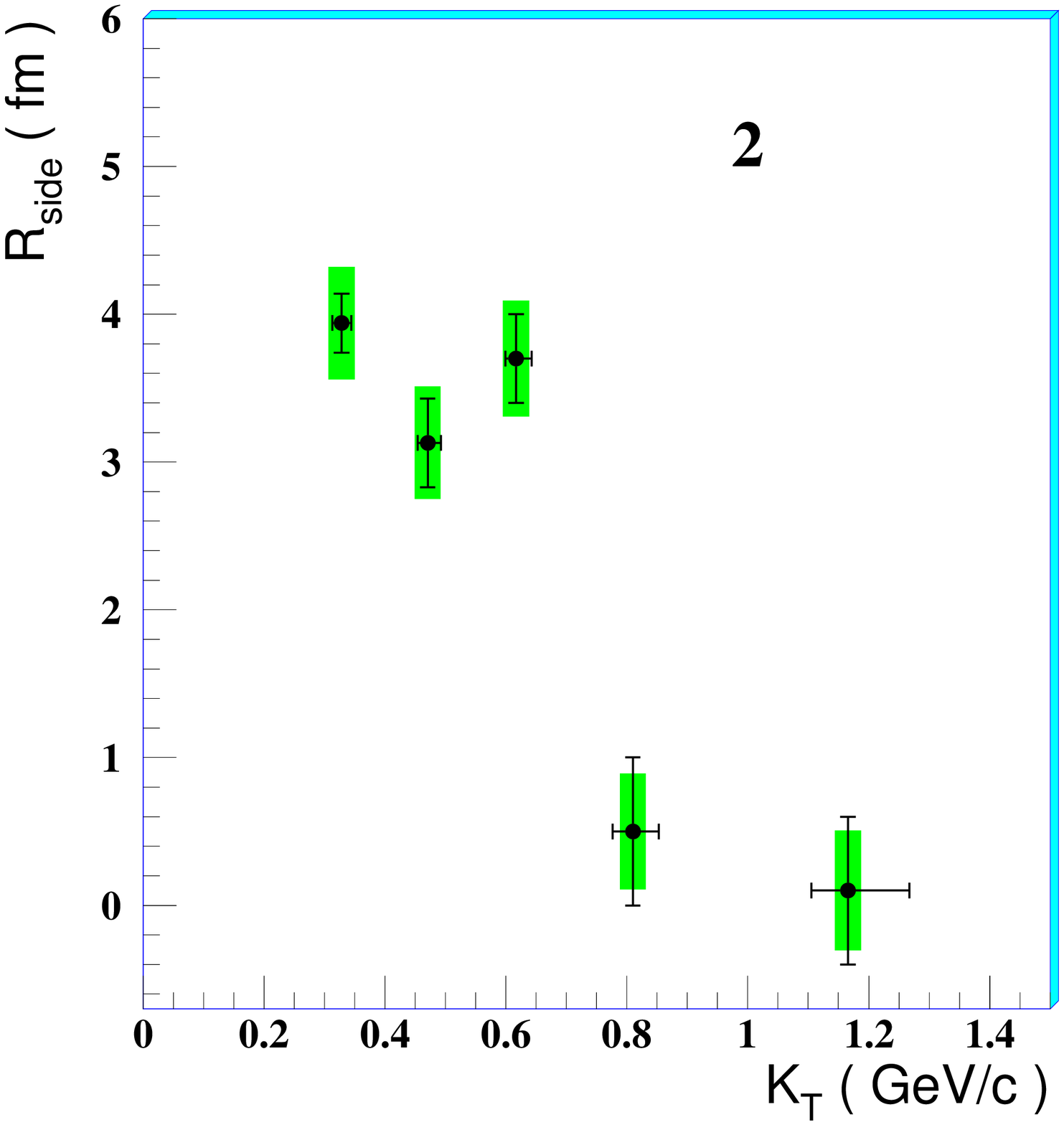}
\includegraphics{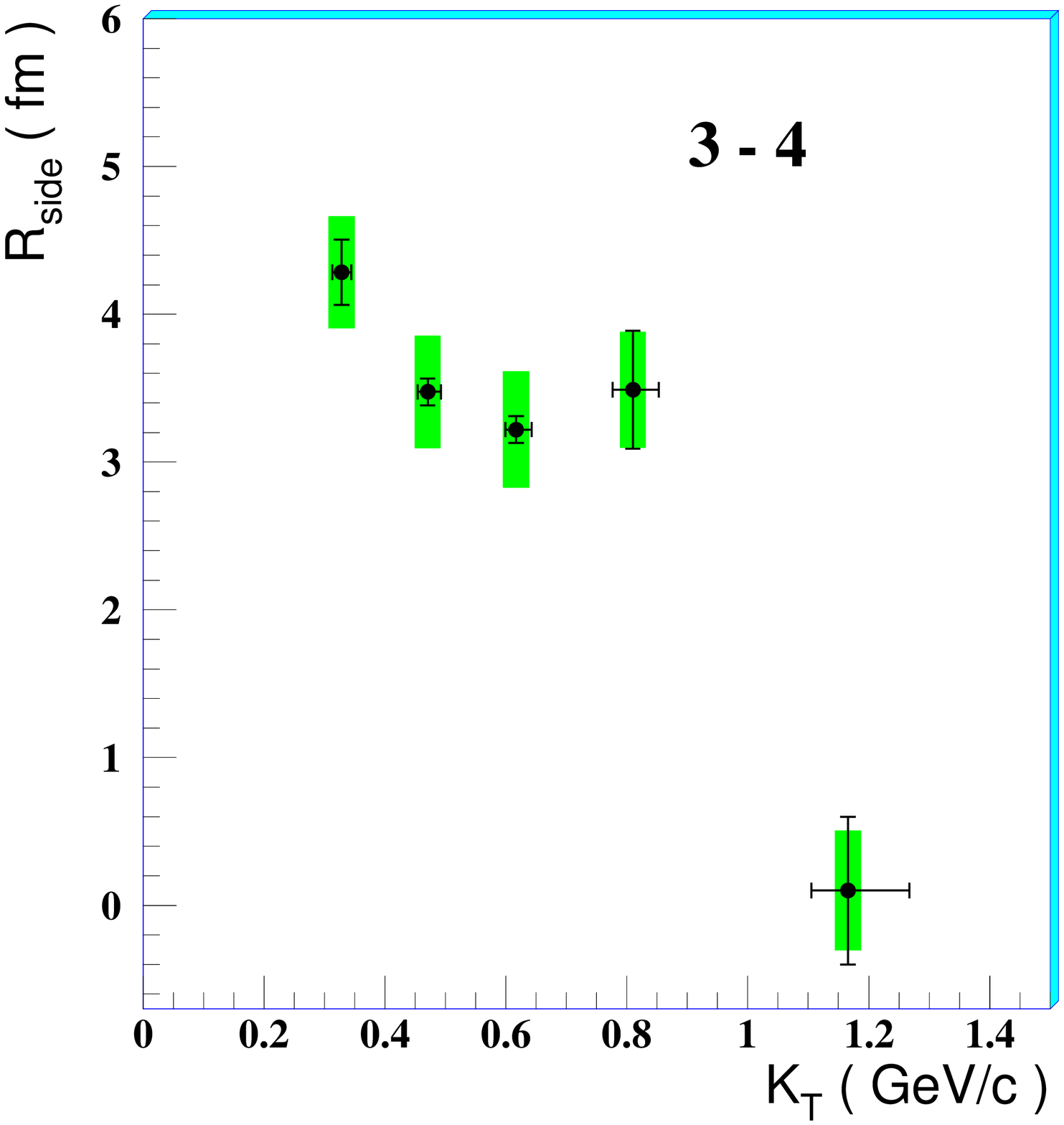}}
\caption{The $R_{side}$\ radius as a function of $\Kt$\ for
         the most central 53\% of the Pb--Pb cross-section (top-left panel) and for individual
         centrality classes 0--1 (top-right), 2 (bottom-left) and 3-4 (bottom right) of table 1.
         Statistical and systematic errors are indicated with line bars and shaded 
         boxes
         (green on-line), respectively.
         }
\label{fig:Rside}
\end{figure}

In figure~\ref{fig:Rside} we plot $\Rs$\ as a function of $\Kt$\ for 
the integrated centrality range and for the three centrality classes used in this HBT analysis.  
In the hydro-dynamical view, the decrease of $\Rs$\ with increasing $\Kt$\ is due to 
the collective expansion in the transverse direction;   
the dependence of $\Rs$\ on the pair momentum $\vec{K}$\ can be parameterized  
by the  formula~\cite{Trovami,Trovami2}  
\begin{equation}
\Rs(\Kt,Y_{\pi\pi})=\frac{R_{G}}{\sqrt{1+M_{\rm T} \frac{\beta_\perp^2}{T}\cosh(Y_{\rm yk}-Y_{\pi\pi})}}
\label{eq:Rs}
\nonumber
\end{equation}
where $M_{\rm T}=\sqrt{\Kt^2+m_{\pi}^2}$, $R_{\rm G}$\  
is equal to the transverse geometric (Gaussian) radius of the source times $1/\sqrt{2}$, $T$\ 
is the freeze-out temperature 
and $\beta_\perp$\ the slope of the (linear) transverse flow velocity profile: 
$\beta_\perp(r)=\beta_\perp\frac{r}{R_{\rm G}}$.  
A fit of equation~\ref{eq:Rs} to the experimental data points provides the model 
parameter $R_{\rm G}$\ and the ratio $\frac{\beta_\perp^2}{T}$; they are given 
in table~\ref{tab:Rs}. 

\begin{table}
\caption{The transverse geometric parameter $R_G$\
 and the ratio $\beta^2_{\perp}/T$, $T$\ being the freeze-out temperature.
\label{tab:Rs}}
\begin{center}
\begin{tabular}{||c|c|c||} \hline \hline
 Centrality & $ R_G \; ({\rm fm})$ &
$\frac{\beta_{\perp}^2}{T} \; ({\rm GeV}^{-1})$ \\ \hline
 ALL & $ 5.7 \pm 0.9 $ & $ 2.9 \pm 1.4 $  \\
 0-1 & $ 4.3 \pm 0.9 $ & $ 2.7 \pm 1.7 $  \\
  2  & $ 6.9 \pm 2.2 $ & $ 5.4 \pm 3.3 $  \\
 3-4 & $ 6.6 \pm 1.5 $ & $ 4.1 \pm 2.6 $  \\ \hline \hline
\end{tabular}
\end{center}
\end{table}

\begin{figure}[hbt]
\centering
\resizebox{0.74\textwidth}{!}{%
\includegraphics{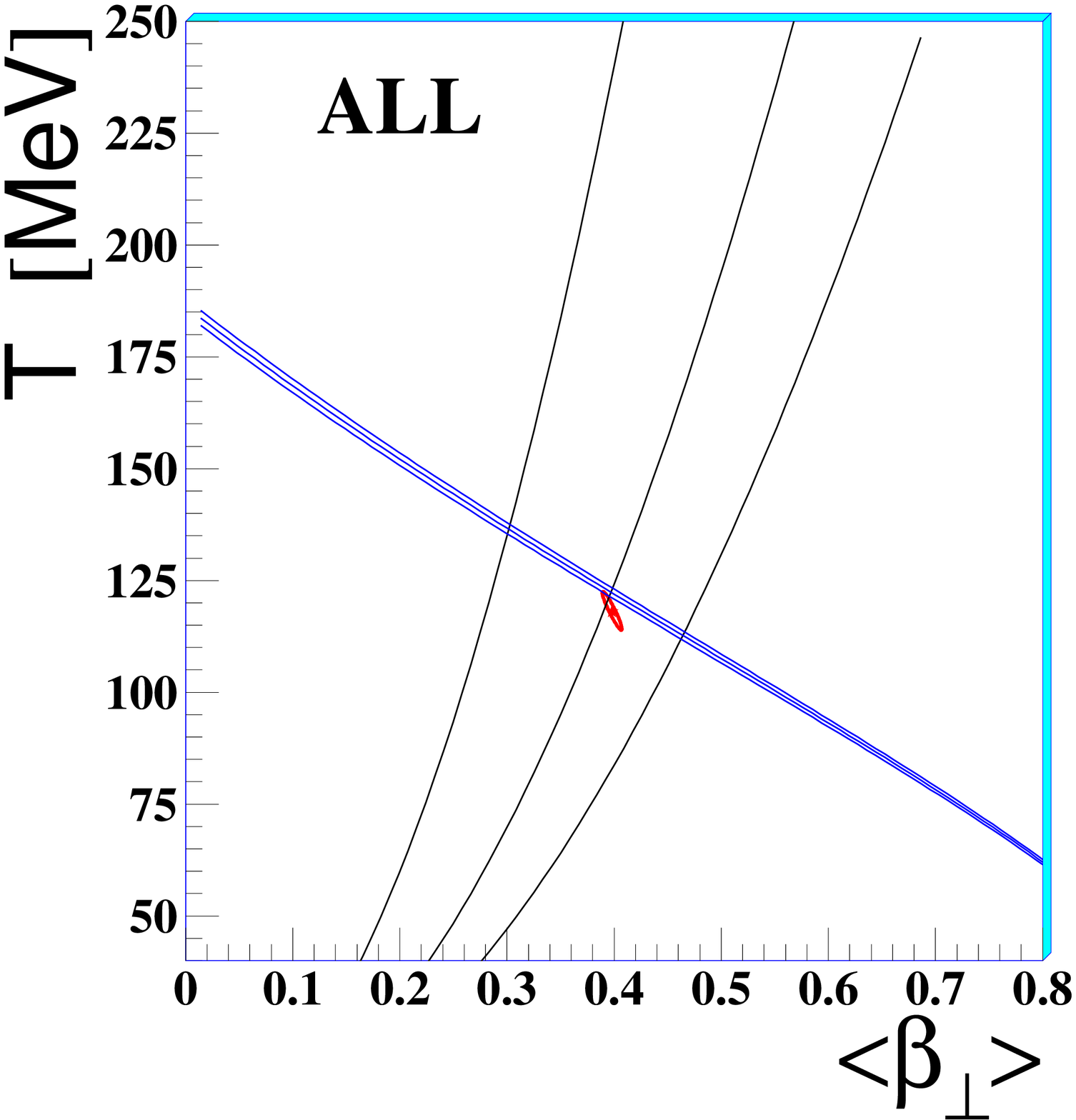}
\includegraphics{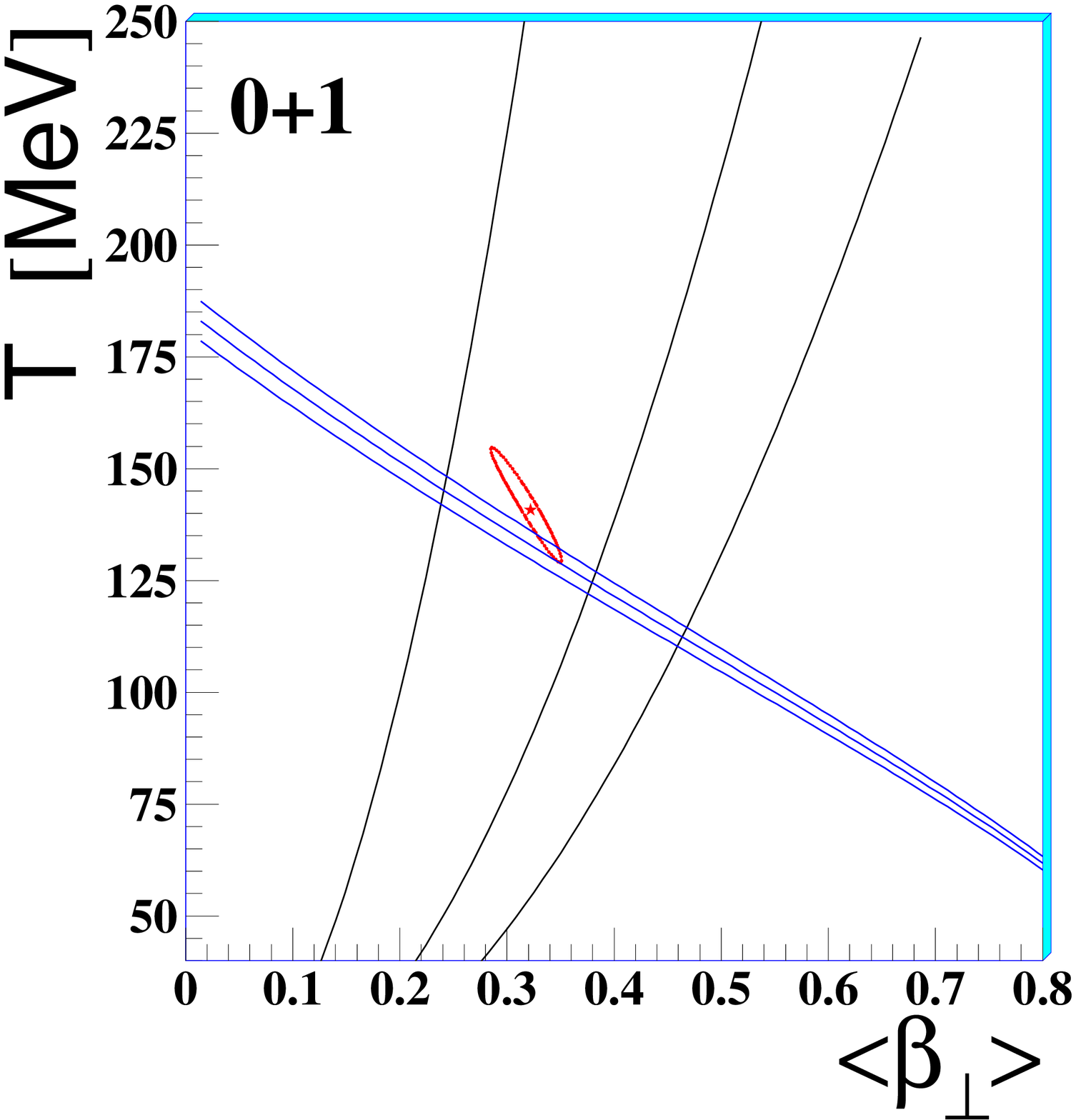}}\\
\resizebox{0.74\textwidth}{!}{%
\includegraphics{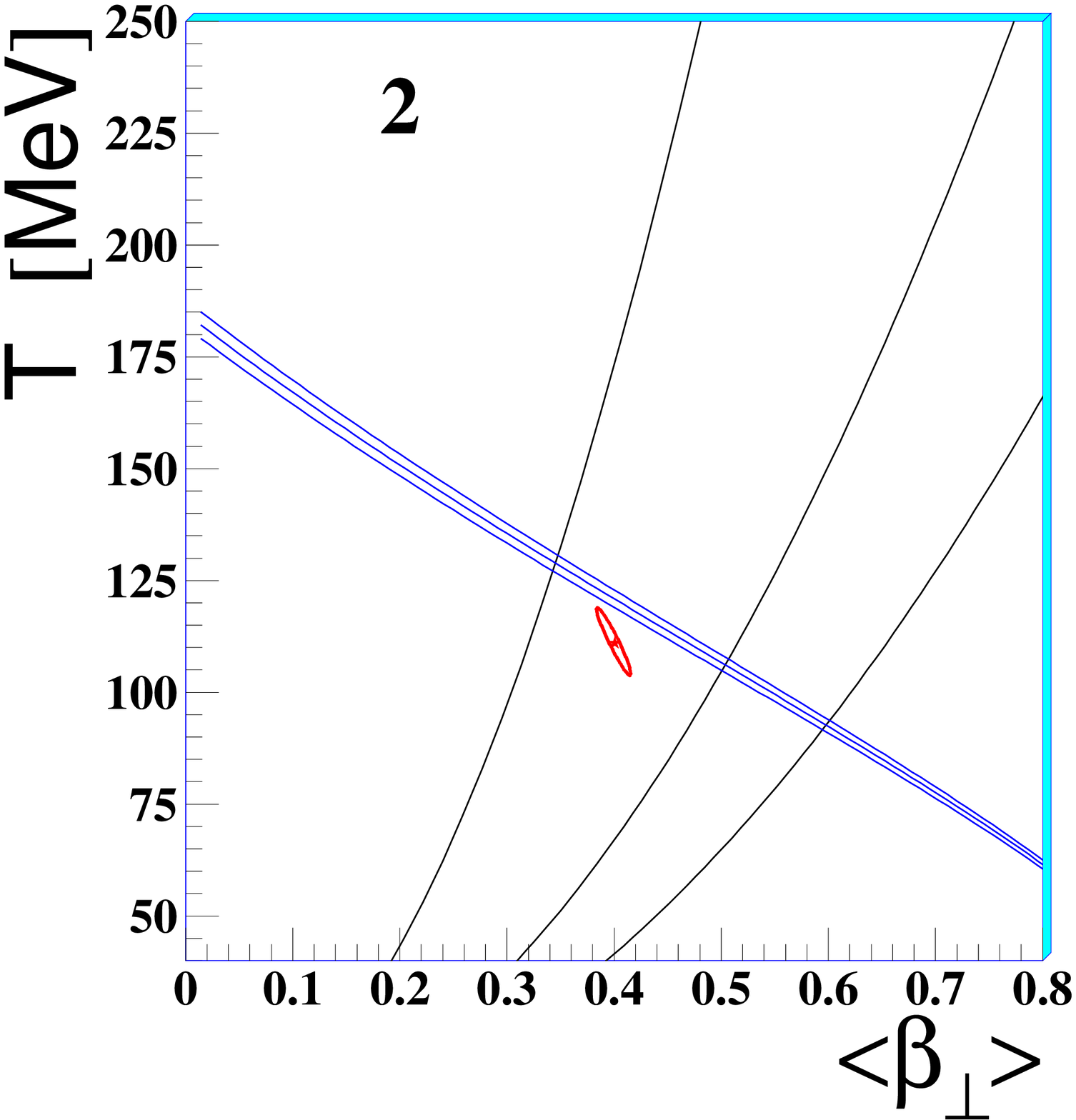}
\includegraphics{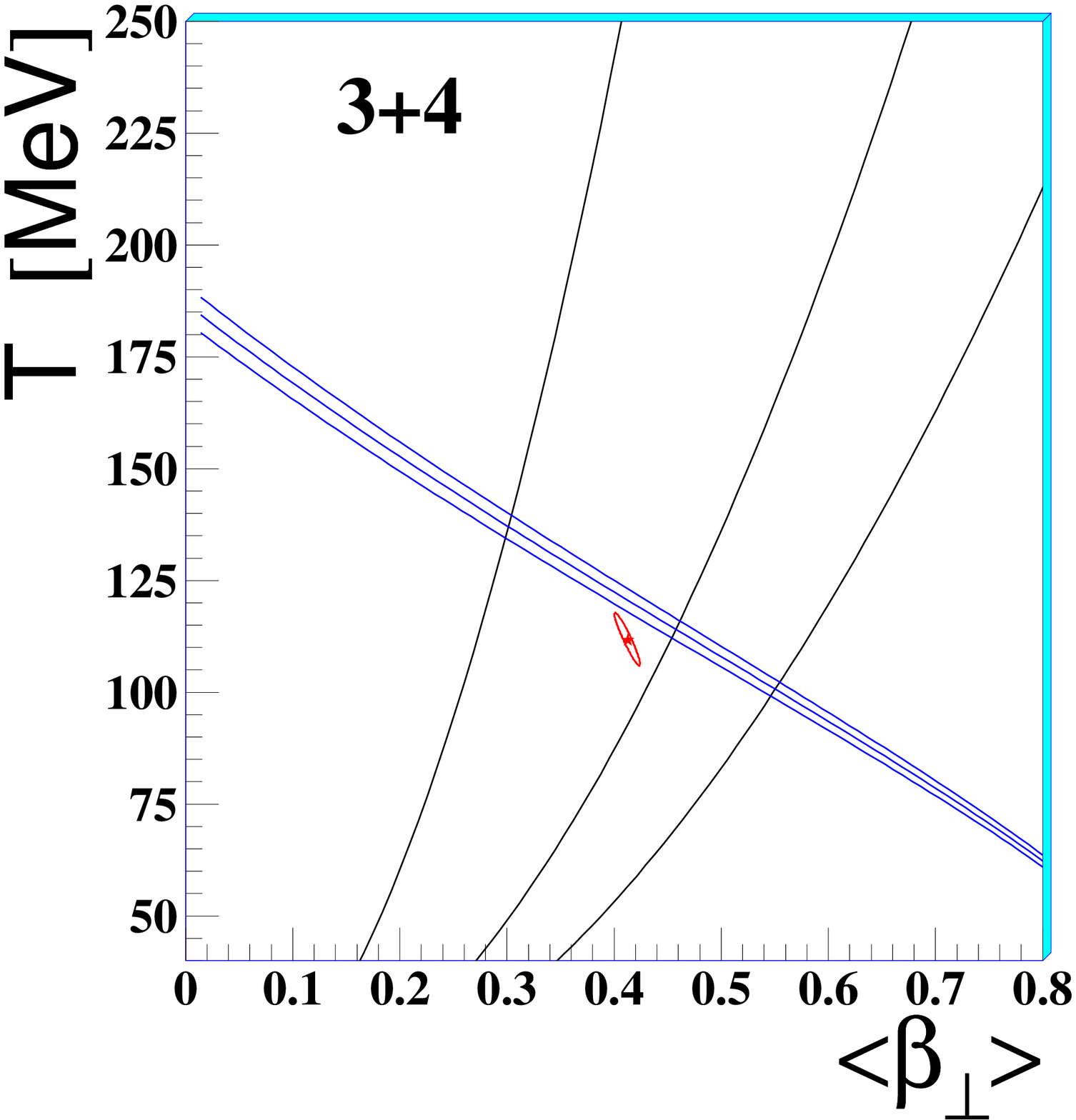}}
\caption{Contour plots at 1$\sigma$\ confidence level (statistical error)
         in the $\Bt$--$T$\ freeze-out
         parameter space, see text for details.  
         }
\label{fig:Contour}
\end{figure}

The measurement of the ratio $\frac{\beta_{\perp}^2}{T} $\ determines an allowed 
region for the pair of variables $\Bt$\ and $T$, the former being 
computed from the linear slope $\beta_{\perp}$\ by assuming a uniform particle 
density of the source.   
Those allowed regions, at the 1$\sigma$\ confidence 
level, are shown in figure~\ref{fig:Contour}; they correspond to  
the wide bands with positive slopes (in black on-line).  

In order to disentangle the two parameters, we followed a technique first used   
by the NA49 Collaboration~\cite{NA49HBT}, where  the measured
single-particle $\mt$\ spectra are exploited. In this approach, the inverse slope parameters 
of the $\mt$\ distributions 
are interpreted according to the blue-shift formula~\ref{eq:BlueShift}, 
as discussed in section~\ref{single}. This independent measurement provides another 
allowed region in the freeze-out parameter space, which corresponds to the narrow 
bands with negative slopes of figure~\ref{fig:Contour} (blue on-line).  
The disentangled values for $T$\ and $\Bt$ 
are reported in table~\ref{tab:Separated}.  
\begin{table}
\caption{The freeze-out temperature $T$\ and the
average transverse flow velocity $\Bt$\ as obtained from the study  
of negatively charged hadrons.  
\label{tab:Separated}}
\begin{center}
\begin{tabular}{||c|c|c||} \hline \hline
 Centrality & $ T ({\rm MeV})$ & $\Bt$ \\ \hline
 ALL & $ 122^{+15}_{-10} $  & $ 0.40^{+0.07}_{-0.10} $ \\
 0-1 & $ 125^{+20}_{-13} $ & $ 0.37^{+0.10}_{-0.15} $ \\
  2  & $ 106^{+22}_{-13} $ & $ 0.50^{+0.10}_{-0.16} $ \\
 3-4 & $ 114^{+22}_{-14} $ & $ 0.45^{+0.10}_{-0.15} $ \\  \hline \hline
\end{tabular}
\end{center}
\end{table}

In figure~\ref{fig:Contour} we also show the 1$\sigma$\ confidence regions obtained 
from the blast-wave analysis of the $\mt$\ spectra of 
singly-strange particles (\PKzS, \PgL\ and \PagL)~\cite{Blast40}. They correspond to the 
small closed contours (red on-line) of figure~\ref{fig:Contour}, with the markers 
indicating the optimal fit locations.   
Our results thus suggest compatible freeze-out conditions for singly-strange 
particles and $\hm$\ (mainly negative pions) in Pb--Pb collisions at 40 $A$\ GeV/$c$. 
On the other hand, the analysis of the transverse mass spectra of multiply-strange 
hyperons (\PgXm\ and \PgOm) 
suggested that for the same colliding system these particles  
may undergo an earlier 
freeze-out than singly-strange particles~\cite{Blast40}.  

Postponing to 
the next section  
a global discussion about these results, we note here
that the most peripheral class 0--1 features a smaller transverse freeze-out radius, 
a lower transverse expansion velocity and a higher temperature.  
This suggests an expansion on a smaller scale; the higher temperature at the freeze-out 
may be interpreted as the remnant of an earlier decoupling of the expanding system.
Such a centrality dependence of the freeze-out parameters  
is well established  at the SPS and RHIC 
both from studies of the transverse mass 
spectra~\cite{Blast40,BlastPaper,STARstrange,PHENIXcen,BRHAMScen}
and from HBT analyses~\cite{WA97HBT,CEREShbt,STARhbt,PHENIXhbt}.  

\subsection{Temporal characterization of the expansion}
Two parameters characterize the temporal evolution of the expansion dynamics, the 
proper time $\tau$\ of the kinetic freeze-out and the duration of the pion 
emission $\Delta\tau$.  
\subsubsection{Proper time of freeze-out}
\begin{figure}[hbt]
\centering
\resizebox{0.70\textwidth}{!}{%
\includegraphics{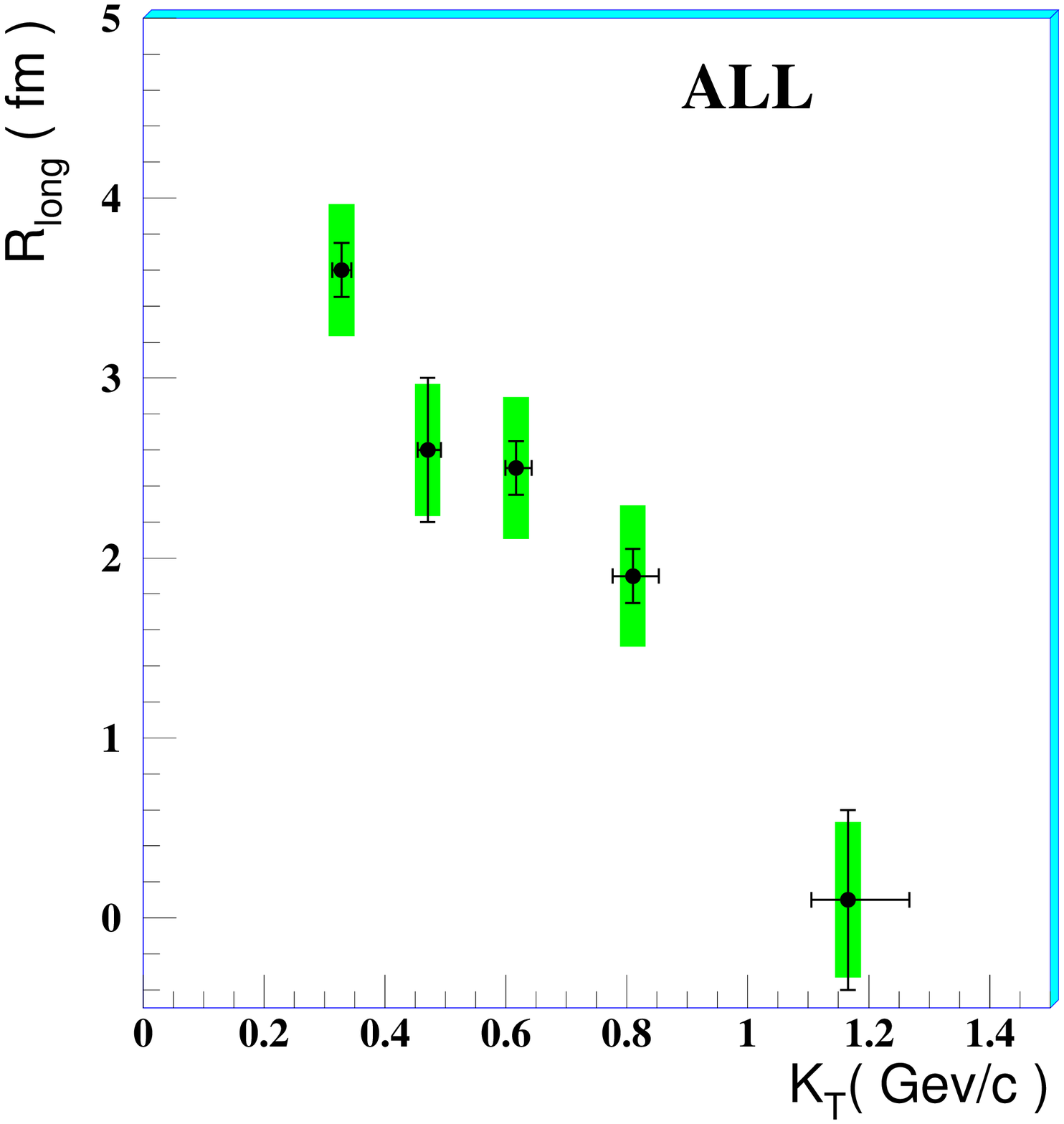}
\includegraphics{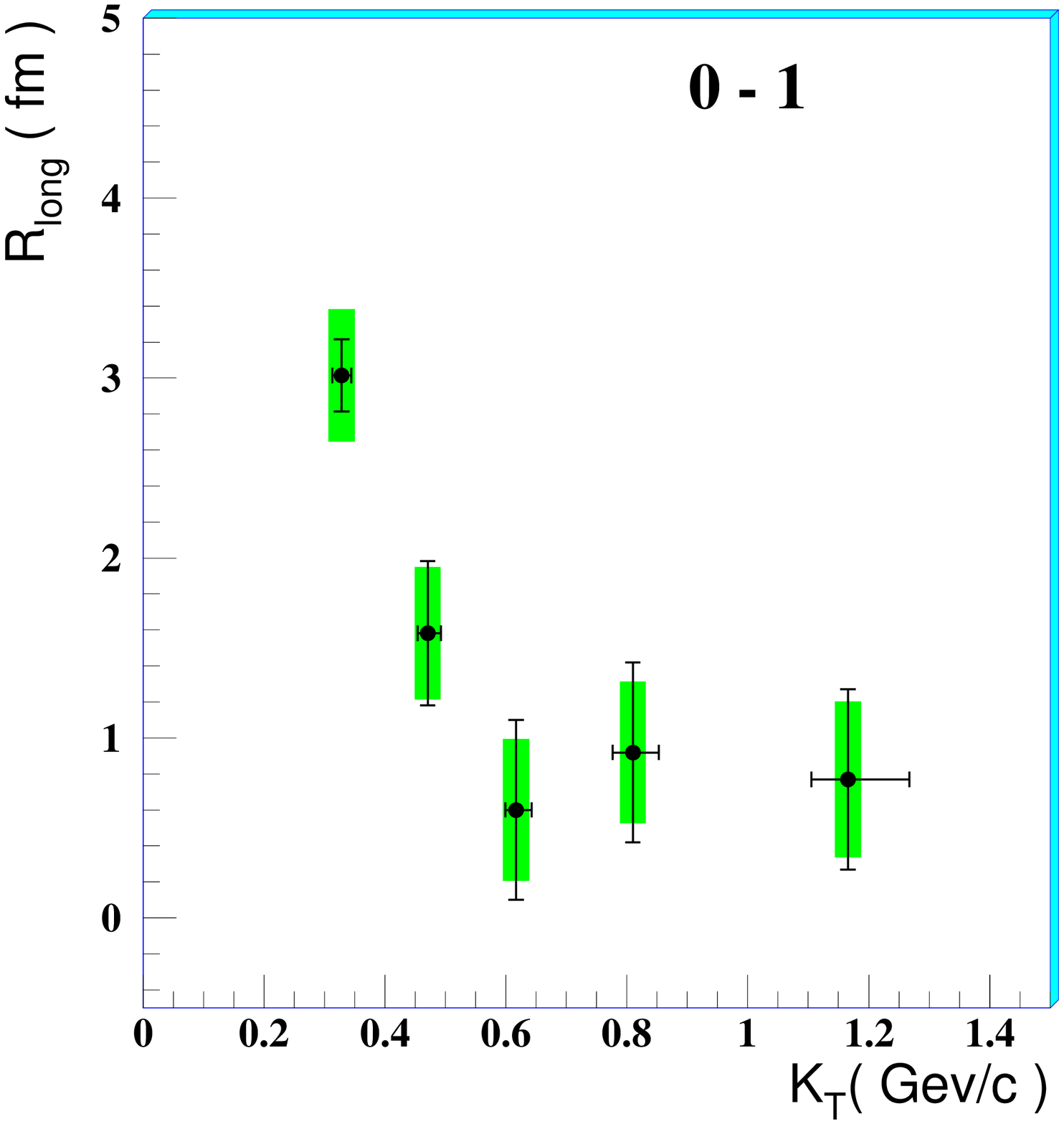}}\\
\resizebox{0.70\textwidth}{!}{%
\includegraphics{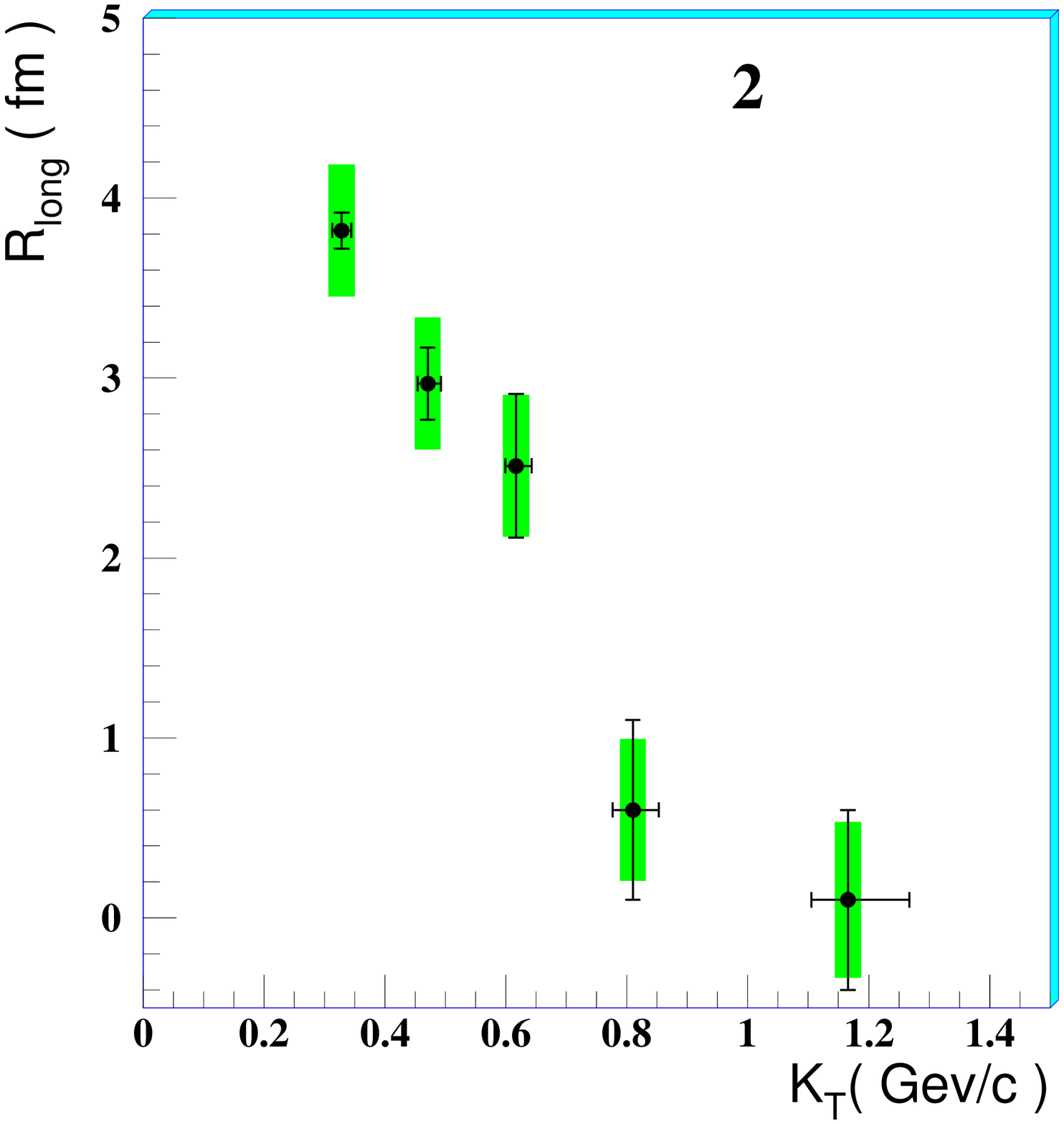}
\includegraphics{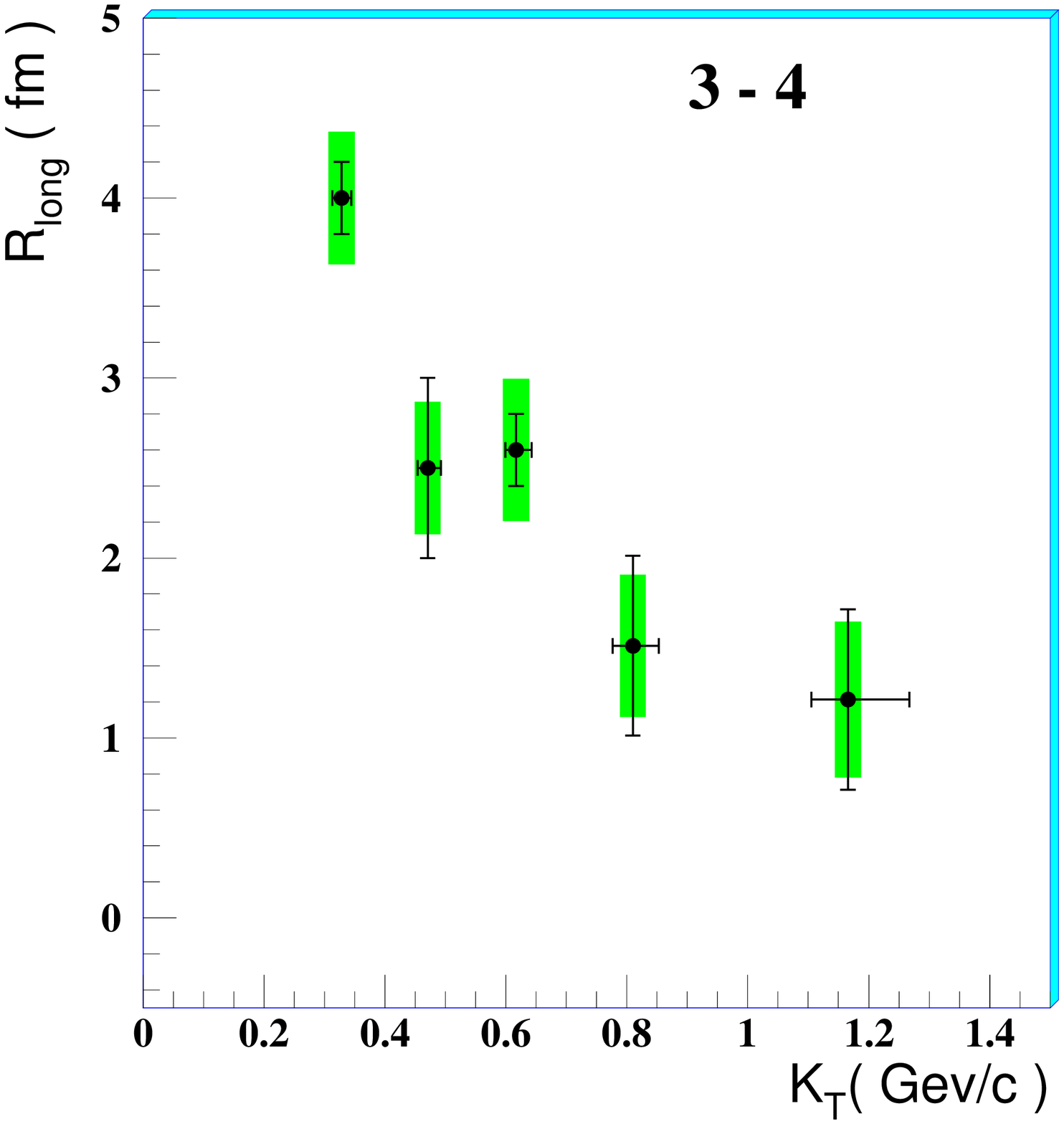}}
\caption{The $R_{long}$\ radius as a function of $\Kt$.    
         Statistical and systematic errors are indicated with line bars and shaded 
         boxes (green on-line), respectively.
         }
\label{fig:Rlong}
\end{figure}

Information about 
the evolution time-scale of the source, or proper
time of freeze-out, can be extracted from the $\vec{K}$-dependence of the $\Rl$\ radius.  
In figure~\ref{fig:Rlong} we show the $\Kt$\ dependence of $\Rl$. 
We have fitted this parameter to a formula first  
suggested by Sinyukov and collaborators~\cite{Sinyukov2}
and then improved by Chapman \etal~\cite{Trovami}: 
\begin{equation}
\Rl(\Kt,Y_{\pi\pi})=\sqrt{\frac{T}{\Mt}}\frac{\tau_{\rm f}}{
\cosh(Y_{\rm yk}-Y_{\pi\pi})}
\label{eq:Rl}
\nonumber
\end{equation}  
This formula assumes an instantaneous freeze-out in proper time (i.e., $\Delta\tau$ = 0); such 
an approximation is justified by the small $\Delta\tau$\ found in the present (see later) and similar 
analyses.  

The results of the bidimensional fits of equation~\ref{eq:Rl} to the experimental data 
are presented in table~\ref{tab:tau}. In the fitting procedure we have used 
the freeze-out temperature determined from the analysis of the transverse dynamics, as 
discussed in section~\ref{transv}. This introduces an error which is kept separate in 
table~\ref{tab:tau} from that due to the $\Rl$\ uncertainties.  
\begin{table}
\caption{The proper time of freeze-out ($\tau$) in integrated centrality class (all) and for 
         individual centrality classes. The second error is associated to the uncertainties on 
         the value of $T$, as discussed in the text.  
\label{tab:tau}}
\begin{center}
\begin{tabular}{||l|l||} \hline \hline
 Centrality &  $\tau$\ (fm/$c$) \\ \hline
 ALL &  $ 6.7\pm0.2 \; ^{+0.3} _{-0.3} $ \\
 0-1 &  $ 5.0\pm0.3 \; ^{+0.3} _{-0.4} $ \\
  2  &  $ 7.2\pm0.2 \; ^{+0.5} _{-0.7} $ \\
 3-4 &  $ 7.5\pm0.3 \; ^{+0.6} _{-0.6} $ \\  \hline \hline
\end{tabular}
\end{center}
\end{table}
As observed for the transverse expansion, the class 0-1 shows an expansion on a smaller scale,   
its freeze-out (proper) time $\tau$\ being significantly shorter than for the two 
most central classes.  
\subsubsection{Mean duration of the pion emission}
\begin{figure}[hbt]
\centering
\resizebox{0.65\textwidth}{!}{%
\includegraphics{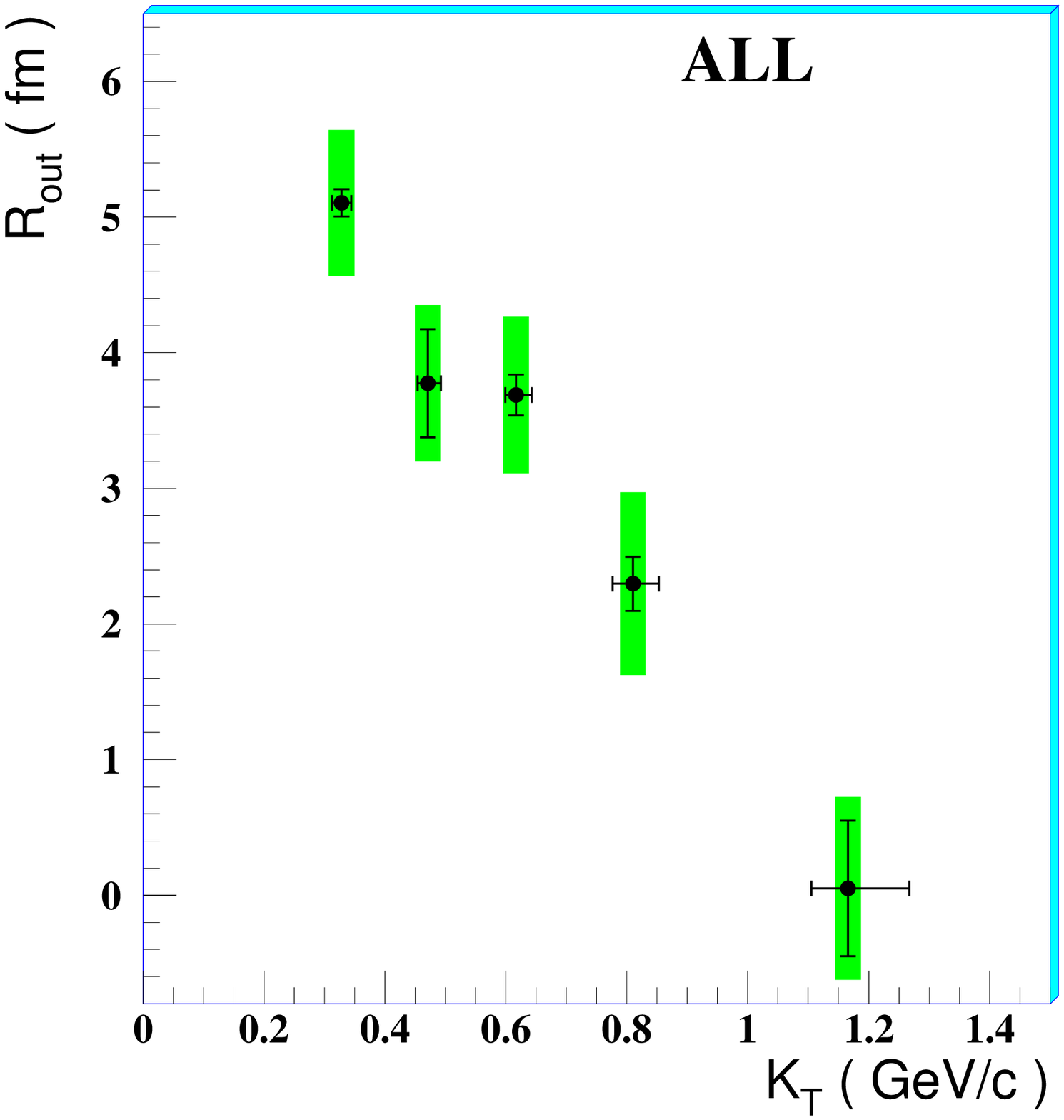}
\includegraphics{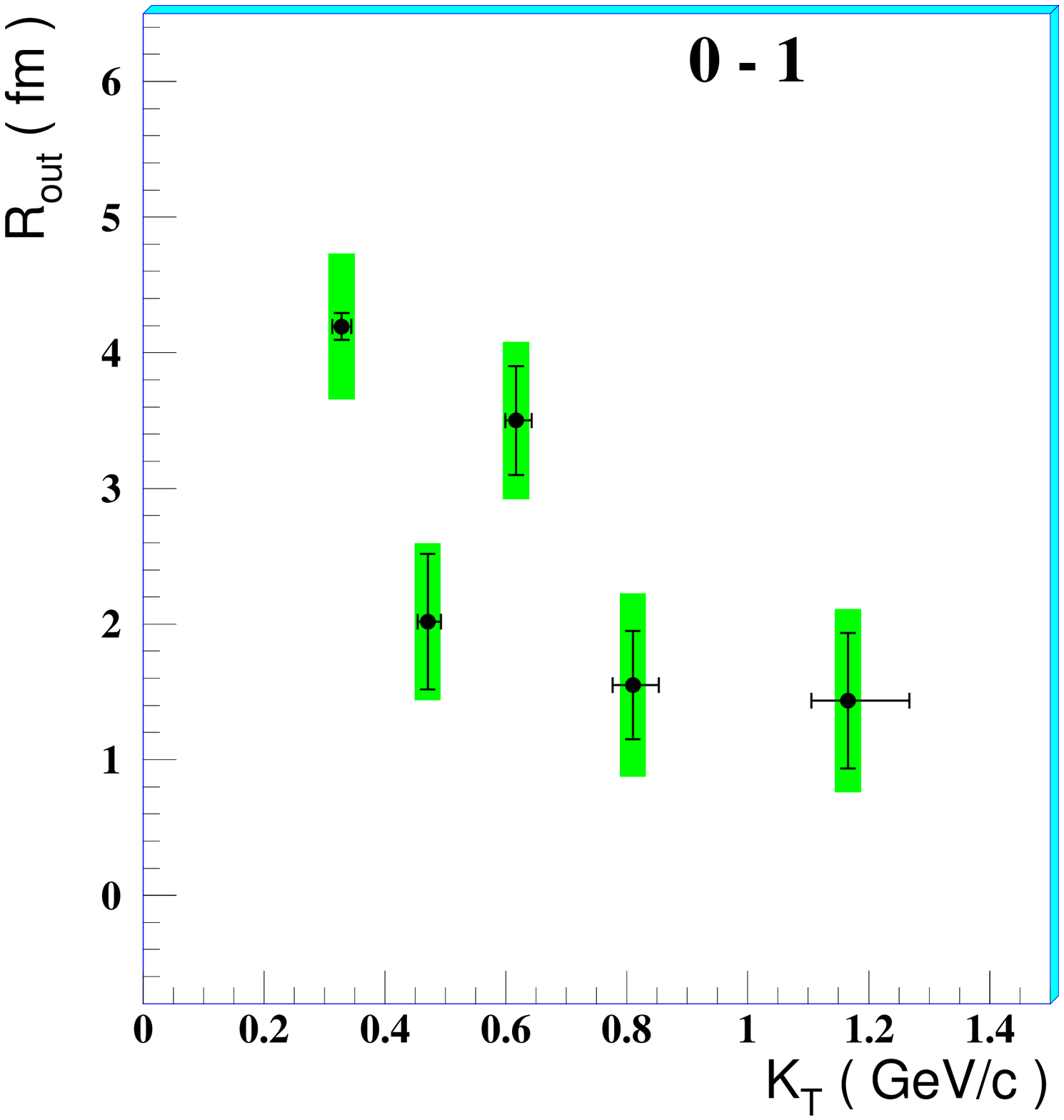}}\\
\resizebox{0.65\textwidth}{!}{%
\includegraphics{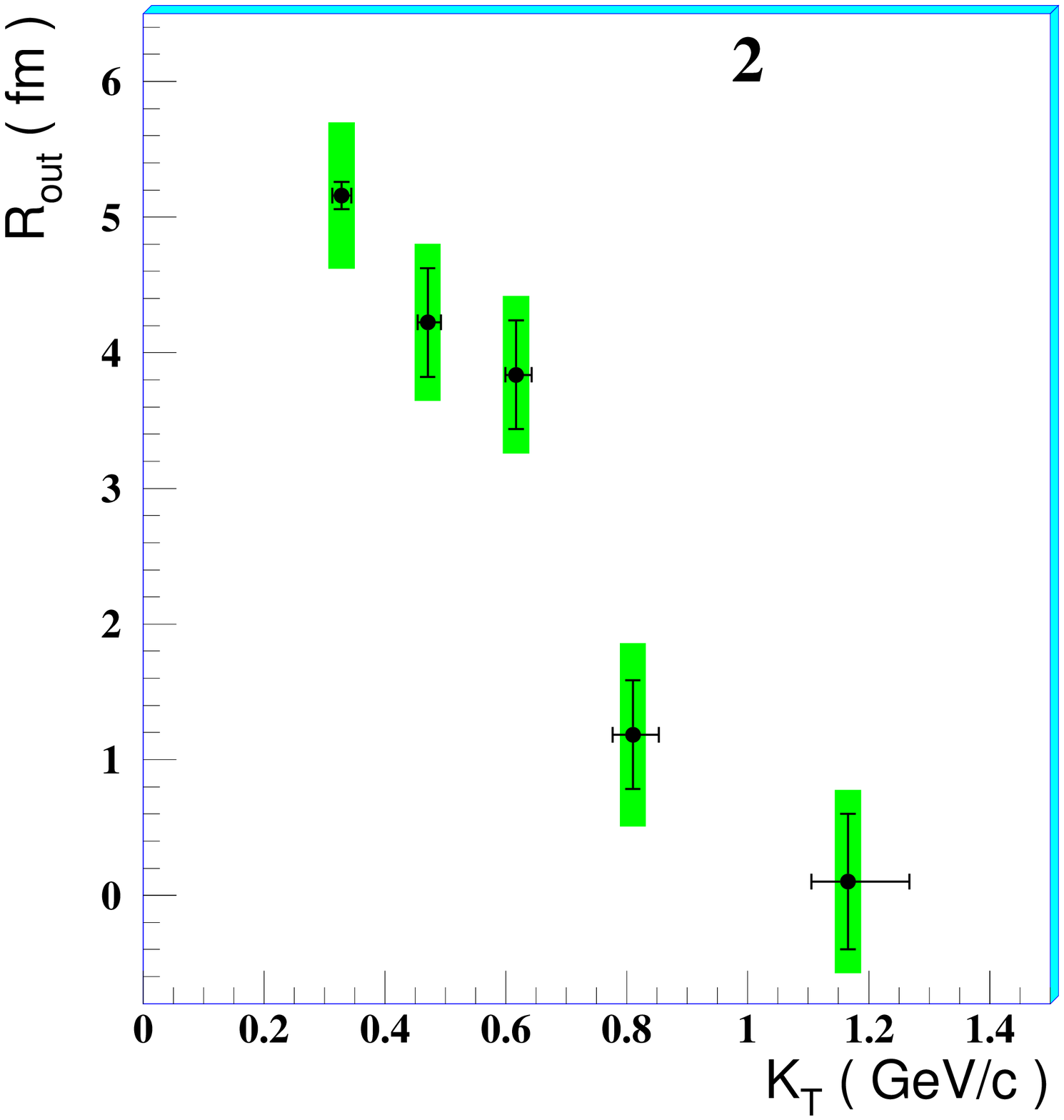}
\includegraphics{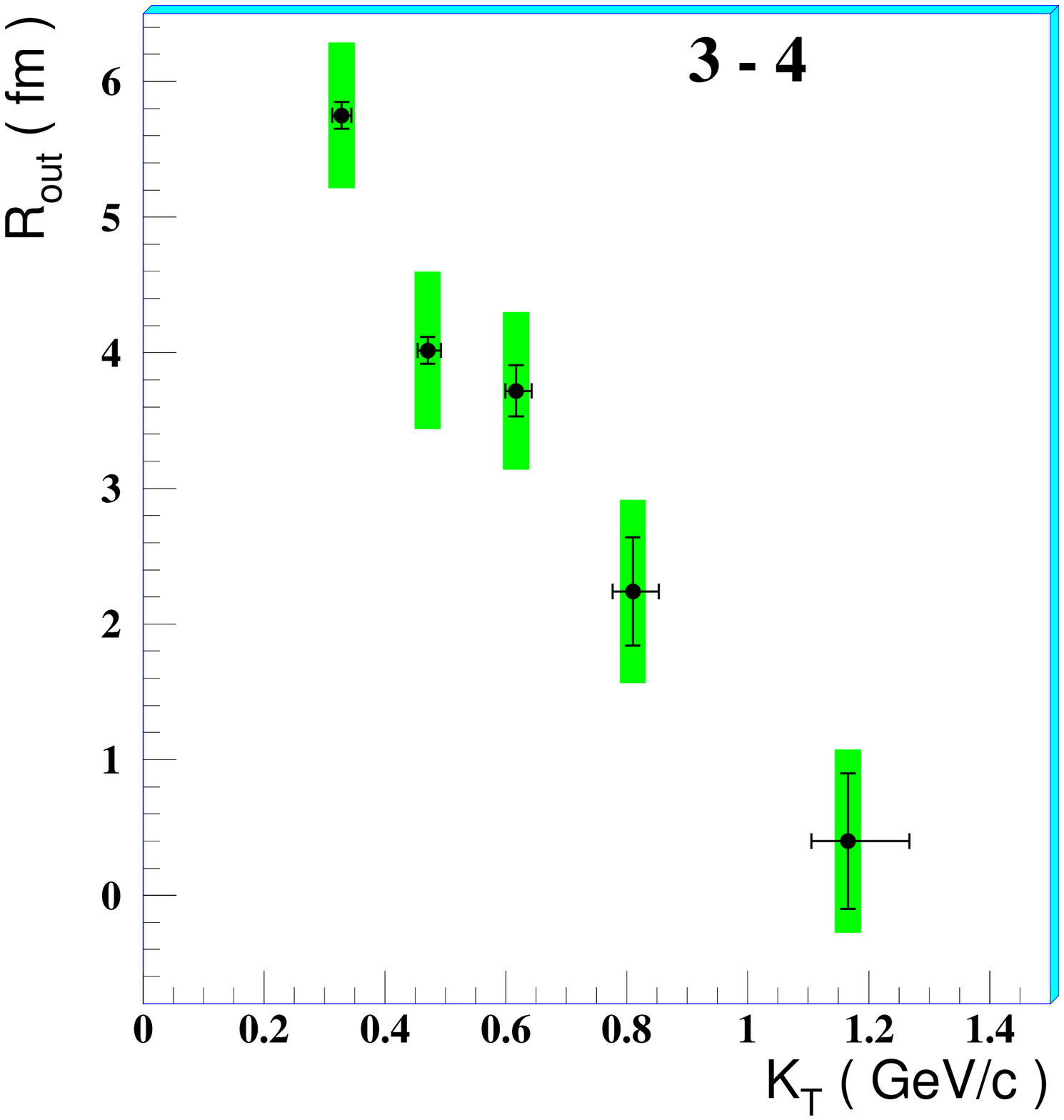}}
\caption{The $R_{out}$\ radius as a function of $\Kt$\ for
         the most central 53\% of the Pb--Pb cross-section (top-left panel) and for individual
         centrality classes 0--1 (top-right), 2 (bottom-left) and 3-4 (bottom right) of table 1.
         Statistical and systematic errors are indicated with line bars and shaded 
         boxes  
         (green on-line), respectively.
         }
\label{fig:Rout}
\end{figure}

\begin{figure}[tb]
\centering
\resizebox{0.35\textwidth}{!}{%
\includegraphics{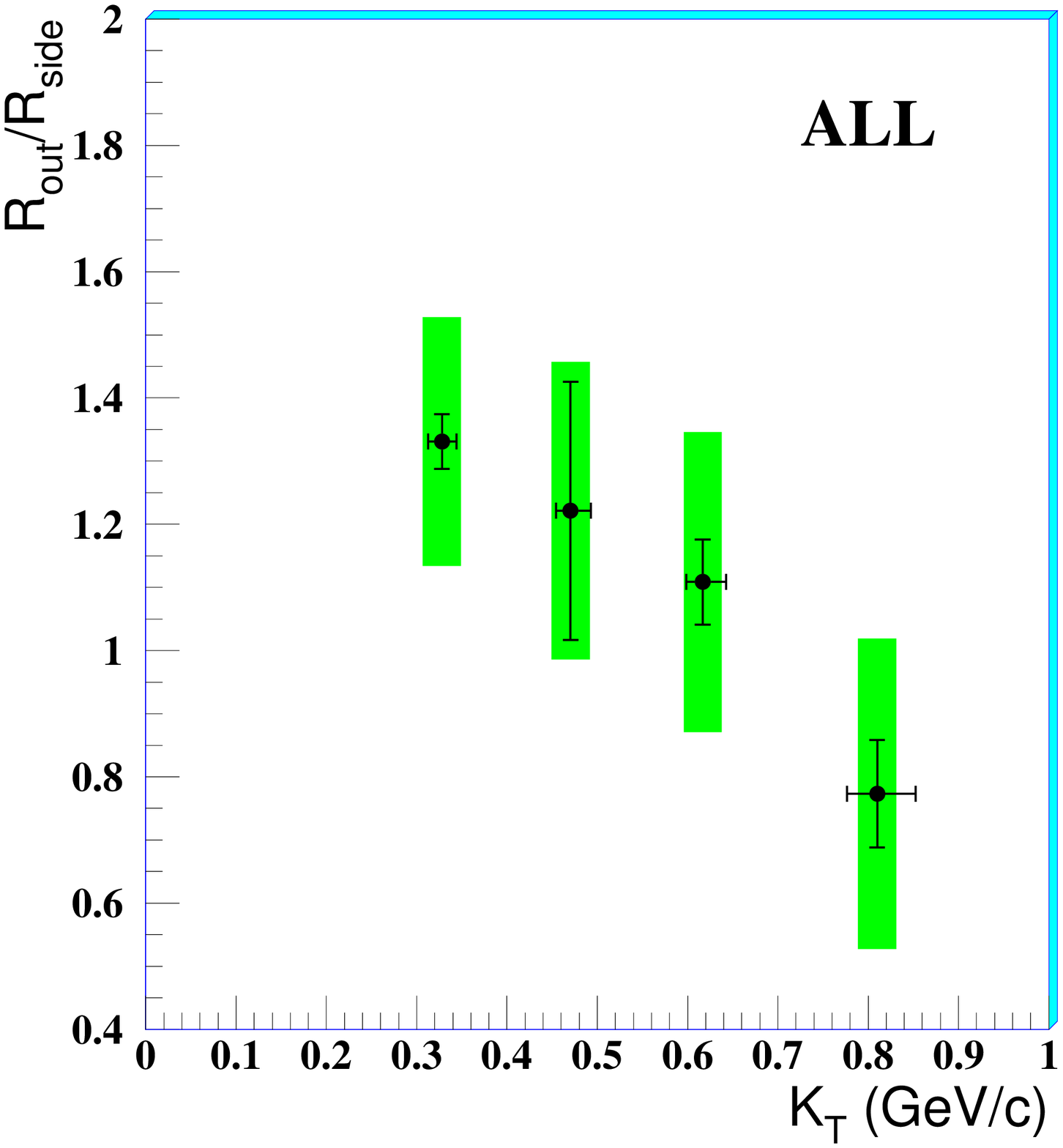}}
\caption{The ratio $R_{\tt out}/R_{\tt side}$\ as a function of $\Kt$\ for the data sample 
         corresponding to the most central 53\% of the Pb--Pb inelastic cross-section.}
\label{fig:ratio}
%
\centering
\resizebox{0.95\textwidth}{!}{%
\includegraphics{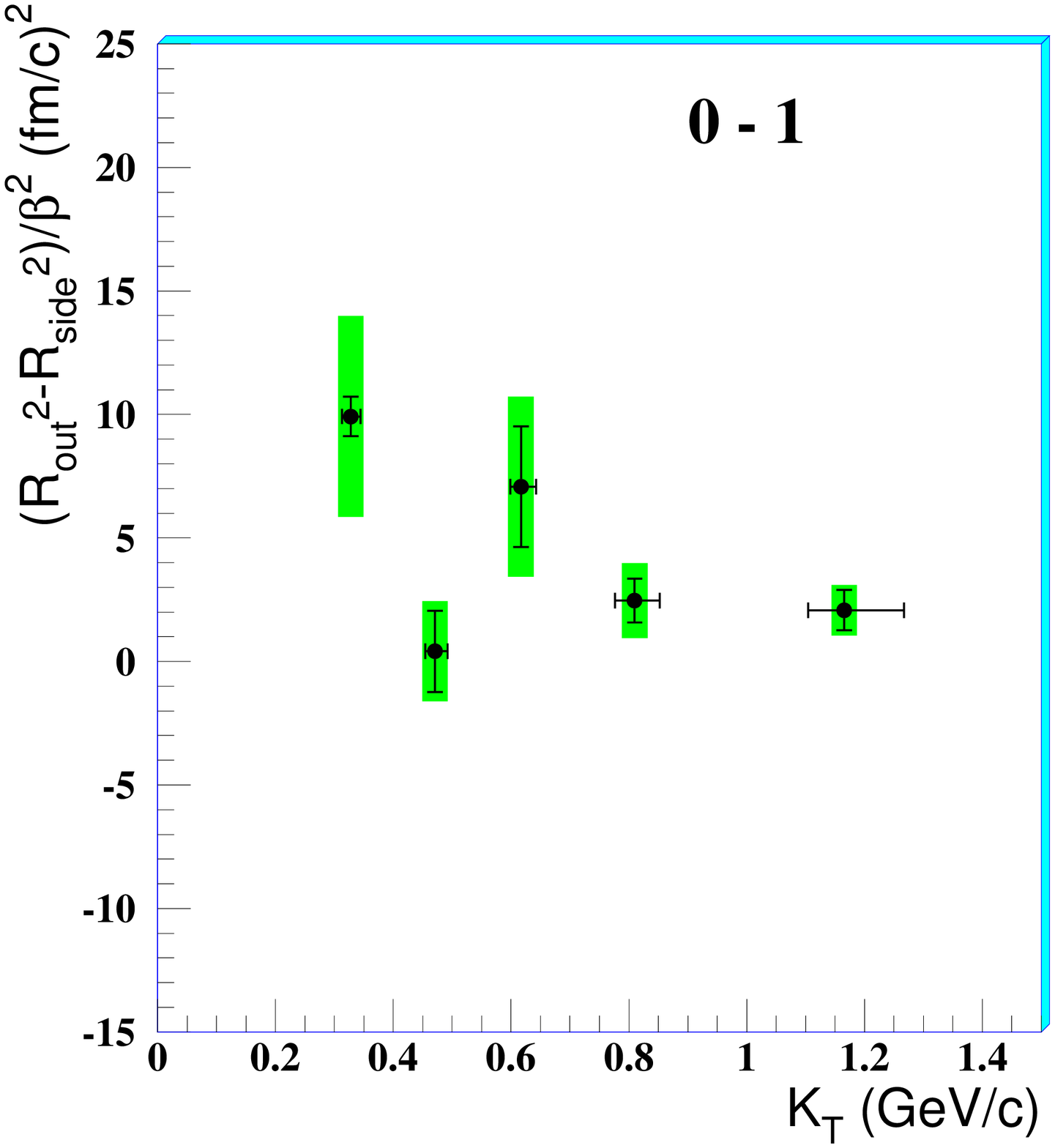}
\includegraphics{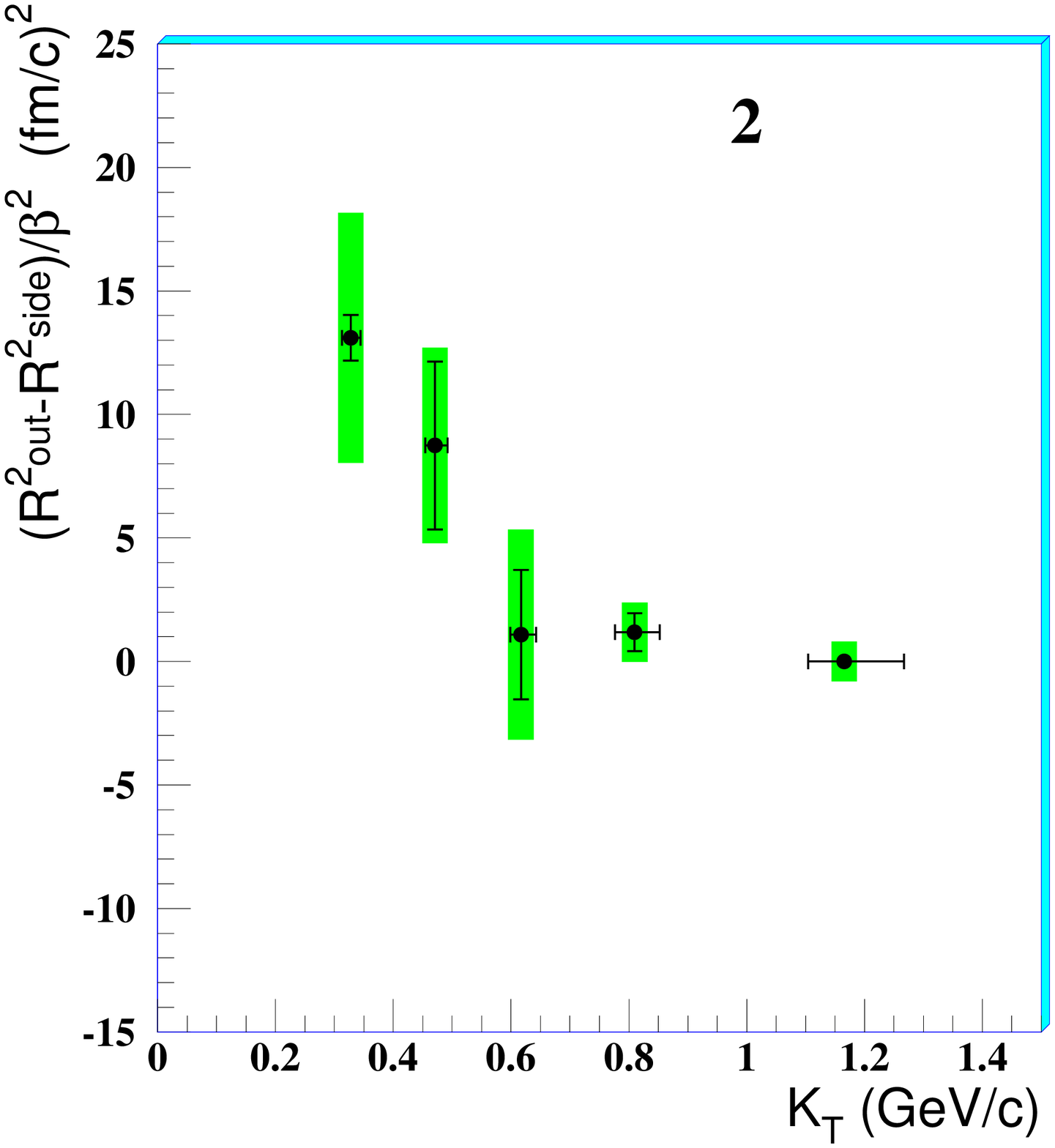}
\includegraphics{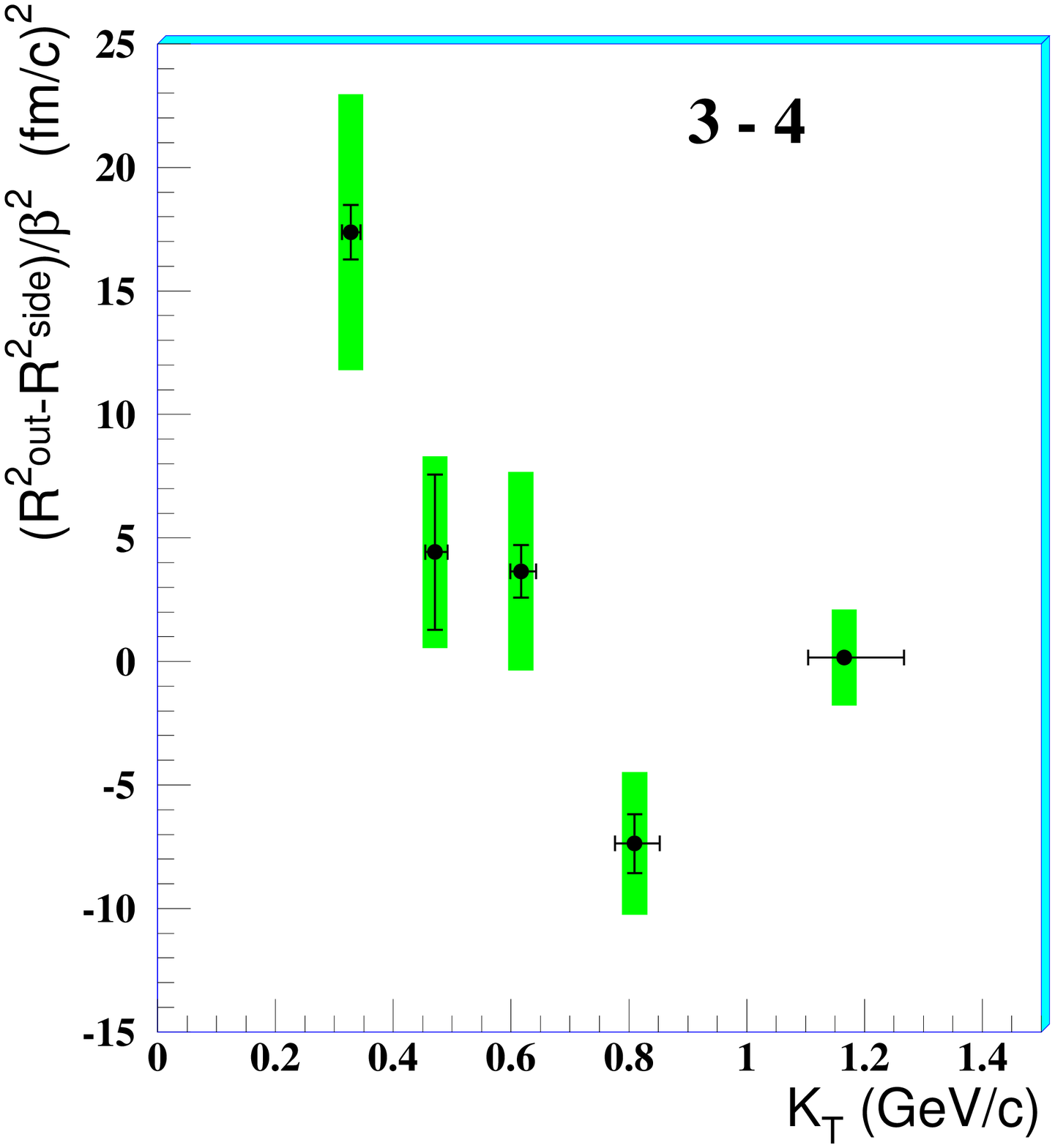}
}
\caption{The ${\Delta\tau}^2$\ parameter of equation~\ref{eq:DeltaTau} 
         as a function of $\Kt$\ for the three centrality classes.  
         Statistical and systematic errors are indicated with line bars and shaded 
         boxes
         (green on-line), respectively.
         }
\label{fig:DeltaT}
\end{figure}

It has been proposed that the existence of a strong first order phase transition and 
an accordingly long-lived mixed phase would be observable by a large outward radius $\Ro$\ 
compared to $\Rs$, indicating a long duration of the pion emission 
$\Delta\tau$~\cite{4,5,6,7,8,9}:  
\begin{equation}
{\Delta\tau}^2=\frac{\Ro^2-\Rs^2}{\beta_{\rm T}^2}
\label{eq:DeltaTau}
\end{equation}
where $\beta_{\rm T}$\ is the average transverse velocity of the pair. 
The dependence of the $\Ro$\ parameter on $\Kt$\ is shown in figure~\ref{fig:Rout}.  
We found that the ratio $\Ro/\Rs$ is compatible with one 
in the explored $\Kt$\ range (figure~\ref{fig:ratio}).   
Values smaller than unity for this ratio can be expected for sources with surface 
dominated emission~\cite{35}, such as emission from an expanding shell. 
We plot  
the quantity 
$\frac{\Ro^2-\Rs^2}{\beta_{\rm T}^2}$\    
for the individual centrality classes in figure~\ref{fig:DeltaT}.  
The data do not support the scenario of a long-lived source, but more likely that of a 
sudden freeze-out, $\Delta\tau$\ being of the order of a few fm/$c$ 
for small $\Kt$\ and nearly zero for high $\Kt$.     

\subsection{Consistency}
In the first three rows of table~\ref{tab:comparison} we have summarized the model parameters 
obtained from the study of $\hm$\ single-particle $\mt$\ spectra and two-particle correlation 
functions for the three centrality classes considered.   
These data provide implicitly a dynamical picture of the collision process.  

To see  
if this picture is self-consistent, we can compare 
the two-dimensional rms width $R^{\rm freeze\, out}_{\rm rms} = \sqrt{2}R_{\rm G}$\ 
with the two-dimensional rms width of a cold lead nucleus  
$R^{\rm Pb}_{\rm rms}=\sqrt{\frac{3}{5}}1.2A^{1/3}\simeq 4.5$\ fm.  
For the most central collisions (class 3-4), the system expands by a factor $\approx$2 or,  
equivalently, by about 5 fm in the transverse direction. If the transverse flow velocity 
is equal to $\beta_{\perp}=\frac{3}{2}\Bt\simeq 0.67$\ at the surface  
during the whole expansion, in a time of $\tau_0 \simeq$ 7.5 fm/$c$\ nuclear matter would
travel over $\tau_0 \beta_{\perp}\simeq$ 5 fm in the transverse direction.  
This is consistent with the previous estimate from the difference 
$R^{\rm freeze\, out}_{\rm rms} - R^{\rm Pb}_{\rm rms}$. 
Therefore, the dynamical description of the system expansion is internally consistent.
Similar pictures can be drawn for the other two classes.   

\begin{table}[hbt]
\begin{center}
\footnotesize{
\begin{tabular}{|l|c|l|c|c|c|c|c|} \hline \hline
 Coll. & $\sqrt{s_{\tt NN}}$ & Centrality & $R_{\rm G}$ (fm) & $T$ (MeV)& $\Bt$ &$\tau$\ (fm/$c$)&$\Delta\tau$ (fm/c)\\ \hline
 NA57 & 8.8 & 0 -- 11 \% & $6.6 \pm 1.5$ & $114^{+22}_{-14}$ & $0.45_{-0.2}^{+0.1}$ & $7.5 \pm 0.7$ &\\
 NA57 & 8.8 & 11 -- 23 \% & $6.9 \pm 2.2$ & $106^{+22}_{-13}$ & $0.50_{-0.2}^{+0.1}$ & $7.2 \pm 0.6$ & $1 - 2$ \\
 NA57 & 8.8 & 23 -- 53 \% & $4.3 \pm 0.9$ & $125^{+20}_{-13}$ & $0.37_{-0.2}^{+0.1}$ & $5.0 \pm 0.5$ & \\  \hline
 CERES & 8.8 &  0 -- 5 \% & $12.1_{-2.6}^{+7.7}$ &  120   & - & $6.6 \pm 0.1$ & - \\
 CERES & 8.8 &  5 -- 10 \% & $9.9_{-2.0}^{+5.2}$ &  120  & - & $6.5 \pm 0.1$ & - \\
 CERES & 8.8 &  10 -- 15 \% & $10.1_{-2.2}^{+6.9}$ &  120 & - & $6.3 \pm 0.1$ & - \\
 CERES & 8.8 &  15 -- 30 \% & $6.0_{-0.7}^{+1.1}$ &  120 & - & $5.9 \pm 0.1$ & - \\ \hline \hline
 CERES & 12.3 &  0 -- 5 \% & $7.2_{-0.6}^{+0.8}$ &  120   & $0.58_{-0.06}^{+0.07}$ & $7.3 \pm 0.1$ & - \\
 CERES & 12.3 &  5 -- 10 \% & $6.2_{-0.4}^{+0.5}$ &  120  & $0.50_{-0.06}^{+0.06}$ & $7.3 \pm 0.1$ & - \\
 CERES & 12.3 &  10 -- 15 \% & $5.9_{-0.4}^{+0.6}$ &  120 & $0.50_{-0.07}^{+0.07}$ & $7.0 \pm 0.1$ & - \\
 CERES & 12.3 &  15 -- 30 \% & $5.3_{-1.0}^{+0.7}$ &  120 & $0.46_{-0.14}^{+0.13}$ & $6.5 \pm 0.2$ & - \\ \hline
 CERES & 17.3 &  0 -- 5 \% & $6.9_{-0.3}^{+0.3}$ &  120   & $0.49_{-0.06}^{+0.06}$ & $8.2 \pm 0.1$ &  \\
 CERES & 17.3 &  5 -- 10 \% & $5.9_{-0.2}^{+0.3}$ &  120  & $0.53_{-0.05}^{+0.06}$ & $7.9 \pm 0.1$ & $ 2 - 3$ \\
 CERES & 17.3 &  10 -- 15 \% & $5.9_{-0.4}^{+0.4}$ &  120 & $0.46_{-0.04}^{+0.04}$ & $7.5 \pm 0.1$ &  \\
 CERES & 17.3 &  15 -- 30 \% & $5.5_{-0.4}^{+0.4}$ &  120 & $0.55_{-0.03}^{+0.03}$ & $7.3 \pm 0.1$ &  \\  \hline
 WA97 & 17.3 &  0 -- 5\% & $5.1 \pm 0.6$ & $120_{-11}^{+15} $ & $0.46_{-0.10}^{+0.07}$ & $5.6 \pm 0.2$ &\\
 WA97 & 17.3 &  5 -- 12\% & $5.0 \pm 0.6$ & $117_{-11}^{+16} $ & $0.48_{-0.11}^{+0.08}$ & $5.6 \pm 0.2$ & $\approx 0$ \\
 WA97 & 17.3 &  12 -- 25\% & $4.6 \pm 0.4$ & $121_{-11}^{+15} $ & $0.47_{-0.10}^{+0.07}$ & $5.1 \pm 0.2$ & \\
 WA97 & 17.3 &  25 -- 40\% & $3.2 \pm 0.3$ & $140_{-13}^{+26} $ & $0.30_{-0.16}^{+0.09}$ & $3.7 \pm 0.2$ &\\ \hline
 NA49 & 17.3 &  0 -- 3\% & $6.5 \pm 0.5$ & $120\pm12$ & $0.55\pm0.12$ & 8 & $ 3 - 4$ \\ \hline
 PHENIX  & 130 &  0 - 30\% & $8.1 \pm 0.3$ &  $125$  & $ 0.51 $ & - &\\ \hline
 STAR  & 200 &  0 - 5\% & $13.3 \pm 0.2$ & $97 \pm 2 $ & $0.59\pm0.05$ & $9.0 \pm0.3$ & $2.83\pm 0.19$ \\
 STAR  & 200 &  5 - 10\% & $12.6 \pm 0.2$ & $98 \pm 2 $& $0.58\pm0.05$ & $8.7 \pm 0.2$ & $2.45 \pm 0.17$\\
 STAR  & 200 &  10 - 20\% & $11.5 \pm 0.2$ & $98 \pm 3$&$0.56\pm0.06$& $8.1 \pm 0.2$ & $2.35 \pm 0.16$ \\
 STAR  & 200 &  20 - 30\% & $10.5 \pm 0.1$ & $100 \pm 2 $ & $0.54\pm0.06$& $7.2 \pm 0.1$ & $2.10 \pm 0.09$ \\
 STAR  & 200 &  30 - 50\% & $8.8 \pm 0.1$ & $108 \pm 2 $ &$0.51\pm0.06$& $5.9 \pm 0.1$ &$1.74 \pm 0.12$ \\
 STAR  & 200 &  50 - 80\% & $6.5 \pm 0.1$ & $113 \pm 2 $ &$0.41\pm0.06$& $4.0 \pm 0.2$ & $1.73 \pm 0.10$ \\ \hline
\end{tabular}
}
\end{center}
\caption{Freeze-out parameters from HBT measurements in heavy ion collisions at SPS and RHIC.
        In the CERES~\cite{CEREShbt} and PHENIX~\cite{PHENIXhbt} analyses the freeze-out temperature
        has been fixed to the quoted values. The reported values of $\Bt$\ for the PHENIX 
        results have been computed from their parameterization of the transverse flow,
        which is assumed to be linear in the rapidity profile.
        In the STAR analysis~\cite{STARhbt}, the values of $\Bt$\ and $T$\
        have been  extracted from blast-wave fits to pion, kaon,
        and proton transverse momentum spectra~\cite{star67} and used to describe HBT radii.
\label{tab:comparison}
        }
\end{table}
\section{Comparison with other experiments}
\begin{figure}[hbt]
\centering
\resizebox{0.56\textwidth}{!}{%
\includegraphics{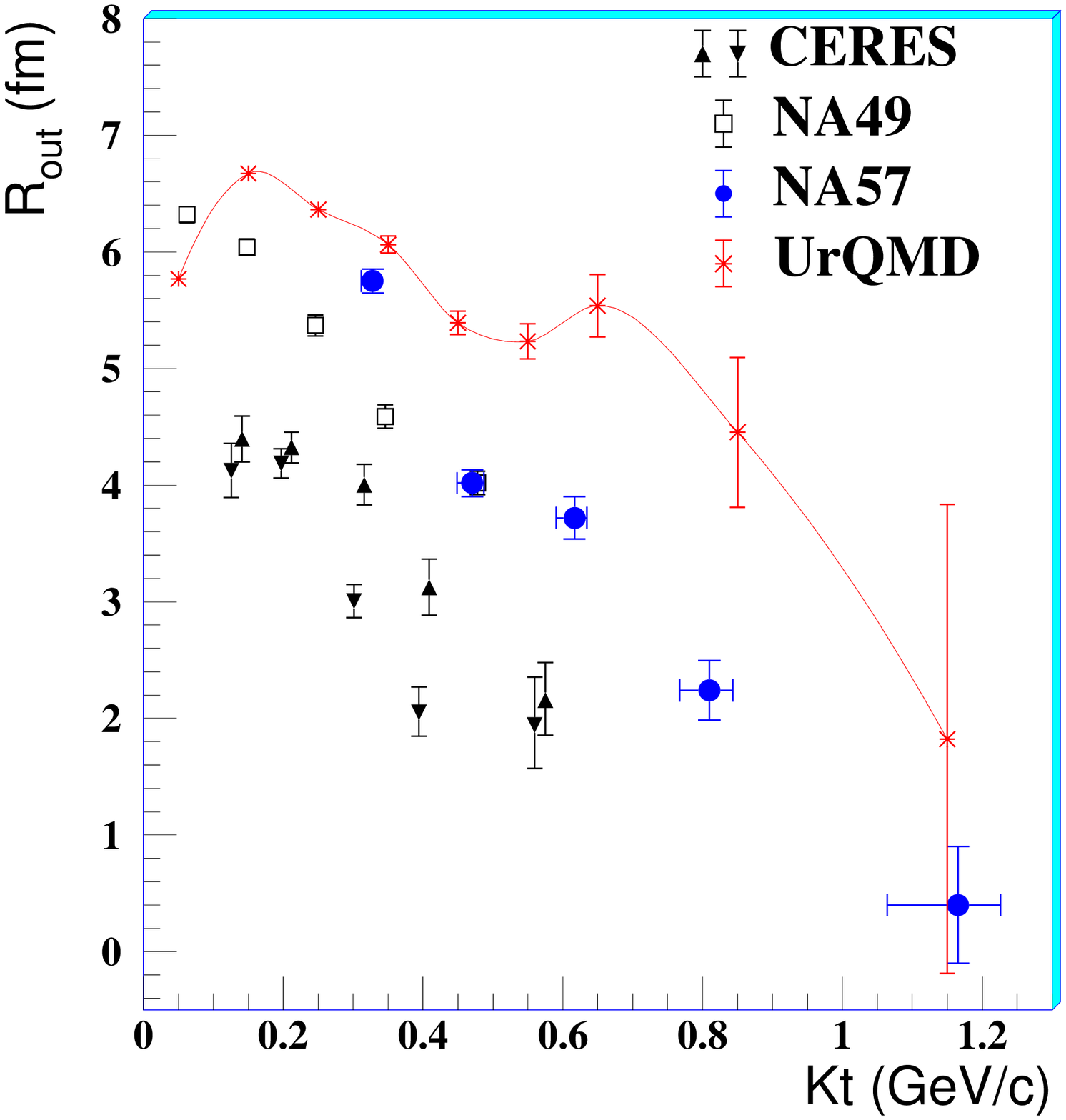}
\includegraphics{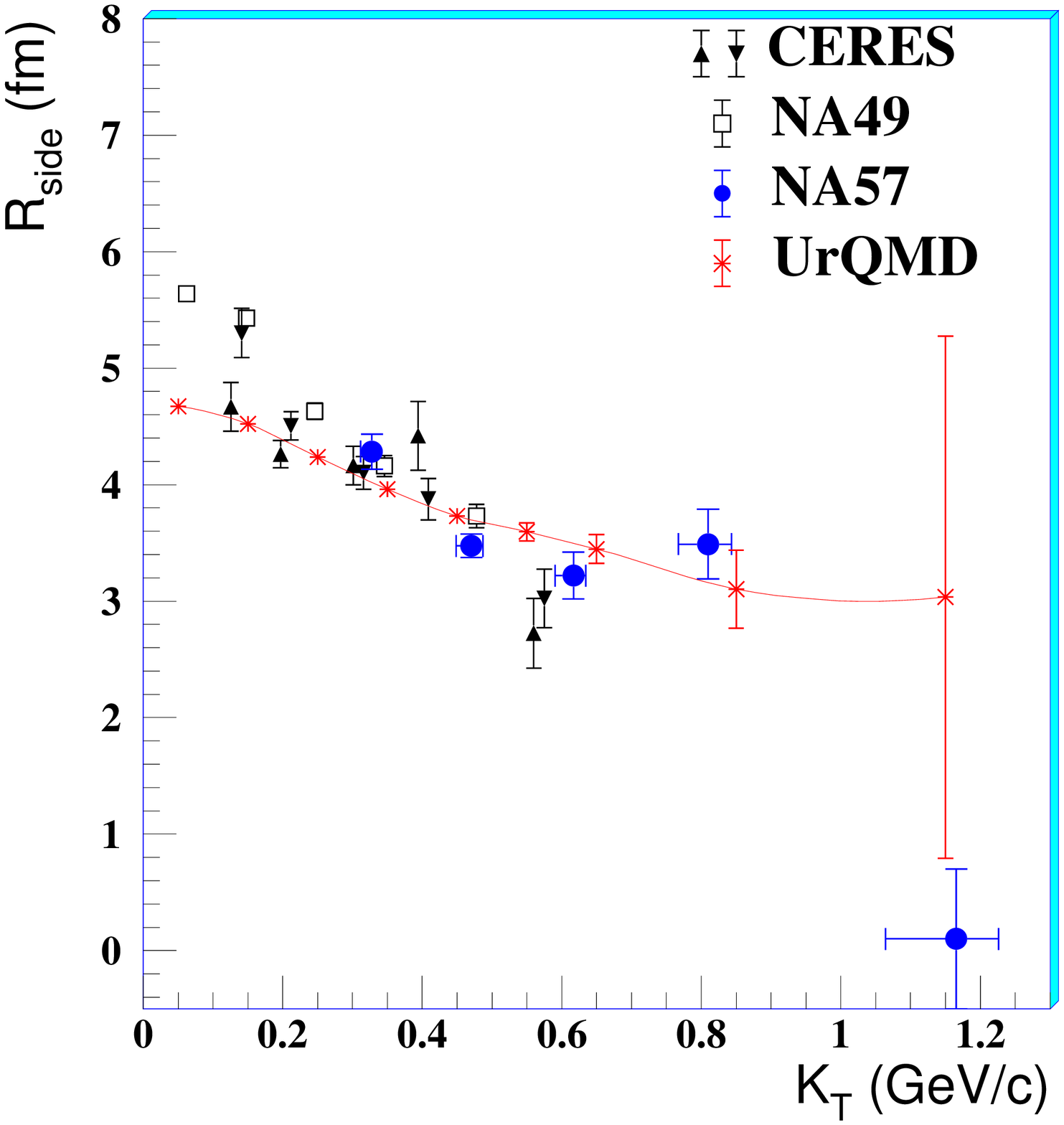}}\\
\resizebox{0.56\textwidth}{!}{%
\includegraphics{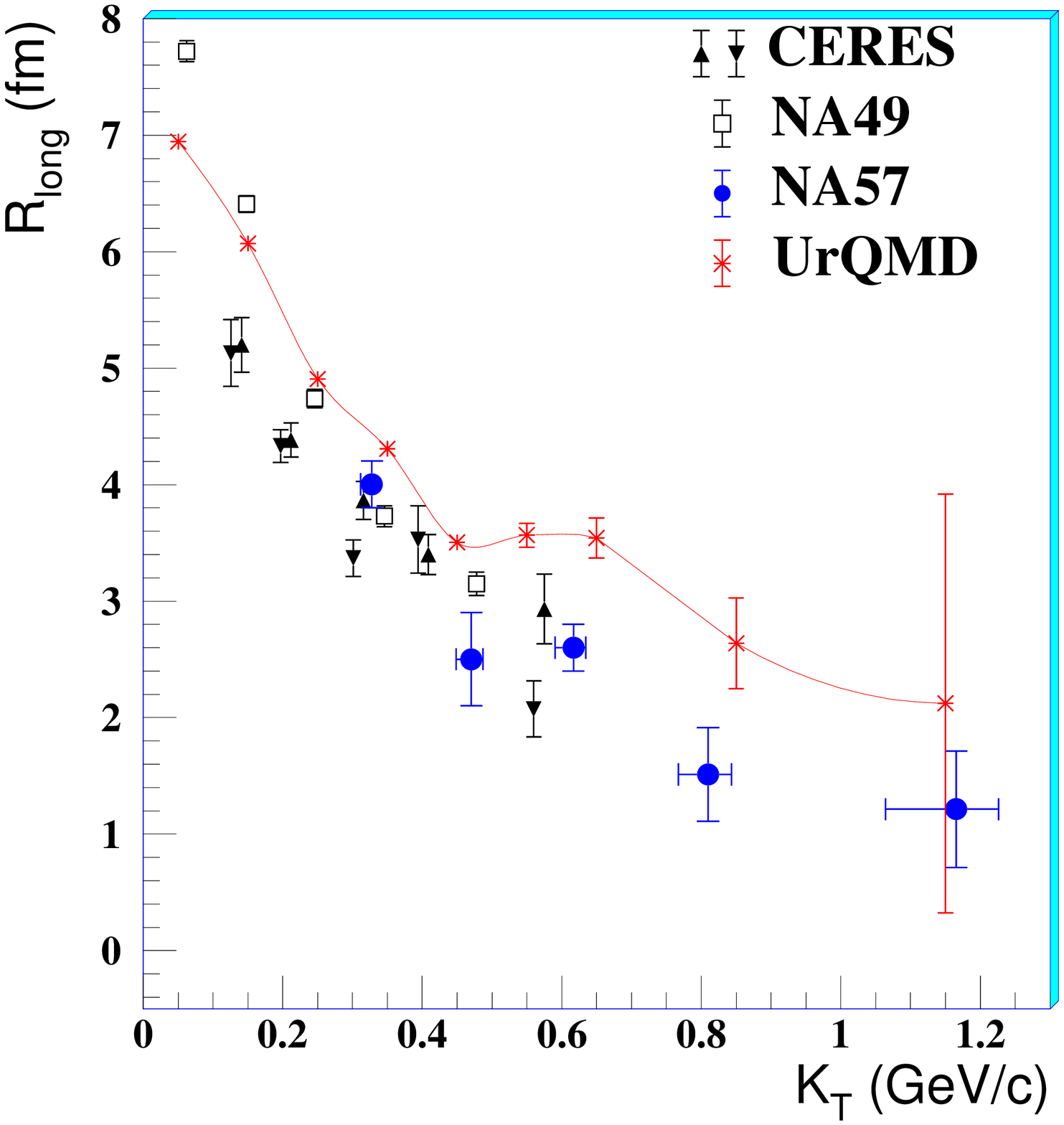}
\includegraphics{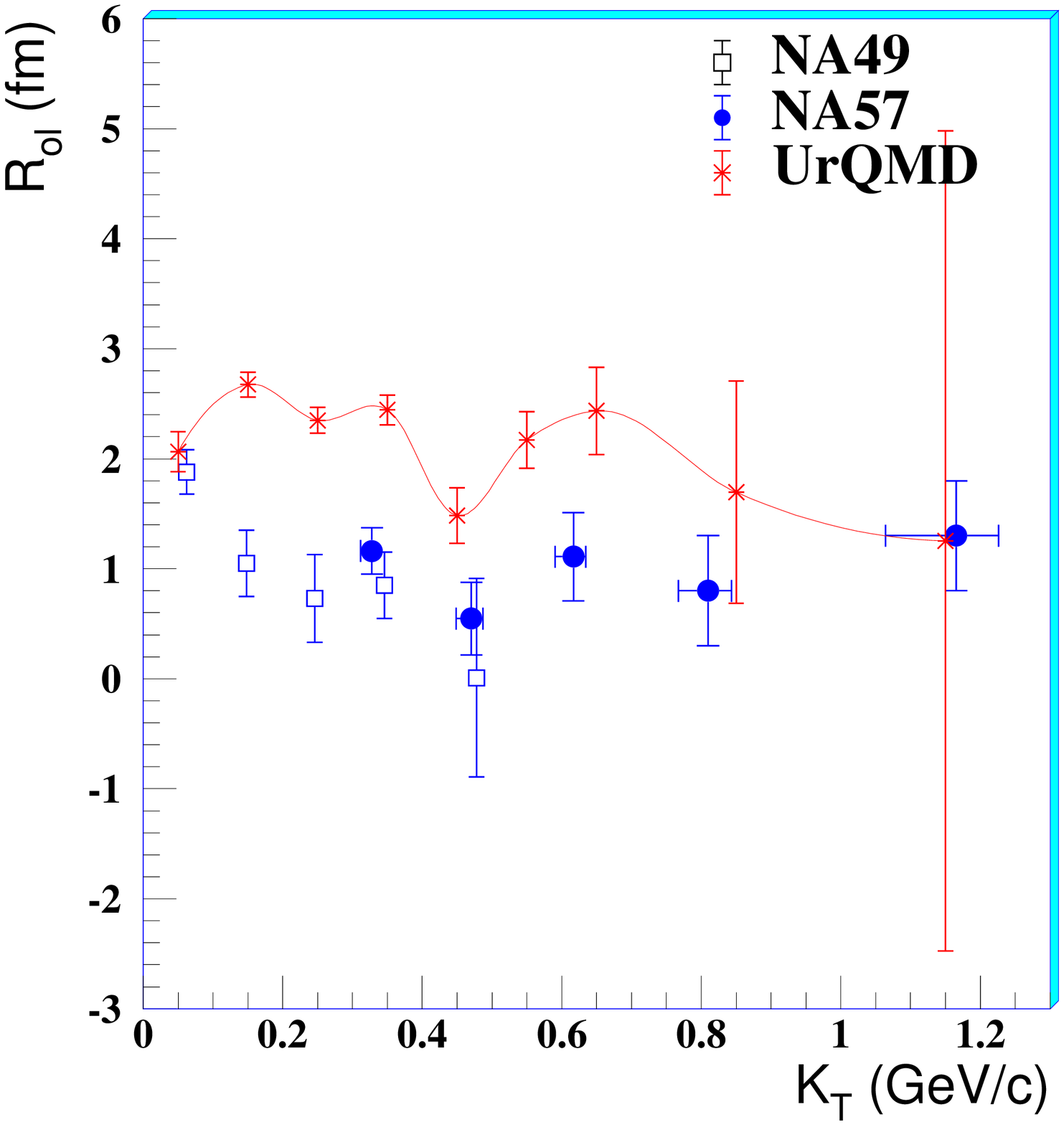}
}
\caption{The HBT radii as a function of $\Kt$\ in Pb--Pb collisions at 40 $A$\ GeV/$c$\ beam momentum. 
         The CERES data set~\cite{CEREShbt} corresponds to the most central 5\% of the 
         inelastic Au--Pb cross-section; 
         those of NA49~\cite{NA49hbt40} and NA57 correspond to the most 
         central 7.2\% and 11\% of the inelastic Pb--Pb cross-section, respectively.  
         }
\label{fig:Comp40}
\end{figure}

In figure~\ref{fig:Comp40} we show HBT radii 
versus $\Kt$\ measured in central Pb--Pb collisions at 40 $A$\ GeV/$c$\ by various experiments.   
The overall agreement of the data is good for the $\Rs$\ and $\Rl$\ radii. A discrepancy is observed for 
the $\Ro$\ radius, with the CERES results significantly smaller than the others.   
The values of the parameter $\Rol$\ depend strongly on the pair rapidity, hence a comparison with 
the CERES data, which are given at slightly backward rapidity, would not be straightforward. 
On the other hand, the agreement between NA49 and NA57 data, both shown at slightly forward rapidity 
over a similar rapidity range, is satisfactory also for the $\Rol$\ parameter.   
NA57 is the sole experiment that explores the high $\Kt$\ region and  our data suggest that the HBT 
radii $\Rs$, $\Ro$\ and $\Rl$\ decrease steadily up to  $\Kt \approx 1.2$\ GeV/$c$.  

Superimposed on the experimental data, we show in figure~\ref{fig:Comp40} 
(with asterisks, connected by lines) results by Qingfeng Li et al.~\cite{Li} 
based on the UrQMD v2.2 transport model~\cite{UrQMD}.  
The radii $\Rl$\ and $\Rs$\ are reasonably well in line with experimental data;   
the predicted $\Ro$\ values 
are larger than those of the experimental data. As a consequence, the extracted quantity 
$\sqrt{\Ro^2-\Rs^2}$\ of the pion emission source (see equation~\ref{eq:DeltaTau}) becomes 
larger than the experimental evaluation.

\begin{figure}[hbt]
\centering
\resizebox{0.46\textwidth}{!}{%
\includegraphics{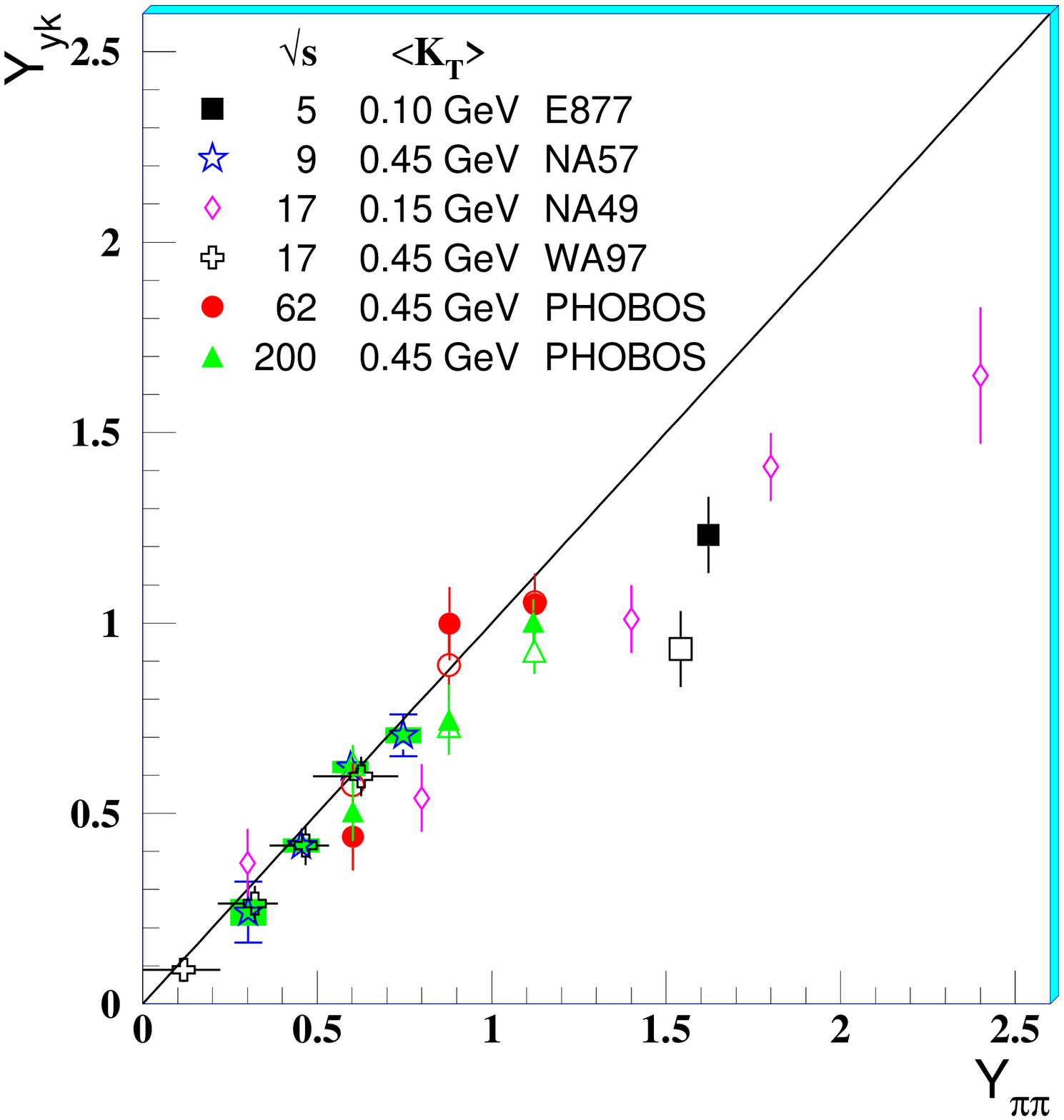}}
\caption{The Yano-Koonin rapidity plotted as a function of the pair rapidity  
for central Pb(Au)--Pb(Au) over a broad range of energies (open
symbols are for \Pgpm--\Pgpm, closed symbols for \Pgpp--\Pgpp).
Both quantities are in the center of mass frame of the colliding system. 
Data of other experiments are taken from references~\cite{E877,NA49HBT,WA97HBT,PHOBOShbt}.}  
\label{fig:CompRap}
\end{figure}

In figure~\ref{fig:CompRap} we show the plot of the Yano-Koonin rapidity versus the pair rapidity,  
both evaluated in the centre-of-mass rest frame, as measured at different energies. 
Figure~\ref{fig:CompRap} reveals a roughly universal dependence of $Y_{\rm yk}$\ on $Y_{\pi\pi}$\ 
for pions from 
central collisions, depending weakly, if at all, on $\sqrt{s_{\rm NN}}$.  
This trend is particularly striking given the very different centre-of-mass projectile
rapidities ($\approx$ 1.57 and 5.5 for $\sqrt{s_{\rm NN}}$ = 5 and 200 GeV, respectively) and
the corresponding widths of the pion distributions ${\rm d}N/{\rm d}y$.

\begin{figure}[hbt]
\centering
\resizebox{0.65\textwidth}{!}{%
\includegraphics{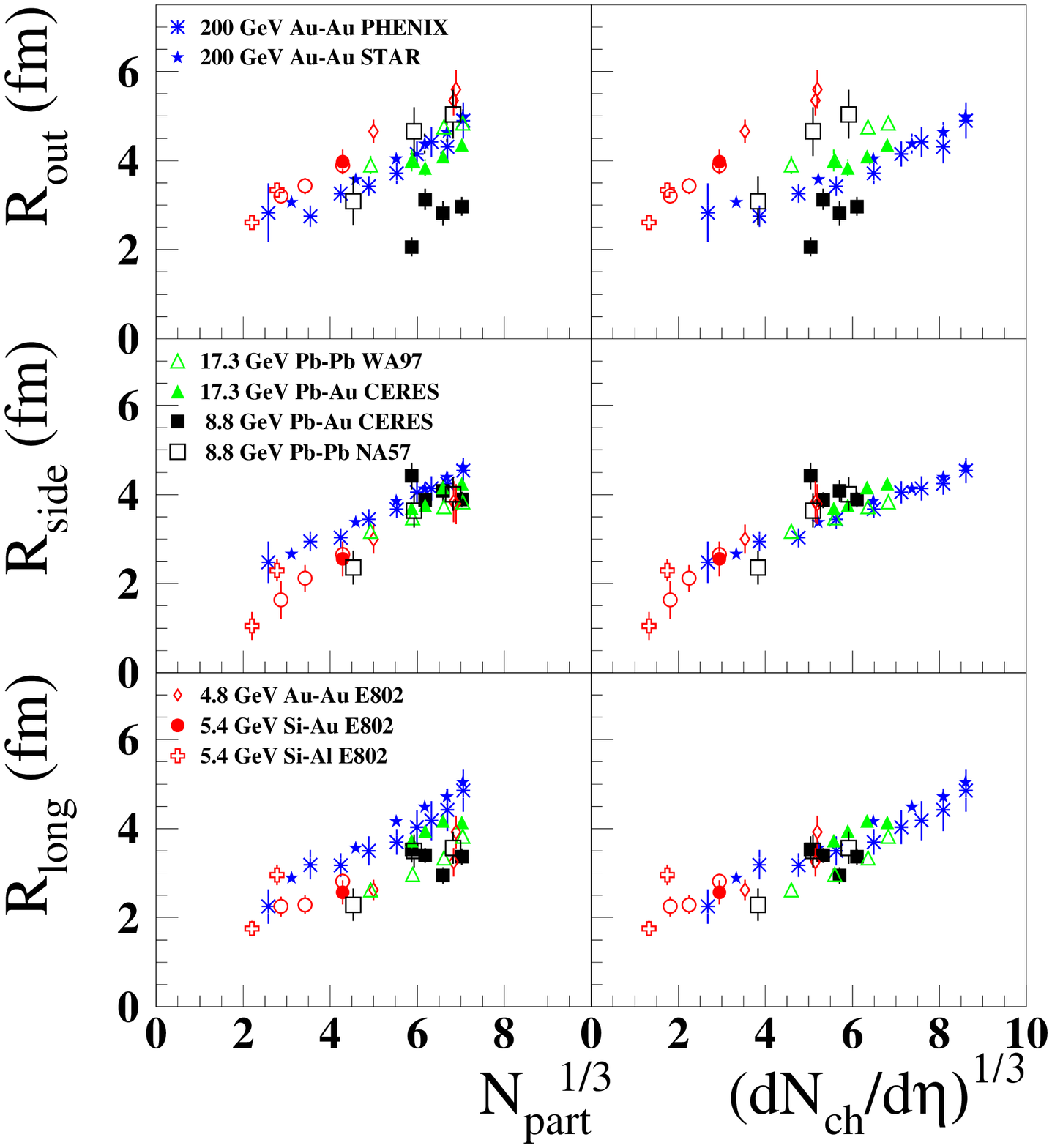}}
\caption{The HBT radius dependence on number of participants (left) and on charged 
particle multiplicity (right). Data are for Pb(Au)--Pb(Au)  collisions
at several values of $\sqrt{s_{\rm NN}}$, and also for Si--A collisions at the lowest energy.  
Average transverse momentum $\langle K_{\rm T} \rangle$\ is $\sim$450 MeV/$c$ for 
the PHENIX data~\cite{PHENIXhbt} and $\sim$390 MeV/$c$\ for the others~\cite{E802,CEREShbt,WA97HBT,STARhbt}.
Open symbols are for \Pgpm--\Pgpm, closed symbols for \Pgpp--\Pgpp, but 
for the STAR, PHENIX and CERES results which combine the two sets. 
Systematic and statistical errors have been added when both available.}
\label{fig:CompEne}
\end{figure}

The centrality dependence of HBT radii is shown in figure~\ref{fig:CompEne}  
for a wide range of collision energies. The left panels show the dependence on the
number of participating nucleons, $N_{\rm part}$.  
All of the radii exhibit a linear scaling in $N_{\rm part}^{1/3}$.  
Only the slope of the $\Rl$\ dependence shows a significant increase  
from the AGS to RHIC, consistent with a lifetime that increases with both centrality and $\sqrt{s_{\rm NN}}$. 
The trend of increasing $\Rl$\ with increasing  $\sqrt{s_{\rm NN}}$ is reversed for 
$\sqrt{s_{\rm NN}} < $ 5 GeV~\cite{86}. We note that $\Ro$\ 
radii from CERES at 40 $A$\ GeV/$c$\ are well below the observed systematics.  

The right panels of figure~\ref{fig:CompEne} show the same radii as a function of 
$(\diffD N_{\rm ch}/\diffD \eta)^{1/3}$.
The primary motivation for exploring the $(\diffD N_{\rm ch}/\diffD \eta)^{1/3}$\ 
dependence is its relation to the final state geometry through the density at freeze-out. 
However, the two scaling quantities are highly correlated. In fact, the values of  
$\diffD N_{\rm ch}/\diffD \eta$\ shown
on the right side of figure~\ref{fig:CompEne} are derived (as suggested in reference~\cite{RevHBT1}) 
from $N_{\rm part}$\ using the $N_{\rm part}^\alpha$\ parameterizations given in~\cite{158}, 
and conversely, the $N_{\rm part}$\ values are often calculated from multiplicity distributions  
using a Glauber model. Given this caveat, the
$\Rs$\ and $\Rl$\ values exhibit a linear dependence on $(\diffD N_{\rm ch}/\diffD \eta)^{1/3}$. 
The 
similar behaviour  
from $\sqrt{s_{\rm NN}}$ of 5 to 200 GeV leads one to
believe that the approximate $N_{\rm part}$\ scaling (initial overlap geometry) is a result
of the scaling with multiplicity (final freeze-out geometry) and not the other way
around.  

In table~\ref{tab:comparison} we present, along with our results, a compilation of results 
on the source parameters obtained at the SPS and RHIC from blast-wave fits to HBT correlation functions.   
At all collision energies 
it is observed that, with increasing centrality:  
(i)~the geometrical transverse dimension of the system at freeze-out, the average transverse flow, 
the proper time of the freeze-out and the duration of pion emission increase; (ii)~the 
thermal freeze-out temperature decreases.   

With increasing $\sqrt{s_{\rm NN}}$, both the proper time of the freeze-out (which can be assumed  
as the duration of the expansion) and the average transverse flow velocity increase, 
thus resulting in a stronger increase of the transverse radius parameter $R_{\rm G}$\ 
at the freeze-out: after a faster expansion which lasts longer, the system ends up showing 
much larger spatial extent.  
The freeze-out temperature at top RHIC energy is also significantly smaller than that measured 
at SPS for a given collision centrality. This can be explained again as due to a 
longer and faster expansion, which causes the system to cool down more effectively.  
On the other hand, 
similar freeze-out emission durations of a few fm/$c$\ are observed independently of the collision energy.  

\section{\label{conc} Conclusions}
The analyses of transverse mass spectra of negatively charged hadrons and 
of Hanbury-Brown and Twiss correlation functions have provided a description of 
the later stages of the expansion dynamics of the system formed in Pb--Pb collisions
at 40 $A$\ GeV/$c$. 
They have been performed 
over a centrality range corresponding to the most central 53\% of the inelastic 
Pb--Pb cross section.  

Based on a model derived from hydro-dynamics, we have determined parameters 
describing the spatial dimension and the dynamical/thermal state of the system, 
namely the transverse radius and the temperature of the fireball at kinetic freeze-out, 
the average velocity of the collective flow in the transverse direction,  
the proper time of the freeze-out and the mean duration of particle emission. 
For central collisions, the system undergoes a strong collective expansion in the transverse 
directions, with an average velocity of about 45\% of the speed of light, which lasts for about 
7 fm/$c$. After this expansion, the system has doubled its transverse size 
and it freezes out at a temperature of about 110 MeV.     
For less central collisions, the expansion proceeds on a smaller scale; the
higher temperature at freeze-out may be interpreted as the remnant of an 
earlier decoupling of the expanding system. 

Concerning the time evolution, we have deduced that $\hm$\ emission  
is a fast process ($\Delta\tau \approx 1-2$\ fm/$c$) starting after $∼$7.5 fm/$c$\ (5 fm/$c$)
in the case of central (semi-peripheral) collisions, that takes places in the bulk 
(i.e. it is not surface dominated).  
This resembles the decoupling process of photons in the early universe.

In the longitudinal direction, the indications are that particles emitted at a given  
rapidity are produced by 
source elements  
moving collectively at the same rapidity. 

The kinetic freeze-out conditions for $\hm$, namely temperature and transverse flow, 
are compatible with those obtained, within the same model, by studying the $\mt$\ 
spectra of singly-strange particles (\PKzS, \PgL\ and \PagL)~\cite{Blast40}.   

Quite universal kinetic freeze-out conditions are observed with increasing $\sqrt{s_{\rm NN}}$\ from SPS 
to RHIC, with moderate increases (by less than 20\%) of the transverse flow velocity and 
of the duration of the expansion. The latter effect is consistent with the lower 
(by $\approx$20\%) 
freeze-out temperatures measured at top RHIC energy ($\sqrt{s_{\rm NN}}=200$\ GeV); 
on the other hand, the combined effect of a faster and longer expansion 
causes the final transverse size at freeze-out to be significantly larger 
(about a factor of two for central collisions)  
at top RHIC energy than at SPS.  
%
%


\section*{References}
\begin{thebibliography}{33}
%
%
\bibitem{lattice} Karsch F 2002 {\it Lect. Notes Phys.} {\bf 583} 209 
\bibitem{QM04-QM05} Ritter H G and Wang X-N (ed) 2004 \JPG {\bf 30} S633-S1430 
({\it Proc. Quark Matter 2004}) \nonum
Cs\"{o}rgo T, D\'{a}vid G, L\'{e}vai P and Papp G (ed) 2006 \NP A {\bf 774} 1-968
({\it Proc. Quark Matter 2005}) 
\bibitem{Blast40} Antinori F {\it et al.} (NA57 Collaboration) 2006 \JPG 32 xxx-yyy
\bibitem{BlastRef} Schnedermann E, Sollfrank J and Heinz U 1993 \PR C
                   {\bf 48} 2462 \nonum
                   Schnedermann E and Heinz U 1994 \PR C {\bf 50} 1675
\bibitem{HBT}  Hanbury-Brown R and Twiss R Q 1954 {\it Phil. Mag.} {\bf 45} 633 \nonum
               Goldhaber G, Goldhaber S, Lee W and Pais A 1960 \PR {\bf 178} 300
\bibitem{enh160} Antinori F {\it et al.} (NA57 Collaboration) 2006 \JPG {\bf 32} 427-441
\bibitem{BlastPaper} Antinori F {\it et al.} (NA57 Collaboration) 2004 \JPG {\bf 30} 823-840 
\bibitem{MANZ} Manzari V {\it et al.} 1999 \JPG {\bf 25} 473   \nonum
               Manzari V {\it et al.} 1999 \NP A {\bf 661} 761c
\bibitem{Multiplicity} Antinori F {\it et al.} (NA57 Collaboration) 2005 \JPG {\bf 31} 321-335
%
\bibitem{RevHBT1} Lisa M, Pratt S, Soltz R and Wiedemann U A 2005 
                  {\it Ann. Rev. Nucl. Part. Sci.} {\bf 55} 357-402, nucl-ex/0505014
\bibitem{RevHBT2} Tomasik B and Wiedemann U A 2003 in {\it Quark Gluon Plasma 3},  
 Hwa R C and Wang X N (ed) {\it World Scientific}, hep-ph/0210250
%
\bibitem{ReviewHydro} Hirano T 2004 \JPG {\bf 30} S845-S851 \nonum
   Torrieri G and Rafelski J \JPG {\bf 30} S557-S564 \nonum
   Heinz U W 2005 \JPG {\bf 31} S717-S724 \nonum
   Csernai L P, Moln\'ar E, Ny\'iri \'A and Tamosiunas K 2005 \JPG {\bf 31} S951-S957 \nonum
   Steinberg P A 2005  \NP A {\bf 752} 423c-432c 
\bibitem{NA49yield} Afanasiev S V {\it et al.} 2002 \PR C {\bf 66} 054902 \nonum
                        Anticic T {\it et al.} 2004 \PR C {\bf 69} 024902
%
\bibitem{NA49HBTkaon} Afanasiev S V \etal (NA49 Collaboration) 2003 \PL B {\bf 557} 157-166
\bibitem{NA35}
Bamberger A \etal (NA35 Collaboration) 1988 \PL B {\bf 203} 320-326 \nonum
Bamberger A \etal (NA35 Collaboration) 1988 \ZP C {\bf 38} 79 \nonum
Alber T \etal (NA35 Collaboration) 1995 \ZP C {\bf 66} 77 \nonum
Alber T \etal (NA35 Collaboration) 1995 \PRL {\bf 74} 1303
\bibitem{NA49HBT} Appelsh\"{a}user H {\it et al.} (NA49 Collaboration) 1998 {\it  The Eur. Phys. J.} C {\bf 2} 661
\bibitem{WA97HBT} Antinori F {\it et al.} (WA97 Collaboration) 2001  \JPG {\bf 27} 2325-2344
\bibitem{ref:17} Boal D, Gelbke C K and Jennings B K 1990 {\it Rev. Mod. Phys.} {\bf 62} 553
\bibitem{GEANT} Brun R, Bruyant F, Maire M, McPherson A C and Zanarini P 1985 {\it GEANT3 User Guide}
 CERN Data Handling Division, DD/EE/84-1, \\ http://wwwinfo.cern.ch/asdoc/geantold/GEANTMAIN.html \nonum
Brun R {\it et al} 1994 {\it GEANT Detector Description and Simulation Tool} CERN Program Library
Long Write-up W5013
\bibitem{Bowler-Sinyukov} Bowler M G 1991  \PL B  {\bf 270} 69 \nonum
        Sinyukov Yu M \etal. 1998 \PL B {\bf 432} 249
\bibitem{CEREShbt}  Adamova D \etal (CERES Collaboration)
                    2003 \NP  A {\bf 714} 124
\bibitem{PHOBOShbt} Back B B \etal (Phobos Collaboration)
                    2006  \PR C {\bf 73} 031901
\bibitem{STARhbt} Adams J \etal (STAR Collaboration) 
 2005 \PR C {\bf 71} 044906
\bibitem{Gamov}  Gamow G 1928 \ZP {\bf 51} 204 \nonum
 Gurney R W and Condon E U 1929 \PR {\bf 33} (1929) 204
\bibitem{Gyulassy} Gyulassy M, Kauffmann S K and Wilson L W 1979 \PR C {\bf 20} 2267
%
\bibitem{Crab} Koonin S E 1977 \PL B {\bf 70} 43 \nonum
 Pratt S \etal 1994 \NP A {\bf 566} 103c
%
\bibitem{Bjorken} Bjorken J D 1983 \PR D {\bf 27} 140
\bibitem{HeinzLisa} U. Heinz U, Hummel A, Lisa M A and Wiedemann U A 2002 \PR C {\bf 66} 044903
\bibitem{YKparam} Yano F and Koonin S 1978 \PL B {\bf 78} 556 \nonum
                  Podgoretskii M I 1983 {\it Sov. J. Nucl. Phys.} {\bf 37} 272
\bibitem{SuperHeinz} Heinz U, Jacak B V 1999 {\it Ann. Rev. Nucl. Part. Sci.} {\bf 49} 529-579
\bibitem{RecProgr2} Heinz U, Tom\'{a}\v{s}ik B, Wiedemann U A
                    and Wu Y -F 1996 \PL B {\bf 382} 181
\bibitem{Anal} Wu Y-F, Heinz U, Tom\'{a}\v{s}ik B and Wiedemann U A 1998 {\it Eur. Phys. J} C {\bf 1} 599 
\bibitem{Trovami} Chapman S, Rayford Nix J and Heinz U 1995 \PR C {\bf 52} 2694  
\bibitem{Trovami2} Sinyukov Yu M 1995 {\it Hot Hadronic Matter: Theory and Experiment} (New York: Plenum)  
%
\bibitem{STARstrange}  Adams J. \etal (STAR Collaboration)  2004 \PRL {\bf 92} 182301
\bibitem{PHENIXcen} Adcox K     \etal (PHENIX Collaboration)  2004 \PR C {\bf 69} 024904
\bibitem{BRHAMScen} Arsene I    \etal (BRHAMS Collaboration) 2005 \PR C {\bf 72} 014908
\bibitem{PHENIXhbt} Adler S S \etal (PHENIX Collaboration)  2004 \PRL {\bf 93} 152302 
%
\bibitem{Sinyukov2}  Makhlin N and Sinyukov Yu M 1988 \ZP  C {\bf 39} 69 \nonum
                     Akkelin S V and Sinyukov Yu M 1995 \PL B {\bf 356} 525
%
\bibitem{4} Bertsch G 1989 \NP A {\bf 498} 173c
\bibitem{5} Pratt S 1986 \PR D {\bf 33} 1314
\bibitem{6} Bertsch G, Gong M and Tohyama M 1988 \PR C {\bf 37} 1896
\bibitem{7} Bertsch G and Brown G E 1989 \PR C {\bf 40}  1830
\bibitem{8} Rischke D 1996 \NP A {\bf 610} 88
\bibitem{9} Rischke D and Gyulassy M 1996 \NP A {\bf 608} 479
\bibitem{35} Heiselberg H and Vischer A P 1998 {\it Eur. Phys. J.} C {\bf 2} 593
%
\bibitem{NA49hbt40} Kniege S \etal (NA49 Collaboration) 2006 
 {\it AIP Conference Proceedings} {\bf 828};  
nucl-ex/0601024 \nonum
Kniege S \etal (NA49 Collaboration) 2004 \JPG 30 S1073-S1078
\bibitem{Li} Qingfeng Li, Bleicher M and St\"{o}cker H 2006 \PR C {\bf 73} 064908 \nonum
Qingfeng Li 2006 {\it private communication}
\bibitem{UrQMD} 
Bass S A \etal (UrQMD-Collaboration) 1998 {\it Prog. Part. Nucl. Phys.} {\bf 41} 255 \nonum
Bleicher M \etal (UrQMD-Collaboration) 1999 \JPG {\bf 25} 1859 \nonum
Bratkovskaya E L, Bleicher M, Reiter M, Soff S, St\"{o}cker H, van Leeuwen M, Bass S A and 
Cassing W 2004 \PR C {\bf 69} 054907\nonum
Zhu X, Bleicher M and St\"{o}cker H 2005 \PR C {\bf 72} 064911 
%
\bibitem{E877} Miskowiec D  \etal (E877 Collaboration) 1996 \NP A {\bf 610} C227 
\bibitem{E802} Ahle L 2002 \etal (E802 Collaboration) \PR C {\bf 66} 054906 
\bibitem{86} Lisa M A \etal (E895 Collaboration) 2000 \PRL {\bf 84} 2798 
\bibitem{158} Adler S S \etal (PHENIX Collaboration) 2005 \PR C {\bf 71} 034908 
%
\bibitem{star67} Adams J \etal (STAR Collaboration) 2004 \PRL {\bf 92} 112301
%
\end{thebibliography}
\end{document}